\documentclass[graybox]{svmult}

\usepackage{setspace}
\usepackage{listings}
\usepackage{ marvosym }
\usepackage{type1cm}        
\usepackage{float}
\usepackage{makeidx}         
\usepackage{graphicx}        
\usepackage{multicol}        
\usepackage[bottom]{footmisc}

\graphicspath{{Figure/}}
\usepackage{ amssymb }

\usepackage{newtxtext}       %
\usepackage{newtxmath}       
\usepackage{algorithm}
\usepackage{algorithmicx}
\usepackage{algpseudocode}
\usepackage{amsmath}
\usepackage{caption}
\floatname{algorithm}{Algorithm}

\renewcommand{\algorithmicensure}{\textbf{Output:}}

\definecolor{codegreen}{rgb}{0,0.6,0}
\definecolor{codegray}{rgb}{0.5,0.5,0.5}
\definecolor{codepurple}{rgb}{0.58,0,0.82}
\definecolor{backcolour}{rgb}{0.95,0.95,0.92}
\floatname{algorithm}{}

\lstdefinestyle{mystyle}{
  backgroundcolor=\color{backcolour},   commentstyle=\color{codegreen},
  keywordstyle=\color{magenta},
  numberstyle=\tiny\color{codegray},
  stringstyle=\color{codepurple},
  basicstyle=\ttfamily\footnotesize,
  breakatwhitespace=false,
  breaklines=true,
  captionpos=b,
  keepspaces=true,
  numbers=left,
  numbersep=5pt,
  showspaces=false,
  showstringspaces=false,
  showtabs=false,
  tabsize=2
}

\makeindex             

\usepackage{nameref}
\begin{document}

\title*{HexGen and Hex2Spline: Polycube-based Hexahedral Mesh
  Generation and Spline Modeling for Isogeometric Analysis Applications
  in LS-DYNA} \titlerunning{HexGen and Hex2Spline: LS-DYNA}
\author{Yuxuan Yu, Xiaodong Wei, Angran Li, Jialei Ginny Liu, Jeffrey
  He and Yongjie Jessica Zhang}
\authorrunning{Y. Yu, X. Wei, A. Li, J. G. Liu, J. He and Y. J. Zhang}
\institute{Y.~Yu \at Department of Mechanical Engineering, Carnegie
  Mellon University, Pittsburgh, PA 15213, USA
  \email{yuxuany1@andrew.cmu.edu} \and X.~Wei \at Institute of
  Mathematics, École Polytechnique Fédérale de Lausanne, 1015
  Lausanne, Switzerland \email{xiaodong.wei@epfl.ch} \and A.~Li \at
  Department of Mechanical Engineering, Carnegie Mellon University,
  Pittsburgh, PA 15213, USA \email{angranl@andrew.cmu.edu} \and
  J. G.~Liu \at Department of Mechanical Engineering, Carnegie Mellon
  University, Pittsburgh, PA 15213, USA \email{jialeil@andrew.cmu.edu}
  \and J.~He \at Department of Mechanical Engineering,
  Northwestern University, Evanston, IL 60208,
  USA\\
  \email{jeffreyhe2022@u.northwestern.edu} \and Y. J.~ Zhang (\Letter)  \at
  Department of Mechanical Engineering, Carnegie Mellon University,
  Pittsburgh, PA 15213, USA \email{jessicaz@andrew.cmu.edu}}
%
%
\maketitle

\abstract{ In this paper, we present two software packages, HexGen and
  Hex2Spline, that seamlessly integrate geometry design with
  isogeometric analysis (IGA) in LS-DYNA. Given a boundary
  representation of a solid model, HexGen creates a hexahedral mesh by
  utilizing a semi-automatic polycube-based mesh generation
  method. Hex2Spline takes the output hexahedral mesh from HexGen as
  the input control mesh and constructs volumetric truncated
  hierarchical splines. Through B\'{e}zier extraction, Hex2Spline
  transfers spline information to LS-DYNA and performs IGA therein. We
  explain the underlying algorithms in each software package and use a
  rod model to explain how to run the software. We also apply our
  software to several other complex models to test its robustness. Our
  goal is to provide a robust volumetric modeling tool and thus
  expand the boundary of IGA to volume-based industrial applications.
  }

\section{Introduction}

Isogeometric analysis (IGA)~\cite{Hughes} is a computational technique
that integrates computer aided design (CAD) with simulation methods
such as finite element analysis (FEA). It adopts the idea of
design-through-analysis and enables direct analysis of the designed
geometry. IGA has many advantages over traditional FEA such as exact,
smooth geometric representation and superior numerical performance.
Many software packages have been developed for IGA and there are
mainly two directions.
The first direction is to incorporate IGA with commercial finite
element software. For example, the user subroutine UEL in Abaqus is
used to define IGA elements and perform IGA in Abaqus
\cite{YicongLai2015,Lai2017}. The second direction is to develop open
software packages. GeoPDEs~\cite{deFalco20111020} and
igatools~\cite{pauletti2015igatools} work on NURBS (Non-Uniform
Rational B-Spline) patches and provide a general framework to
implement IGA methods. PetIGA~\cite{Dalcin2016} is another framework
for IGA based on PETSc~\cite{balay1997efficient}. These software
packages help boost the use of IGA in engineering
applications. However, these packages are
analysis-oriented. Currently, there is no available toolkit from the
geometric modeling side, especially for volume
parameterization. Therefore, the motivation of our work is to develop
a geometric modeling tool to bridge the gap between geometric design
and IGA analysis.

There are two major challenges in volume parameterization, control
mesh generation and volumetric spline construction. A control mesh is
generally an unstructured hexahedral (hex) mesh. Various strategies
have been proposed in the literature~\cite{ref:zhangbook}
for unstructured hex mesh generation, such as grid-based or octree-based \cite{schneiders1996grid, S97}, medial surface
\cite{Price1995,price1997hexahedral},
plastering~\cite{Blacker1991,SKOB06}, whisker weaving~\cite{FM99} and vector field-based methods \cite{nieser2011cubecover}.
These methods have created hex meshes for certain geometries, but are
not robust and reliable for arbitrary geometries. The polycube-based
method~\cite{Tarini2004,Gregson2011} is another appealing approach for
all-hex meshing. A smooth harmonic field \cite{LeiLiu2012a} was used to
generate polycubes for arbitrary genus geometries. Boolean operations \cite{LZH2014} were introduced to cope with arbitrary genus
geometries. In \cite{Liu2015}, polycube structure was generated based on
the skeleton branches of a geometric model. For these methods, the hex
mesh quality is directly affected by the polycube structure and
mapping distortion. Computing the polycube structure with a
low-distortion mapping remains an open problem for arbitrary
geometries. It is essential to improve the mesh quality for analysis
by using methods such as pillowing, smoothing and
optimization~\cite{qian2010quality,YZhang2009c,wenyan2011b,qian2012automatic}.
Pillowing is a sheet insertion technique that eliminates the
situations where two neighboring hex elements share more than one
face. Smoothing and optimization are used to further improve mesh
quality by relocating vertices. In our software, we implement all the
above mentioned methods for quality improvement.

The second ingredient in volume parameterization is volumetric spline
construction. Several algorithms have been developed.  The initial
development of IGA was based on NURBS. Since it adopts a global
tensor-product structure, it does not support local
refinement. T-splines were initially developed to support local
refinement for surfaces~\cite{ref:sederberg03,Sederberg1}. For solid models, the rational T-spline basis functions were used to convert unstructured hex meshes to solid T-splines \cite{Wenyan2011a}. Boolean operations~\cite{LZH2014} and
skeletons~\cite{Liu2015} are used to create hex meshes, which are
later converted to T-meshes. However, local refinement using T-splines
requires extensive mesh manipulation to satisfy desired properties
such as linear independence. Hierarchical spline is an alternative to
T-spline to avoid this issue. Several techniques were then developed
based on hierarchical B-splines
(HB-splines)~\cite{ref:forsey88,ref:vuong11}, such as truncated
hierarchical B-splines
(THB-splines)~\cite{ref_giannelli12,Giannelli2014}.

In this paper, we integrate our semi-automatic polycube-based mesh
generation with the volumetric truncated hierarchical spline
construction (TH-spline3D)~\cite{wei17a} to perform IGA on volumetric
models in LS-DYNA.
The developed software packages feature: 1) semi-automatic
polycube-based all-hex mesh generation from a CAD model;
2) TH-spline3D construction on hex meshes; and 3) B\'{e}zier extraction
for LS-DYNA.
We first overview the entire pipeline and explain the algorithm behind
each module of the pipeline.  We then provide various examples to
explain how to run the software package. The main objective of the
software package is to make our pipeline accessible to industrial and
academic communities who are interested in real-world engineering
applications. Our software favors versatility over efficiency. We will
use a concrete example to go through all the steps in running the
software. In particular, when user intervention is needed, we will
explain details of the involved manual work.

The paper is outlined as follows. In Section 2, we overview the
pipeline. In Section 3, we present the HexGen software package that
conducts semi-automatic polycube-based all-hex mesh generation from a
CAD file. In Section 4, we talk about Hex2Spline that constructs
TH-spline3D on hex meshes and performs B\'{e}zier extraction for IGA
in LS-DYNA. Finally, in Section 5, we demonstrate several complex
models using our software package.

\section{Pipeline design}


%

\begin{figure}[!htb]
\center{\includegraphics[width=\linewidth]{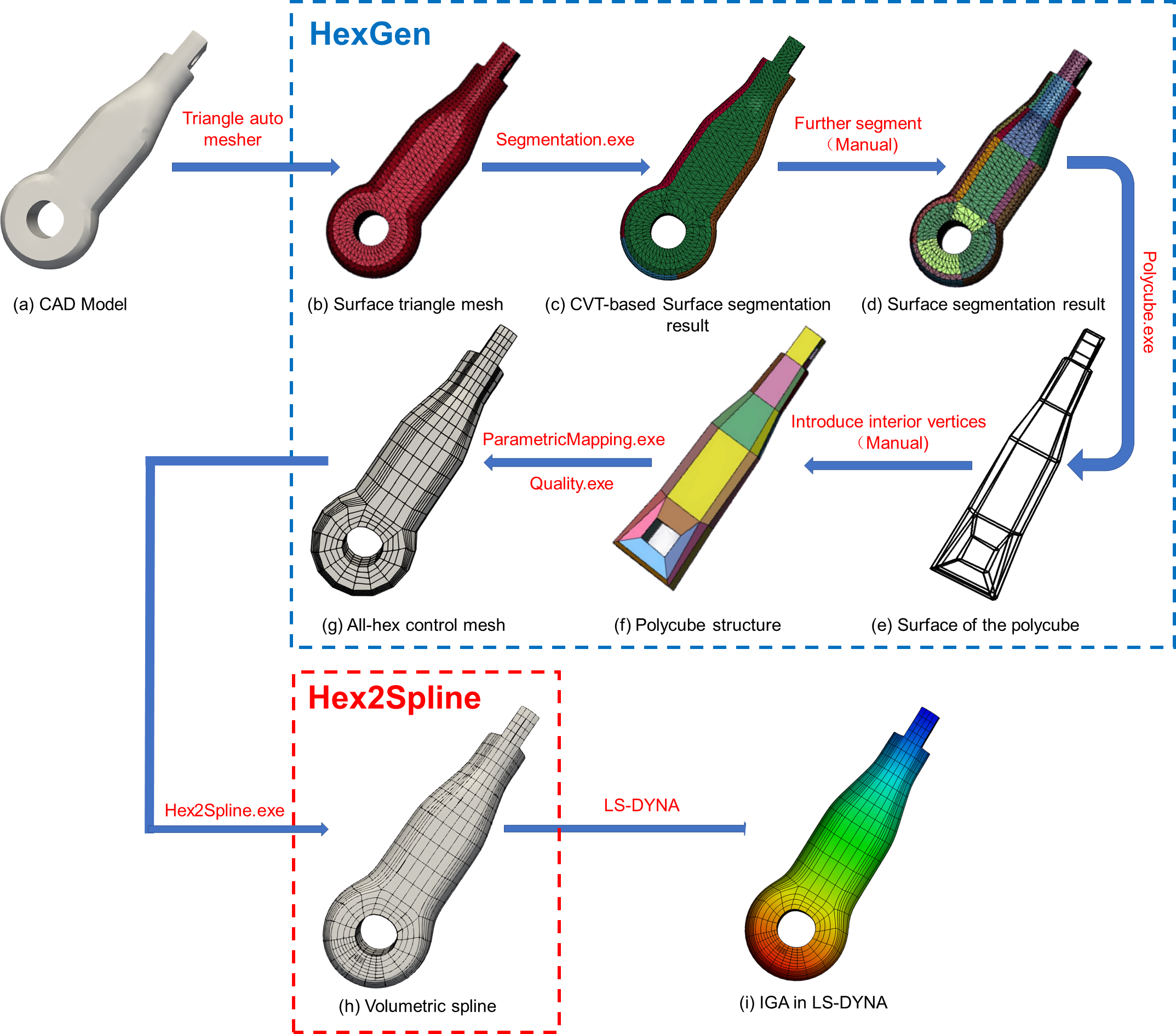}
  \caption{\label{fig:Pipeline_Structure}The HexGen software package
    and the Hex2Spline software package. The red text on each arrow shows the operation between two steps. The manual work is involved in
    \textbf{further segmentation} and \textbf{introducing interior
      vertices}.}}
\end{figure}

Our pipeline incorporates two software packages to bridge the gap
between the input CAD model with IGA in LS-DYNA, as shown in
Fig.~\ref{fig:Pipeline_Structure}.
We first use the HexGen software package to build an all-hex mesh for the CAD model. With a high quality all-hex mesh generated, we then use the Hex2Spline software package to construct
TH-spline3D and extract B\'{e}zier information for LS-DYNA.

As shown in Fig.~\ref{fig:Pipeline_Structure}, we first generate a
triangle mesh from the CAD model by using the free software
LS-PrePost, which is a pre and post-processor for LS-DYNA. Then we use
centroidal Voronoi tessellation (CVT) segmentation~\cite{HZ2015CMAME}
to create a polycube structure~\cite{Tarini2004}, which is used to
generate all-hex meshes via parametric mapping~\cite{floater1997} and
octree subdivision~\cite{wenyan2011b}. The quality of the all-hex mesh
is evaluated to ensure that the resulting volumetric spline model can
be used in IGA.  In case that a poor quality hex mesh is generated,
the program has several quality improvement functions, including
pillowing~\cite{YZhang2009c}, smoothing, and
optimization~\cite{qian2012automatic}. Each quality improvement
function can be run independently and one can use these functions to
improve the mesh quality.

Once a good quality hex mesh is obtained, one can run the Hex2Spline
program to build volumetric splines. In particular, TH-spline3D is
built on the unstructured hex mesh and it also supports local
refinement. The Hex2Spline can output the B\'ezier extraction
information of TH-spline3D in a format that can be imported into
LS-DYNA to perform IGA.
Currently, our software only has a command-line interface (CLI). Users
need to specify necessary options via the command line to run the
software.
In Sections~\ref{sec:hexg-polyc-based}
and~\ref{sec:hex2spl-unstr-spline}, we will explain the algorithms
implemented in each software as well as how to run the software in
detail.

\section{HexGen: Polycube-based hex mesh generation}
\label{sec:hexg-polyc-based}


Surface segmentation, polycube construction, parametric mapping, and
octree subdivision are used together in the HexGen software package to
construct an all-hex mesh from the boundary representation given by
the input CAD model. Given an triangle mesh generated from the CAD
model, we first use surface segmentation to divide the mesh into
several surface patches that satisfy the polycube structure
constraints, which will be discussed in
Section~\ref{sec:surface-partition}. Then, the corner vertices, edges
and face information of each surface patch are extracted from the
surface segmentation result to construct a polycube structure. Each
component of the polycube structure is topologically equivalent to a
cube. Finally, we generate the all-hex mesh through parametric mapping
and octree subdivision. Quality improvement techniques can be used to
further improve the mesh quality.

In this section, we introduce the main algorithm for each module of
the HexGen software package, namely surface segmentation, polycube
construction, parametric mapping and octree subdivision, and quality
improvement. We use a rod model (see Fig.~\ref{fig:Pipeline_Structure}) to
explain how to run CLI for each module. We also discuss the user
intervention that is involved in the semi-automatic polycube-based hex
mesh generation.

\subsection{Surface segmentation}
\label{sec:surface-partition}

The surface segmentation in the pipeline framework is implemented
based on CVT segmentation~\cite{HZ2015CMAME}. CVT segmentation is used
to classify vertices into different groups by minimizing an energy
function. Each group is called a Voronoi region $\{ {V_j}\} $ and it
has a corresponding center called a generator $\{ {g_j}\}$. The
Voronoi region and its corresponding generator are updated iteratively
in the minimization process. In~\cite{HZ2015CMAME}, each element of
the surface triangle mesh is assigned to one of the six Voronoi
regions $\{ {V_j}\} _{j = 1}^6$ based on the normal vector
${\varkappa_{_{\mathcal{T}(i)}}}$ of the surface, where
${{\mathcal{T}(i)}}$ is the $i^{th}$ element of the surface
triangle mesh ${{\mathcal{T}}}$. The initial generators of the Voronoi
regions are the three principal normal vectors and their opposite
normals vectors ($ \pm X$, $ \pm Y$, $ \pm Z$). Two energy functions
and their corresponding distance functions are used together
in~\cite{HZ2015CMAME}. The classical energy function and its
corresponding distance function provide initial Voronoi regions and
generators. Then the harmonic boundary-enhanced (HBE) energy function
and its corresponding distance function are applied to eliminate
non-monotone boundaries. The detailed definitions of energy functions
and their corresponding distance functions are described in
\cite{HZ2015CMAME}.
Here, we summarize the surface segmentation process in\nameref{alg:1}.

\begin{algorithm}[H]
  \caption{\textbf{ Surface Segmentation Algorithm}}
  \label{alg:1}
  \begin{algorithmic}[1]
    \Require Manifold triangular surface mesh $\mathcal{T}$, weighting
    factor $\omega$
    \Ensure Manifold triangular surface mesh including
    segmentation information

    \State Calculate the unit normal ${\varkappa_{_{\mathcal{T}(i)}}}$
    of the triangle mesh \State Use six principal axes ($ \pm X$,
    $ \pm Y$, $ \pm Z$) as the initial generators
    $\{{g_j}\} _{j = 1}^6$
    \renewcommand{\algorithmicensure}{\textbf{CVT step:}}
    \Ensure
    \While { classical energy not converge}
    \State Associate ${\varkappa_{_{\mathcal{T}(i)}}}$ with $\{ {g_j}\} _{j = 1}^6$ by using classical distance functions
    \For{each unit normal ${\varkappa_{_{\mathcal{T}(i)}}}$ in group $\{ {V_j}\} $}
    \State
    Update generators $ {g_j }$ based on classical energy function
    \EndFor
    \EndWhile
    \renewcommand{\algorithmicensure}{\textbf{HBE CVT step:}}
    \Ensure
    \State Use CVT results as the input for HBE CVT
    \While {HBE energy not converge}
    \State Associate ${\varkappa_{_{\mathcal{T}(i)}}}$ with
    $\{ {g_j}\} _{j = 1}^6$ by using the HBE distance functions  controlled by weighting
    factor $\omega$
    \For{each unit normal ${\varkappa_{_{_{\mathcal{T}(i)}}}}$ in group
      $\{ {V_j}\} $}
    \State Update generators ${g_j}$ based
    on HBE energy function
    \EndFor
    \EndWhile
  \end{algorithmic}
\end{algorithm}

Through the above pseudocode in\nameref{alg:1}, we describe two energy
minimization processes, which are combined together to yield a
monotone segmentation. When we use the HBE distance function to define
Voronoi regions, we use a weighting factor $\omega$ to control the
balance between the classical distance and the boundary-enhanced term
(see Eq.~4 in~\cite{HZ2015CMAME}).

Based on\nameref{alg:1}, we implement and organize the code into a CLI
program (Segmentation.exe), which can segment a given triangle mesh
into 6 Voronoi regions. Users can give options through the command
line to run Segmentation.exe.
Taking the rod model as an example, we first generate a triangle
mesh from its CAD model by using LS-PrePost. Then we segment the
triangle mesh by running the following command:
\begin{lstlisting}[style=mystyle]
Segmentation.exe -i rod_tri.k -o rod_initial_write.k -m rod_manual.txt -l 0.1
\end{lstlisting}
There are four options used in the command:
\begin{itemize}
    \item \textbf{-i}: Surface triangle mesh of the input geometry (rod\_tri.k);
    \item \textbf{-o}: Output segmentation result (rod\_initial\_write.k);
    \item \textbf{-m}: Input file with user intervention (rod\_manual.txt); and
    \item \textbf{-l}: Weighting factor $\omega$ used in
      \textbf{HBE distance function}.
\end{itemize}
The input and output are .k files, which can be read by LS-PrePost. We
refer readers to~\cite{manual2007version}, which explains the file
format.  We use \textbf{-l} to control the balance between the
distance and the boundary-enhanced term. The weighting factor $\omega$
can be assigned any arbitrary positive value; however, to obtain the
best segmentation behavior, $\omega$ must take small value. We find
that when $\omega=0.1$, the segmentation result of the rod model has
fewer zig-zags and outliers. Users need to do a trial and error to
obtain a good weighting factor. Note that zig-zags and outliers may
still exist regardless of the choice of $\omega$.
To fix this issue, user intervention is needed to prepare a file that
stores the correct segmentation result for such
elements. Segmentation.exe can read this file through option
\textbf{-m} to improve the segmentation result. The snippets of the
input text file for the rod model are given in Appendix A1.


Once we get the initial segmentation result, we need
to further segment each Voronoi region into several patches to satisfy
the topological constraints for polycube construction (see
Fig.~\ref{fig:Pipeline_Structure}(d)).
The following three conditions should be satisfied during the further
segmentation: 1) two patches with opposite orientations (e.g., +X and
-X) cannot share a boundary; 2) each corner vertex must be shared by
more than two patches; and 3) each patch must have four boundaries.
Note that we define the corner vertex as a vertex locating at the
corner of the cubic region in the model.

The further segmentation is done manually by using the patch ID
reassigning function in LS-PrePost. The detailed operation is shown in
Appendix A2.


In addition to the issue with zig-zags and outliers, the algorithm has several limitations. For example, \nameref{alg:1} cannot guarantee a good quality polycube structure,
which will affect the quality of hex mesh. Elements with small or
negative scaled Jacobian in a hex mesh may appear. Some adjustments on the polycube structure and quality improvement are needed as a follow-up step.
\subsection{Polycube construction}
\label{sec:poly_construction}

In this section, we discuss the detailed algorithm of polycube
construction using the segmented triangle mesh. A polycube
consisting of multiple cubes is topologically equivalent to the
original geometry. Several automatic polycube construction algorithms
have been proposed in the
literature~\cite{He2009369,LinJ2008,HZ2015CMAME}, but it is
challenging to generalize these methods to general CAD models.
To achieve versatility for real industrial applications, we develop a
semi-automatic polycube construction software based on the segmented
surface. However, for some complex geometries, it may slow down the
process because of the potentially heavy user intervention.

The key information we need for a polycube is its corners and the
connectivity relationship among them. For the surface of polycube, we
can automatically get the corners and build their connectivity based
on the segmentation result by using~\nameref{alg:2}. However, it is
usually difficult to obtain inner vertices and their connectivity as
we only have a surface input without any information about the
interior volume. Indeed, this is also the place that involves user
intervention, where we use LS-PrePost to manually build the interior
connectivity. The detailed operation is shown in Appendix A3. As the auxiliary information for this user intervention,
\nameref{alg:2} will output corners and connectivity of the segmented
surface patches into .k file.
Finally, the generated polycube structure is the cubic regions
splitting the volumetric domain of the geometry.

\begin{algorithm}[H]
  \caption{\textbf{Polycube Boundary Surface Construction Algorithm}}
  \label{alg:2}
  \begin{algorithmic}[1]
    \Require Manifold triangular surface mesh including
    segmentation information
        \Ensure The boundary surface of the polycube structure
    \renewcommand{\algorithmicensure}{\textbf{Computing the corners
     step:}}
    \Ensure
    \For{each vertex $v_i$}
    \State Get the number of patches $n$ which the vertex $v_i$ surrounding
    \If{$\mathbf{n}\geq 3$}
       \State Mark the vertex as a corner $v_i^c$
    \EndIf
    \EndFor
    \State Output file including corner coordinates
    \renewcommand{\algorithmicensure}{\textbf{Computing the
       index array step:}}
    \Ensure
    \For{each patches $\{ {S_j}\} $}
    \State Find its four corners $v_i^c$ which define a quad $\mathcal{Q}_j$
    \EndFor
    \State Extract edge information from $\mathcal{Q}_j$
    \State Output files of connectivity relationship including edges and faces
  \end{algorithmic}
\end{algorithm}
We implement and organize the code into a CLI program (PolyCube.exe)
based on~\nameref{alg:2}. For the rod model, we run the following
command to extract the corners and their connectivity for the boundary surface of its polycube:
\begin{lstlisting}[style=mystyle]
PolyCube.exe -i rod_initial_read.k -o rod_polycube_structure.k -c 1
\end{lstlisting}
There are four options used in the command:
\begin{itemize}
    \item \textbf{-i}: Surface triangle mesh with the segmentation
      information (rod\_initial\_read.k);
    \item \textbf{-o}: Polycube surface connectivity
      (rod\_polycube\_structure.k) for polycube construction in
      LS-PrePost; and
    \item \textbf{-c}: Control indicator if some additional file needs to be output
        \begin{itemize}
    \item -c 0: No output; and
    \item -c 1: Output corner points, edges, faces of polycube structure.
\end{itemize}
\end{itemize}

\begin{figure}[htp]
\centering
\begin{tabular}{ccc}
\includegraphics[height=0.46\linewidth]{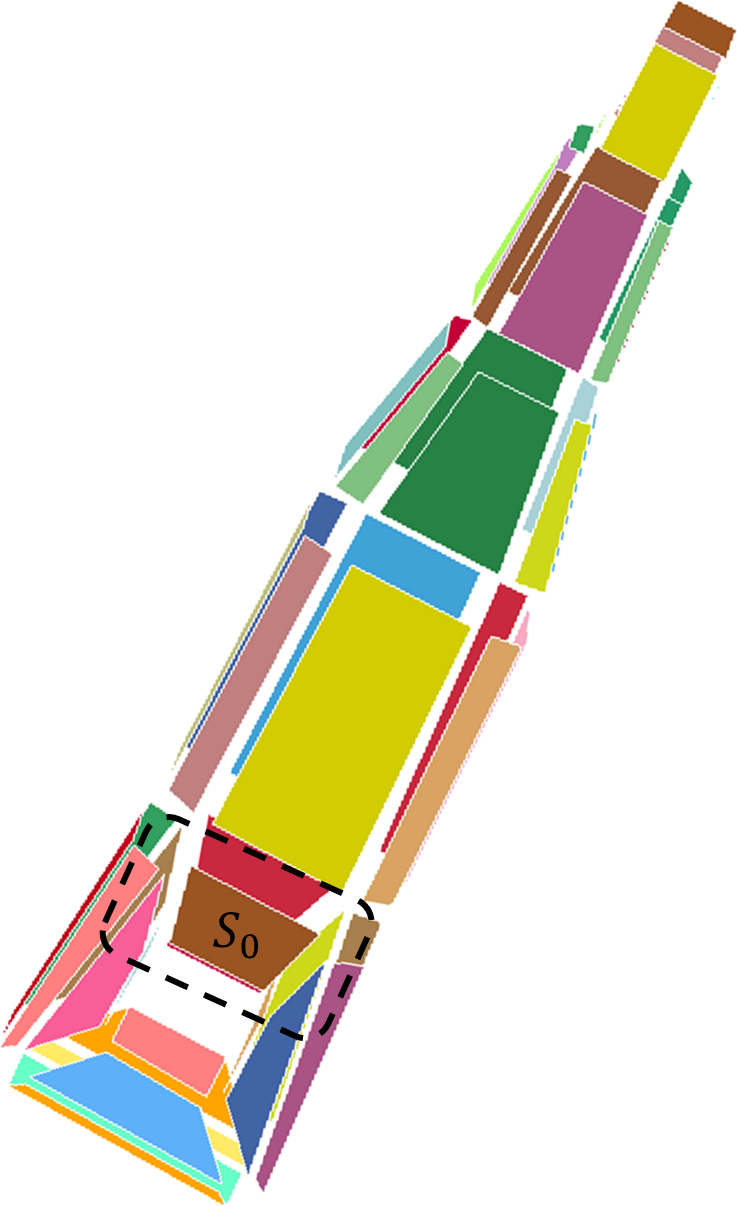}&
\includegraphics[height=0.44\linewidth]{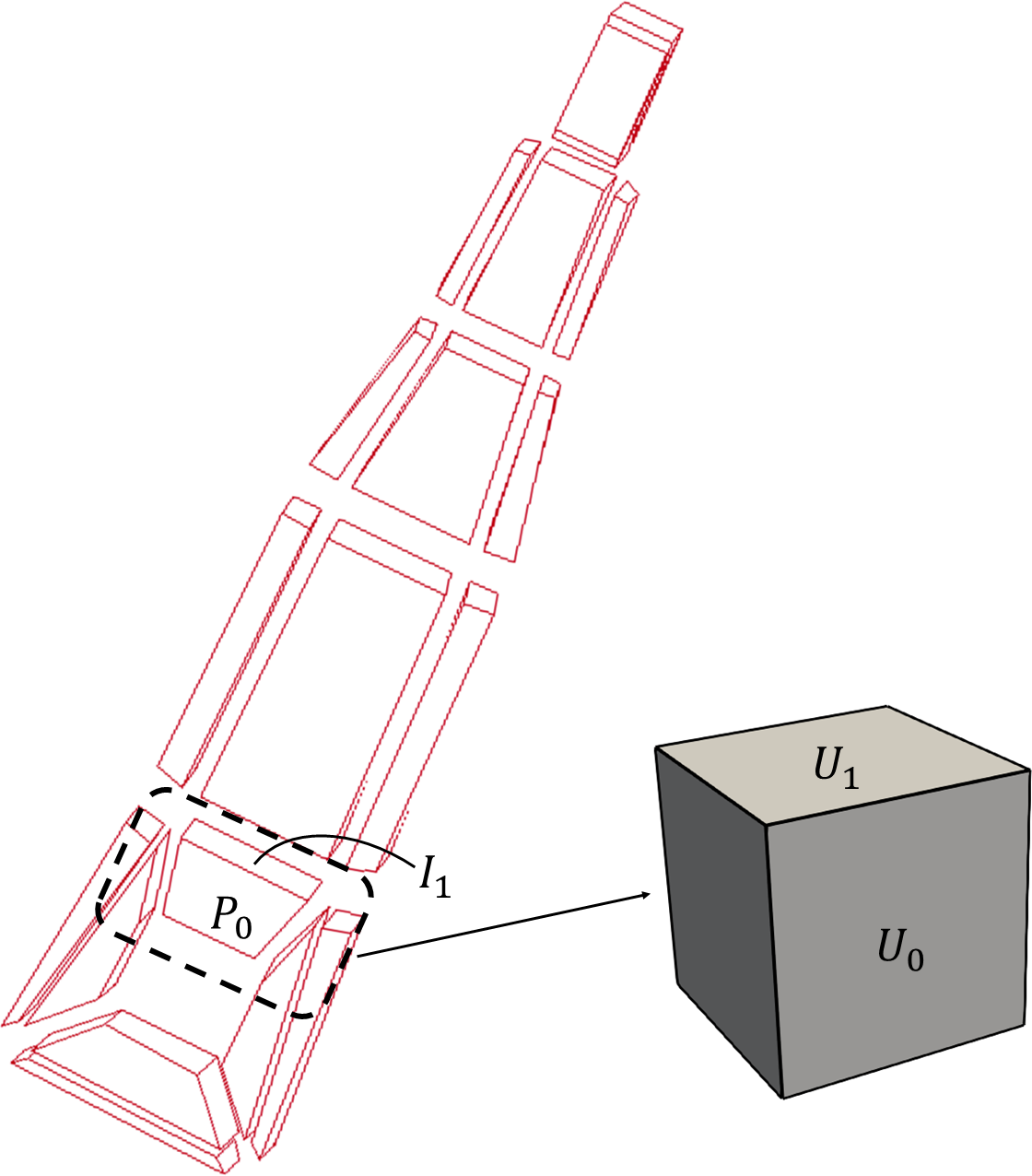}&\\
(a) & (b)\\
\end{tabular}
\caption{The polycube construction and the parametric mapping
  process. (a) The boundary surface of the polycube generated
  by~\nameref{alg:2}; (b) The interior corners and connectivity
  manually created in LS-PrePost to generate polycube structure. We
  split the polycube into multiple cubes and  use each individual cube
  as the parametric domain. $S_0$, $P_0$ and $U_0$ are used for
  parametric mapping. $I_1$ and $U_1$ are used for linear
  interpolation to create interior vertices of the mesh. }
    \label{fig:rod_polycube_construction}
\end{figure}

The output .k file contains the corners and their connectivity for the
boundary surface of the polycube (see
Fig.~\ref{fig:rod_polycube_construction}(a)). Users need to import it
into LS-PrePost and manually create interior corners and corresponding
connectivity (see Fig.~\ref{fig:rod_polycube_construction}(b)) to
build a polycube structure (see
Fig.~\ref{fig:Pipeline_Structure}(f)). We also provide option
\textbf{-c} to output the corners, edges, and faces of the polycube
structure if users intend to use other software to build a polycube
structure. Users can find their file format in Appendix A4.

\subsection{Parametric mapping and octree subdivision}
\label{sec:octree_subdivision}

After the polycube is constructed, we need to build a bijective
mapping between the input triangle mesh and the boundary surface of
the polycube structure. In our software, we implement the same idea as
in \cite{Liu2015} to use a unit cube as the parametric domain for
polycube structure. As a result, we can construct a generalized
polycube structure (see Fig.~\ref{fig:rod_polycube_construction}(b)) that
can align with the given geometry better and generate a high quality
hex mesh.

Through the pseudocode in~\nameref{alg:3}, we describe how the
segmented surface mesh, the polycube structure and the unit cube are combined
to create a (volume) parametric mapping and octree subdivision. Let
$\{S_i\}_{i=1}^N$ be the segmented surface patches coming from the
segmentation result (see Fig.~\ref{fig:rod_polycube_construction}(a)).
Each segmented surface patch corresponds to one boundary surface of
the polycube $P_i$ $(1\leq i \leq N)$ (see
Fig.~\ref{fig:rod_polycube_construction}(b)), where $N$ is the number of the
boundary surface. There are also interior surfaces, denoted by
$I_j$ $(1\leq j \leq M)$, where $M$ is the number of the interior
surface. The union of $\{P_i\}_{i=1}^N$ and $\{I_j\}_{j=1}^M$ is the
set of surfaces of the polycube structure. For the parametric domain,
let $\{U_k\}_{k=1}^6$ denote the six surface patches of the unit cube
(see Fig.~\ref{fig:rod_polycube_construction}(b)).


Each cubic region in the polycube structure represents one volumetric
region of the geometry and has a unit cube as its parametric
domain. Fig.~\ref{fig:rod_polycube_construction}(b) shows the example
of one cubic region and its corresponding volume domain of the
geometry marked in the dashed rectangle. Therefore, for each cube in
the polycube structure, we can find its boundary surface $P_i$ and map
the segmented surface patch $S_i$ to its corresponding parametric
surface $U_k$ of the unit cube.  To map $S_i$ to $U_k$, we first map
its corresponding boundary edges of $S_i$ to the boundary edges of
$U_k$. Then we get the parameterization of $S_i$ by using the
cotangent Laplace operator to compute the harmonic
function~\cite{wenyan2011b,eck1995multiresolution}. Note that for an
interior surface $I_j$ of the polycube structure, we skip the
parametric mapping step.

An all-hex mesh can be obtained from this surface parameterization
combined with the octree subdivision. We generate the hex element for
each cubic region in the following process. To obtain vertex
coordinates on the segmented patch $S_i$, we first subdivide the unit
cube (see Fig.~\ref{fig:rod_polycube_construction}(b)) recursively to
get their parametric coordinates. The physical coordinates can be
obtained by using the parametric mapping, which has a one-to-one
correspondence between the parametric domain $U_k$ and the physical
domain $S_i$. To obtain the vertices on the interior surface of the
cubic region, we skip the parametric mapping step and directly use the
linear interpolation to calculate the physical
coordinates. Fig.~\ref{fig:rod_polycube_construction} shows the
example of the rod model. A composition of mappings among $S_0$, $P_0$
and $U_0$ is done to build parametric mapping and obtain vertex
coordinates on the surface $S_0$. $I_1$ and $U_1$ are combined for
linear interpolation to obtain the vertices on the interior surface of
the cubic region. Finally, the vertices inside the cubic region are
calculated by linear interpolation. The entire all-hex mesh is built
by going through all the cubic regions.

Based on~\nameref{alg:3}, we implement and organize the code into a
CLI program ( ParametricMapping.exe) that can generate an all-hex mesh
by combining parametric mapping with the octree subdivision. Here, we
run the following command to generate an all-hex mesh for the rod model:
\begin{lstlisting}[style=mystyle]
ParametricMapping.exe -i rod_indexPatch_read.k -p
rod_polycube_structure.k -o rod_hex.vtk -s 2
\end{lstlisting}
There are three options used in the command:
\begin{itemize}
    \item \textbf{-i}: Surface triangle mesh of the input geometry
      with segmentation information (rod\_indexPatch\_read.k);
    \item \textbf{-o}: Unstructured hex mesh (rod\_hex.vtk);
    \item \textbf{-p}: Polycube structure
      (rod\_polycube\_structure.k); and
    \item \textbf{-s}: Octree level.
\end{itemize}
We use \textbf{-i} to set the segmentation file generated in
Section~\ref{sec:surface-partition} and use \textbf{-p} to set the
polycube structure created in
Section~\ref{sec:poly_construction}. Option \textbf{-s} is used to set
the level of recursive subdivision to be applied. There is no
subdivision if we set \textbf{-s} to be 0. In the rod model, we set
\textbf{-s} to be 2 to create a level-2 all-hex mesh.  The output
all-hex mesh is stored in the VTK format (see
Fig.~\ref{fig:Pipeline_Structure}(g)) and it can be visualized in
Paraview~\cite{ahrens2005paraview}.




\begin{algorithm}[H]
  \caption{ \textbf{Parametric Mapping Algorithm}}
  \label{alg:3}
        \begin{algorithmic}[1]
            \Require Segmented triangle mesh
            $\mathcal{T}=\{S_i\}_{i=1}^N$, polycube structure
            \Ensure All-hex mesh
            \State Find boundary surfaces $\{P_i\}_{i=1}^N$ and
            interior surfaces $\{I_j\}_{j=1}^M$ in the polycube structure
            \renewcommand{\algorithmicensure}{\textbf{Surface parameterization step:}}
            \Ensure

            \For{each cube in the polycube structure}
               \State Create a unit cube $\{U_k\}_{k=1}^6$ as the
               parametric domain
               \For{each surface in the cube}
               \If {it is a boundary surface $P_i$}

               \State Get the surface parameterization $f: S_i \to U_k \subset \mathbb{R}^2$
               \EndIf
               \EndFor
               \EndFor
               \renewcommand{\algorithmicensure}{\textbf{Parametric mapping and octree subdivision step:}}
               \Ensure
               \For{each cube in the polycube structure}
               \State Subdivide the unit cube $\{U_k\}_{k=1}^6$ recursively to get parametric coordinates $v_{_{para}}$
               \For{each surface in the cube}
               \If {it is a boundary surface $P_i$}
               \State Obtain physical coordinates using $f^{-1}(v_{_{para}})$
               \ElsIf {it is an interior surface $I_j$}
               \State Obtain physical coordinates using linear interpolation
               \EndIf
               \EndFor
                \State Obtain interior vertices in the cubic region using linear interpolation
                \EndFor
            \State Combine hex elements from each cubic region
        \end{algorithmic}
    \end{algorithm}

\subsection{Quality improvement}

If the quality of the hex mesh is not satisfactory, quality
improvement needs to be applied to the hex mesh. We integrate three
quality improvement techniques in the software package, namely
pillowing, smoothing and optimization. Users can improve mesh quality
through the command line options before building volumetric splines.

We first use pillowing to insert one layer around the
boundary~\cite{wenyan2011b}. By using the pillowing technique, we
ensure that each element has at most one face on the boundary, which
can help improve the mesh quality around the boundary. After
pillowing, smoothing and optimization~\cite{wenyan2011b} are used to
further improve mesh quality. For smoothing, different relocation
methods are applied to three types of vertices: vertices on sharp
edges on the boundary, vertices on the boundary surface, and interior
vertices. For each sharp-edge vertex, we first detect its two
neighboring vertices on the curve, and then calculate their middle
point. For each vertex on the boundary surface, we calculate the area
center of its neighboring boundary quadrilaterals (quads).  For each interior vertex, we calculate the weighted volume center of its neighboring hex elements as the new position. We relocate a vertex in an iterative way. Each time the vertex moves only a small step towards the new position and this movement is done only if the new location results in an improved local Jacobian.

\begin{figure}[htp]
\centering
\begin{tabular}{ccc}
\includegraphics[width=0.3\linewidth]{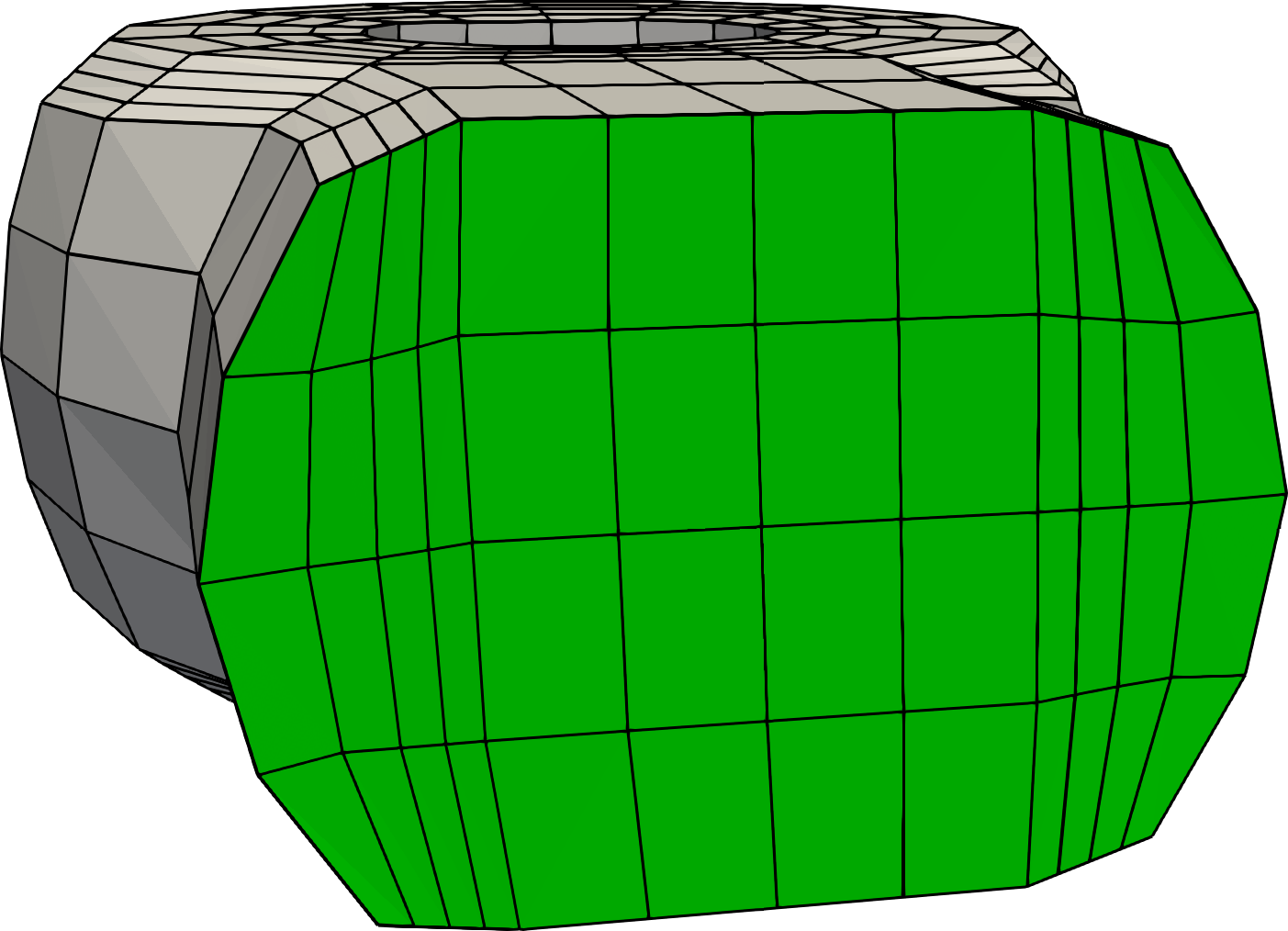}&\hspace{0.1\linewidth}&
\includegraphics[width=0.3\linewidth]{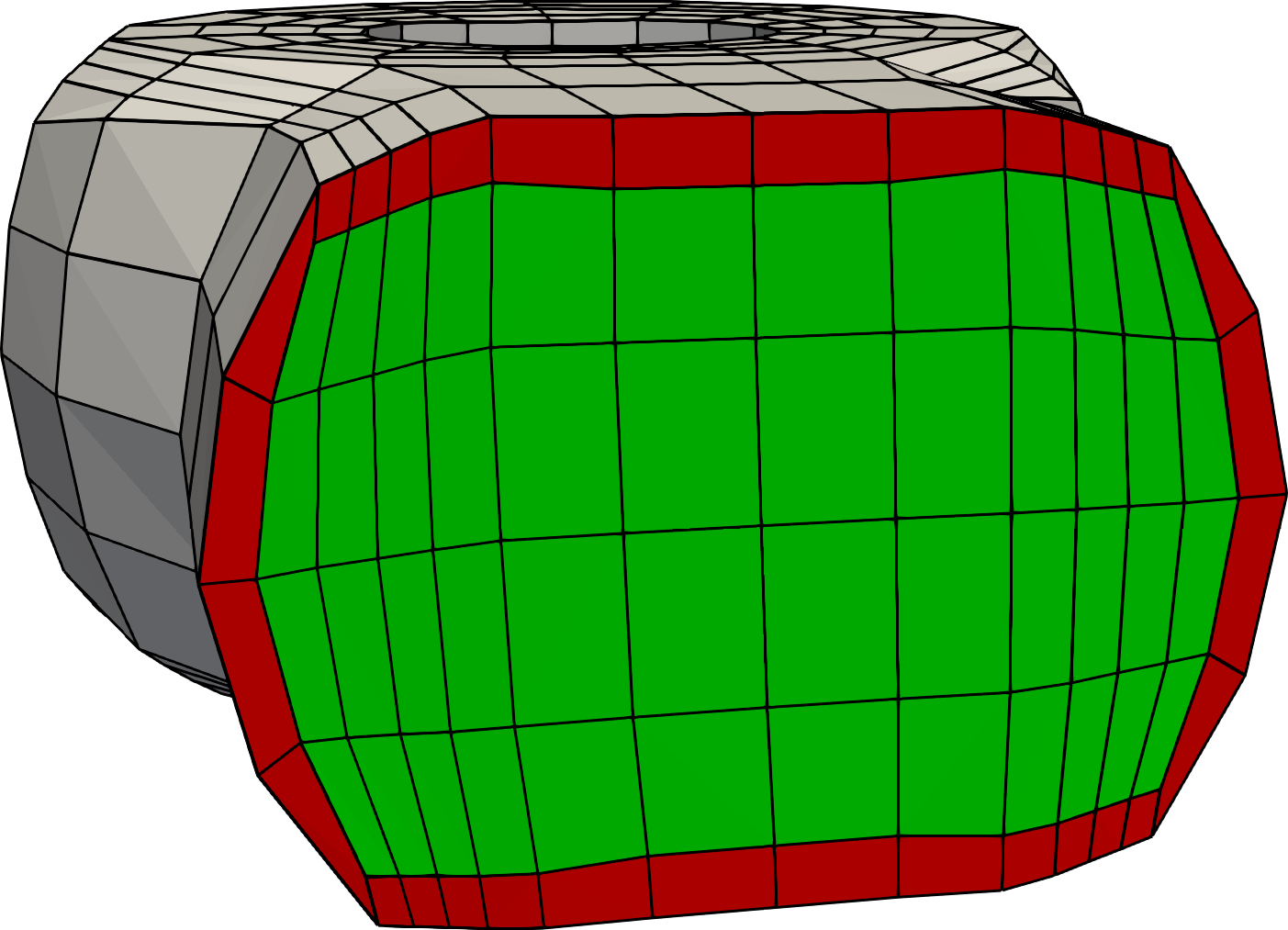}\\
(a)& \hspace{0.1\linewidth} &(b) \\
\end{tabular}
\caption{Mesh quality improvement. (a) The original mesh before quality improvement; (b) The mesh after pillowing, smoothing and optimization. Part of the mesh is removed to show the interior. The elements on the original cross section are labeled in green color while the inserted boundary layer is labeled in red color. }
    \label{fig:pillow_smooth}
\end{figure}

If there are still poor quality elements after smoothing, we run the
optimization whose objective function is the Jacobian. Each vertex is
then moved toward an optimal position that maximizes the worst
Jacobian. We present \nameref{alg:4} for quality improvement.  Here,
we show how to improve mesh quality for the rod model.
We first run the following command to perform pillowing on the rod model:
\begin{lstlisting}[style=mystyle]
Quality.exe -i rod_hex.vtk -Q -m 1 -n 1 -o rod_hex_pillow.vtk
\end{lstlisting}
There are four options used in the command:
\begin{itemize}
    \item \textbf{-i}: Unstructured hex mesh (rod\_hex.vtk);
    \item \textbf{-o}: The hex mesh after quality improvement  (rod\_hex\_pillow.vtk);
    \item \textbf{-m}: Improvement method. Pillowing when \textbf{-m
        1}; and
     \item \textbf{-n}: Number of pillowing layer.
\end{itemize}
Option \textbf{-n} allows users to specify the number of layers to be
inserted. With \textbf{-n 1}, we insert one layer around the boundary,
which is enough to ensure each element have at most one face on the
boundary. The result is shown in Fig.~\ref{fig:pillow_smooth}(a).

After pillowing, we can use the following command to smooth the mesh:
\begin{lstlisting}[style=mystyle]
Quality.exe -i rod_hex_pillow.vtk -Q -m 2 -p 0.001 -n 50 -s 2 -o rod_hex_pillow_lap.vtk
\end{lstlisting}
There are seven options used in the command:
\begin{itemize}
    \item \textbf{-i}: The input unstructured hex mesh in the vtk format (rod\_hex\_pillow.vtk);
    \item \textbf{-o}: The output hex mesh after quality improvement  (rod\_hex\_pillow\_lap.vtk);
    \item \textbf{-m}: Improvement method. Smoothing when \textbf{-m
        2}; and
    \item \textbf{-s}: Sharp feature preservation
    \begin{itemize}
        \item \textbf{-s 0}: No sharp features are preserved;
        \item \textbf{-s 1}: Detect sharp features automatically, and
          set tolerance \textbf{-t}; and
        \item \textbf{-s 2}: Manually select sharp feature points and store the indices in the "sharp.txt" file.
        \end{itemize}
    \item \textbf{-t}: Tolerance for automatically detecting sharp
      features;
   \item \textbf{-p}:  Step size for smoothing; and
      \item \textbf{-n}: Number of steps for smoothing.
\end{itemize}
By using the above command, we relocate a vertex only if the new
location will improve the local scaled Jacobian. Option \textbf{-s} is
used to preserve sharp features. Here, sharp feature detection is only
based on the mesh normal information. Therefore, it is not robust for
complex geometries and manual work is needed to adjust sharp
features. When we use automatic detection with option \textbf{-s 1},
we need to set a tolerance \textbf{-t}. There is no typical number for
this option. We need to do a trial and error to get the optimal
value. For the rod model, we set it to be 0.8. However, some sharp
features may not be detected regardless of the tolerance. User
intervention is needed with the option \textbf{-s 2} if automatic
detection is not satisfactory. Through this command line option,
Quality.exe will read an input file that includes the user-defined
sharp features. The snippets of the related file is shown in Appendix A5.

Then we run
the optimization step by using the command:
\begin{lstlisting}[style=mystyle]
Quality.exe -i rod_hex_pillow_smooth.vtk -Q -m 3 -p 0.001 -n 15 -o rod_hex_pillow_smooth_opt.vtk
\end{lstlisting}
There are five options used in the command:
\begin{itemize}
    \item \textbf{-i}: Unstructured hex mesh (rod\_hex\_pillow\_smooth.vtk);
    \item \textbf{-o}: The hex mesh after quality improvement  (rod\_hex\_pillow\_smooth\_opt.vtk);
    \item \textbf{-t}: Tolerance related to sharp feature preservation;
    \item \textbf{-m}: Improvement method. Optimization when \textbf{-m 3};
     \item \textbf{-p}:  Step size for optimization; and
      \item \textbf{-n}: Number of steps for optimization.
\end{itemize}
The quality improvement result for the rod model is shown in Fig.~\ref{fig:pillow_smooth} with a boundary layer created using pillowing, followed up by smoothing and optimization.

\begin{algorithm}
  \caption{\textbf{Quality Improvement Algorithm} }
  \label{alg:4}
        \begin{algorithmic}[1]
            \Require Hex mesh, step number $n$
            \Ensure Hex mesh with improved quality
            \renewcommand{\algorithmicensure}{\textbf{Pillowing:}}
            \Ensure
            \State Insert an outer layer to the input mesh
            \renewcommand{\algorithmicensure}{\textbf{Smoothing:}}
            \Ensure
            \While { iteration < $n$}
            \For{each vertex $v_i$}
            \If {$v_i$ is on a sharp edge}
            \If { improving the local Jacobian}
            \State Relocate $v_i$ a small step towards the middle of its neighboring
            vertices
            \EndIf
            \ElsIf {$v_i$ is on the boundary surface}
            \If { improving the local Jacobian}
            \State Relocate $v_i$ a small step towards the the area center of its
            neighboring boundary quads
            \EndIf
            \ElsIf {$v_i$ is an interior vertex }
            \If { improving the local Jacobian}
            \State Relocate $v_i$ a small step towards the weighted
            volume center of its neighboring elements
            \EndIf
            \EndIf
            \EndFor
            \EndWhile
            \renewcommand{\algorithmicensure}{\textbf{Optimization:}}
            \Ensure
            \While { iteration < $n$}
            \For{each negative Jacobian element}
            \If { improving the local Jacobian}
            \State Relocate $v_i$ to maximize the
             Jacobian
            \EndIf
            \EndFor
            \State Relocate vertices where Jacobian is minimum to
            maximize the worst Jacobian
            \EndWhile
        \end{algorithmic}
    \end{algorithm}

\clearpage
\section{Hex2Spline: Unstructured spline construction}
\label{sec:hex2spl-unstr-spline}




With the generated hex mesh as the input control mesh, now we present
the Hex2Spline software package to build TH-spline3D. TH-spline3D can
define spline functions on arbitrarily unstructured hex meshes. It
further supports local refinement for adaptive IGA. Hex2Spline can
output the B\'{e}zier information of constructed volumetric splines,
which can be easily used in LS-DYNA or any other existing IGA
frameworks. In the following, we introduce the main algorithm for each
component of the Hex2Spline software package, including blending
functions on an unstructured hex mesh, TH-spline3D with local
refinement, and B\'{e}zier extraction.

\subsection{Blending functions on an unstructured hex mesh}
\label{sec:bf_ie}

In this section, we describe how to build blending functions on an
all-hex mesh. Hex2Spline supports arbitrarily unstructured all-hex
mesh. In the following, we denote $\Omega_e$ as an hex element indexed
by $e$.
There are three types of elements in the hex mesh: boundary elements,
interior regular elements and interior irregular elements.  The
element is defined as a boundary element if it contains a boundary
vertex; otherwise, it is an interior element. For an interior element,
if it contains an extraordinary edge\footnote{An extraordinary edge is
  an interior edge shared by other than four hexahedral elements.}, we
call it an irregular element; otherwise, it is a regular element. In
the following, we discuss the main algorithm for building blending
functions on boundary elements and interior irregular elements. A
regular interior element is a special case of an irregular interior
element whose blending functions are merely tricubic B-splines. Note
that the following construction only applies to tricubic splines with
uniform knot intervals (i.e., the same knot interval for every edge).

\nameref{alg:bzmat1} shows the pseudocode to obtain the blending
functions defined on an interior irregular element. They are obtained
through the B\'{e}zier extraction matrix $\mathbf{M}$. $\mathbf{M}$
can be obtained by computing the 64 B\'{e}zier control points
$\mathbf{Q}_e$ from a local spline control mesh $\mathcal{N}$ that
consists of $\Omega_e$ and its one-ring neighborhood. Each of these
B\'{e}zier points is obtained by a convex combination of the vertices
in the local control mesh, and we have
$\mathbf{Q}_e=\mathbf{M}\mathbf{P}$, where $\mathbf{P}$ is the vector
of vertices in the local control mesh. Then the blending functions on
$\Omega_e$ are defined by using the transpose of $\mathbf{M}$, that
is, $\mathbf{B}_e=\mathbf{M}^T\mathbf{b}$, where
\begin{equation*}
\begin{aligned}
\mathbf{b}=&[N_{0}(u)N_{0}(v)N_{0}(w),\ldots,N_{0}(u)N_{1}(v)N_{0}(w),N_{1}(u)N_{1}(v)N_{0}(w),\\
&\ldots,N_{0}(u)N_{0}(v)N_{1}(w),N_{1}(u)N_{0}(v)N_{1}(w),\ldots,N_{3}(u)N_{3}(v)N_{3}(w)]^T,
\end{aligned}
\end{equation*}
is the vector of 64 tricubic Bernstein polynomials. Each univariate
cubic Bernstein polynomial is given as
$N_k(t)=\dbinom{3}{k}(1-t)^{3-k}t^k$ ($k=0,\ldots,3$).  Readers can
refer to~\cite{wei17a} for the coefficients of $\mathbf{M}$. Note that
we only need $\mathbf{M}$ to define $\mathbf{B}_e$, and we do not
actually compute B\'{e}zier points.


\begin{algorithm}[H]
\caption{ \textbf{Blending Functions Algorithm (Interior)}}
\begin{algorithmic}[1]
\Require{An interior element $\Omega_e$ and its local control mesh $\mathcal{N}$}
\Ensure{The blending functions $\mathbf{B}_e$}
\renewcommand{\algorithmicensure}{\textbf{Obtain the B\'{e}zier transformation matrix  $\mathbf{M}_{64\times N}$:}}
\Ensure
\For{each B\'{e}zier point $Q_{e,i}$ of $\Omega_e$, where $i=0,\ldots,63$}
\If { $Q_{e,i}$ is body point}
\State Compute its coordinates based on $\mathbf{P}$
\ElsIf { $Q_{e,i}$ is face, edge, corner point}
\State Compute its coordinates by averaging the nearest body points
\EndIf
\EndFor
\State Ensemble matrix $\mathbf{M}$ such that $\mathbf{Q}_e=\mathbf{M}\mathbf{P}$
\renewcommand{\algorithmicensure}{\textbf{Obtain the blending functions $\mathbf{B}_e$:}}
\Ensure
\State $\mathbf{B}_e=\mathbf{M}^T\mathbf{b}$
\end{algorithmic}
\label{alg:bzmat1}
\end{algorithm}

\nameref{alg:bzmat2} shows the pseudocode to define blending functions
on a boundary element. The body B\'{e}zier points and the B\'{e}zier
points on the interior corners, edges or faces are obtained the same
way as in \nameref{alg:bzmat1}, while the B\'{e}zier points on the
boundary is defined using only the boundary quadrilateral mesh. The 16
B\'{e}zier points on a boundary face can be obtained by convex
combinations of the vertices on the local quad control mesh. The
detailed computation method is explained in \cite{wei17a}. We finally
get all the B\'{e}zier points as $\mathbf{Q}_e=\mathbf{M}\mathbf{P}$.
The blending functions are then defined by
$\mathbf{B}_e=\mathbf{M}^T\mathbf{b}$. Note that~\nameref{alg:bzmat2}
can also be used to preserve sharp features in the mesh by adjusting
the B\'{e}zier extraction matrix $\mathbf{M}$. Reader can refer to
\cite{wei17a} for details.

\begin{algorithm}[H]
\caption{\textbf{Blending Functions Algorithm (Boundary)}}
\begin{algorithmic}[1]
\Require{A boundary element $\Omega_e$ and its local control mesh $\mathcal{N}$; sharp feature information}
\Ensure{The blending functions $\mathbf{B}_e$}
\renewcommand{\algorithmicensure}{\textbf{Obtain the B\'{e}zier transformation matrix  $\mathbf{M}_{64\times N}$:}}
\Ensure
\For{each B\'{e}zier point $Q_{e,i}$ of $\Omega_e$, where $i=0,\ldots,63$}
\If {$Q_{e,i}$ corresponds to a sharp corner $P_k$}
\State $Q_{e,i}=P_k$
\ElsIf { $Q_{e,i}$ is on a sharp edge}
\State Compute its coordinates as a convex combination of the two end points of the sharp edge
\ElsIf { $Q_{e,i}$ is on the boundary}
\If { $Q_{e,i}$ is a face point}
\State Compute its coordinates based on $P_k$
\ElsIf { $Q_{e,i}$ is an edge or a corner point}
\State Compute its coordinates by averaging the nearest boundary face points
\EndIf
\ElsIf{ $Q_{e,i}$ is a body point or on an interior surface}
\State Calculate $Q_{e,i}$ the same as \nameref{alg:bzmat1}
\EndIf
\EndFor
\State Ensemble matrix $\mathbf{M}$ such that $\mathbf{Q}_e=\mathbf{M}\mathbf{P}$
\renewcommand{\algorithmicensure}{\textbf{Obtain the blending functions $\mathbf{B}_e$:}}
\Ensure
\State $\mathbf{B}_e=\mathbf{M}^T\mathbf{b}$
\end{algorithmic}
\label{alg:bzmat2}
\end{algorithm}

\subsection{TH-spline3D for local refinement}
The next step is to introduce local refinement to achieve
computational efficiency and accuracy. TH-spline3D employs a
hierarchical structure and uses the truncation mechanism to perform
local refinement. Global refinement is also supported since it is a
special case of local refinement, and it is done by Catmull-Clark
subdivision for solids~\cite{ref:bajaj02,ref:burkhart10}.

\nameref{alg:THSpline3D_1} shows the pseudocode to construct
TH-spline3D based on the blending functions developed in
Section~\ref{sec:bf_ie}. Locally refined meshes as well as spline
functions between different levels are related through Catmull-Clark
subdivision for solids.


The program allows users to specify a list of target elements to be
locally refined. This is enabled by reading a series of user-defined
files to the program, each of which contains indices of target
elements at a certain level. The input mesh is treated as the level-0
mesh by default, one needs to provide a file named "lev\_rfid.txt" to
refine certain level-0 elements. As a result, a level-1 mesh is
generated that can be used to define multi-level local refinement. One can check the hierarchical control meshes and add more elements in the "lev\_rfid.txt" file to further refine the mesh.

The remaining procedure to construct TH-spline3D is automatic and can
be divided into three steps: i) refine target elements by
Catmull-Clark subdivision for solids, ii) select certain blending
functions from hierarchical meshes, and iii) truncate some blending
functions on the coarse mesh. Then with the help of
\nameref{alg:bzmat1} and~\nameref{alg:bzmat2}, we can construct
TH-spline3D on hierarchical control meshes. Readers can refer
to~\cite{wei17a} for details.

\begin{algorithm}[H]
\caption{\textbf{TH-spline3D Algorithm}}
\begin{algorithmic}[1]
\Require{Initial control points $\mathbf{P}^0$ and their associated blending functions $\mathcal{B}^0$, element $\Omega_e$ and its local control mesh $\mathcal{N}$}
\Ensure{The truncated blending functions $\mathcal{B}_{\text{TH-spline3D}}$}
\renewcommand{\algorithmicensure}{\textbf{Refined meshes via local refinement:}}
\Ensure
\If{local refinement is needed}
\State Generate mesh $\mathcal{M}^{\ell+1}$ from mesh $\mathcal{M}^\ell$ based on
Catmull-Clark subdivision~\cite{ref:bajaj02,ref:burkhart10}
\EndIf
\State Ensemble matrix $\mathbf{C}$ so that $\mathbf{P}^{\ell+1}=\mathbf{C}\mathbf{P}^{\ell}$
\renewcommand{\algorithmicensure}{\textbf{Construct the hierarchical blending functions:}}
\Ensure
\State Compute the B\'{e}zier control points on hierarchical structure $\mathbf{Q}^{\ell+1}=\mathbf{M}\mathbf{\mathbf{P}^{\ell+1}}$
\State Construct the hierarchical B-splines on the hierarchical control meshes by using \nameref{alg:bzmat1} or \nameref{alg:bzmat2}
\renewcommand{\algorithmicensure}{\textbf{TH-spline3D Construction:}}
\Ensure
\State Select the
blending function to be active
\State Truncate chosen blending functions
\State Collect all the blending functions ($\mathcal{B}_{\text{TH-spline3D}}^{\ell+1}$) up to Level $\ell+1$
\end{algorithmic}
\label{alg:THSpline3D_1}
\end{algorithm}


\subsection{B\'{e}zier extraction for LS-DYNA}
After blending functions are defined based on Bernstein polynomials,
the B\'{e}zier information of the constructed volumetric splines can
be written in the BEXT file for LS-DYNA. The program will also output
files for the visualization of B\'{e}zier mesh in
Paraview~\cite{ahrens2005paraview}.  The BEXT file contains all the
control points and the B\'{e}zier extraction matrix $\mathbf{M}^{T}$
for each B\'{e}zier element. We reduce the file size by using both
sparse and dense formats to write $\mathbf{M}^{T}$. The matrix is
output row by row. In a sparse format, only non-zeros of a row are
output, where an index is paired with each non-zero coefficient to
indicate its column location in the matrix. On the other hand, an
entire row is output in the dense format without additional column
indices. The choice of the two formats depend on the number of
non-zeros in a row. The sparse format is favored when the row only has
a few non-zeros. The snippets of the BEXT format file is shown in Appendix A6.

\subsection{Applying Hex2Spline to the rod model}
Based on the above algorithms, we implement and organize
the code into a CLI program (Hex2Spline.exe) that can construct
TH-spline3D on an unstructured hex mesh and extract B\'{e}zier
information for analysis. During the spline construction, users can
specify if refinement is needed. In the end, Hex2Spline generates the
BEXT file for LS-DYNA. The program also supports sharp feature
preservation. All the available options for this program are explained
as follows:
\begin{itemize}
    \item \textbf{-i}: Unstructured hex mesh (rod\_hex.vtk);
    \item \textbf{-o}: The BEXT file for LS-DYNA (rod\_hex\_BEXT.txt);
    \item \textbf{-S}: Spline construction mode;
    \item \textbf{-s}: Sharp feature preservation;
    \begin{itemize}
        \item \textbf{-s 0}: No sharp features need to be preserved;
        \item \textbf{-s 1}: Detect sharp feature automatically, and
          set tolerance \textbf{-t}; and
        \item \textbf{-s 2}: Manually select sharp feature points and store them in "sharp.txt".
        \end{itemize}
      \item \textbf{-g}: Set the level of global refinement;
      \item \textbf{-l}: Enable local refinement; and
      \item \textbf{-t}: Tolerance related to sharp feature preservation.
\end{itemize}
Here, we apply local refinement to create a hierarchical mesh of the
rod model (see Fig.~\ref{fig:Pipeline_Structure}(g)) by using the command:
\begin{lstlisting}[style=mystyle]
Hex2Spline.exe -i rod_hex.vtk -S -s 2 -l -o rod_hex.BEXT
\end{lstlisting}
In the command, we use the option \textbf{-l} to switch on the local
refinement mode and construct TH-spline3D with local
refinement. Unlike global refinement, users need to prepare a
"lev\_rfid.txt" file to specify indices of target
elements. Fig.~\ref{fig:refinement_rod}(b) shows the spline
construction with one level local refinement. Users can perform
further refinements level by level. For example, users can edit the
"lev\_rfid.txt" file to include more elements and use the same command
to perform two levels of local refinement and the result is shown in
Fig.~\ref{fig:refinement_rod}(c):
\begin{lstlisting}[style=mystyle]
Hex2Spline.exe -i rod_hex.vtk -S -s 2 -l 2
\end{lstlisting}

\begin{figure}[htp]
\centering
\begin{tabular}{ccc}
\includegraphics[width=0.3\linewidth]{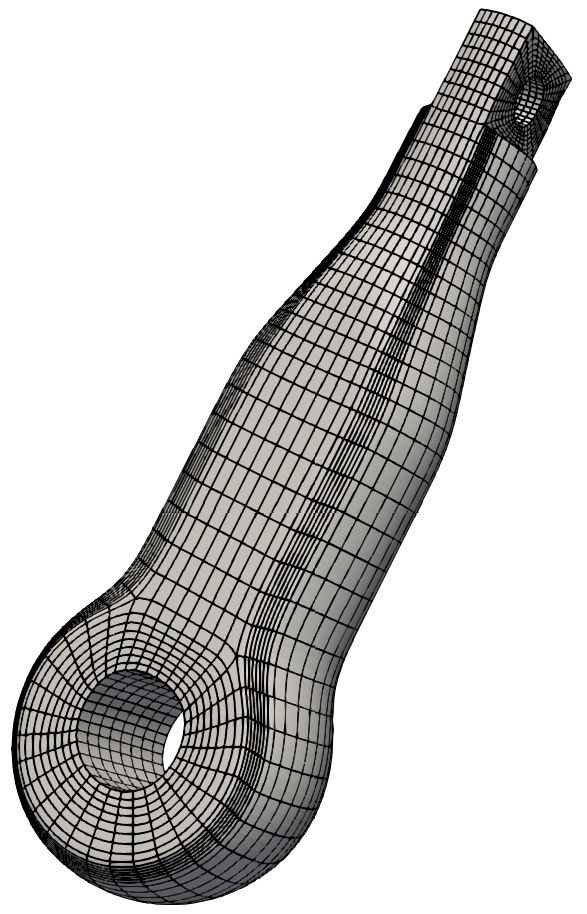}&
\includegraphics[width=0.3\linewidth]{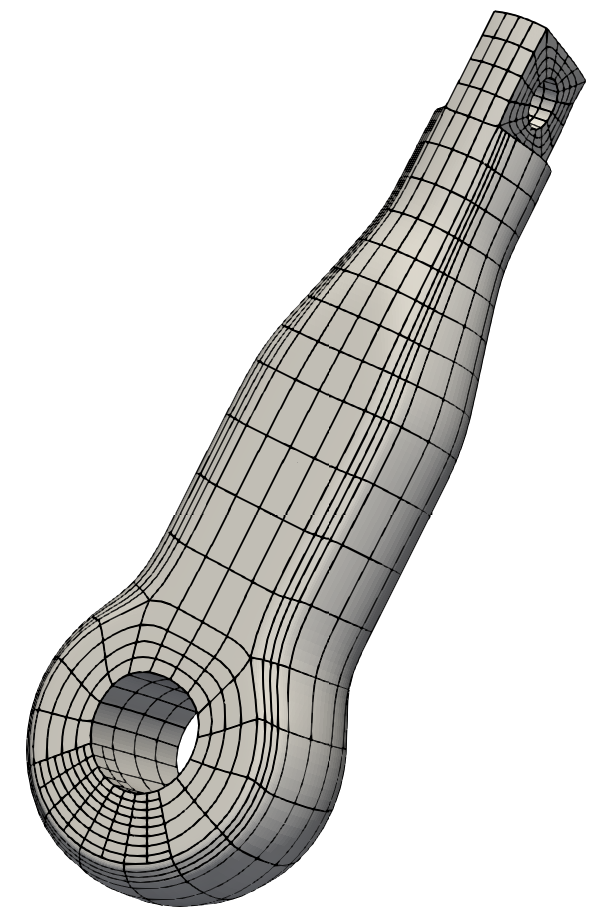}&
\includegraphics[width=0.3\linewidth]{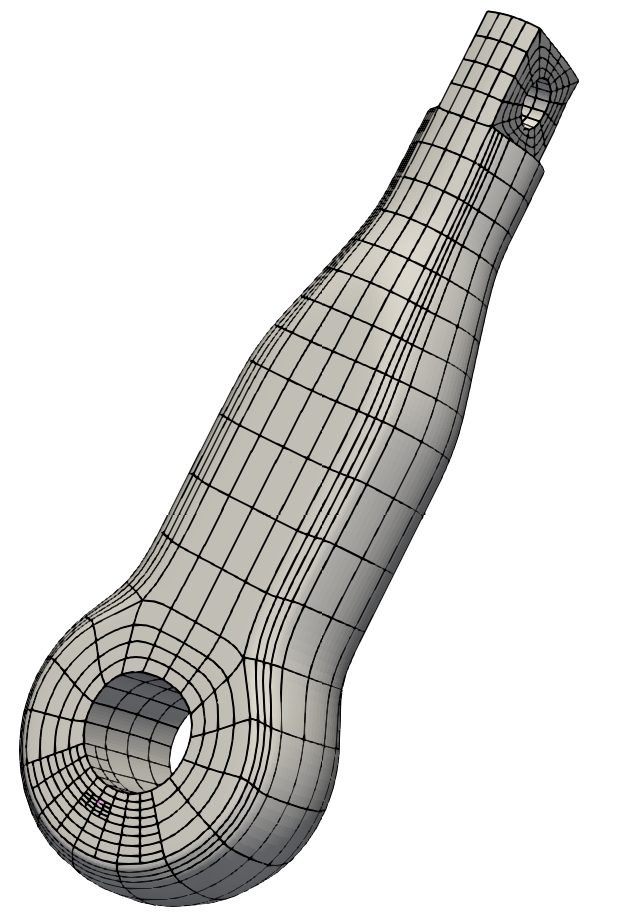}\\
(a) & (b) & (c)\\
\end{tabular}
\caption{The visualization of the output B\'{e}zier mesh with: (a)
  global refinement; (b) one level local refinement; and (c) two levels of
  local refinement.}
    \label{fig:refinement_rod}
\end{figure}

Users can also use the following command to perform spline construction
with one level global refinement and the result is shown in Fig.~\ref{fig:refinement_rod}(a):
\begin{lstlisting}[style=mystyle]
Hex2Spline.exe -i rod_hex.vtk -o rod_hex -S -s 2 -g 1
\end{lstlisting}
Here
we use the option \textbf{-g} to switch on global refinement mode and
set the argument to 1 to construct spline with one level global
refinement.

\section{Applications using HexGen and Hex2Spline}
The algorithms discussed in Sections~\ref{sec:hexg-polyc-based}
and~\ref{sec:hex2spl-unstr-spline} are implemented in C++. The Eigen
library~\cite{eigenweb} and Intel MKL~\cite{intel-alt} are used for
matrix and vector operations and numerical linear algebra. We also
take advantage of openMP to support multi-threading computation. We
use a compiler-independent building system (CMake) and a
version-control system (Git) to support software development. We have
compiled the source code into two software packages,
\begin{itemize}
    \item \textbf{Hex2Gen software package}:
    \begin{itemize}
    \item \textbf{Segmentation module (Segmentation.exe);}
    \item \textbf{Polycube construction module (Polycube.exe);}
    \item \textbf{All-hex mesh generation module
        (ParametricMapping.exe); and}
    \item \textbf{Quality improvement module (Quality.exe).}
\end{itemize}
    \item \textbf{Hex2Spline software package}:
    \begin{itemize}
    \item \textbf{Volumetric spline construction module (Hex2Spline.exe).}
    \end{itemize}
\end{itemize}
The software is open-source and can be found in the following Github
link (https://github.com/yu-yuxuan/HexGen\_Hex2Spline).

\begin{figure}[htp]
\centering
\begin{tabular}{ccc}
\includegraphics[height=0.3\linewidth]{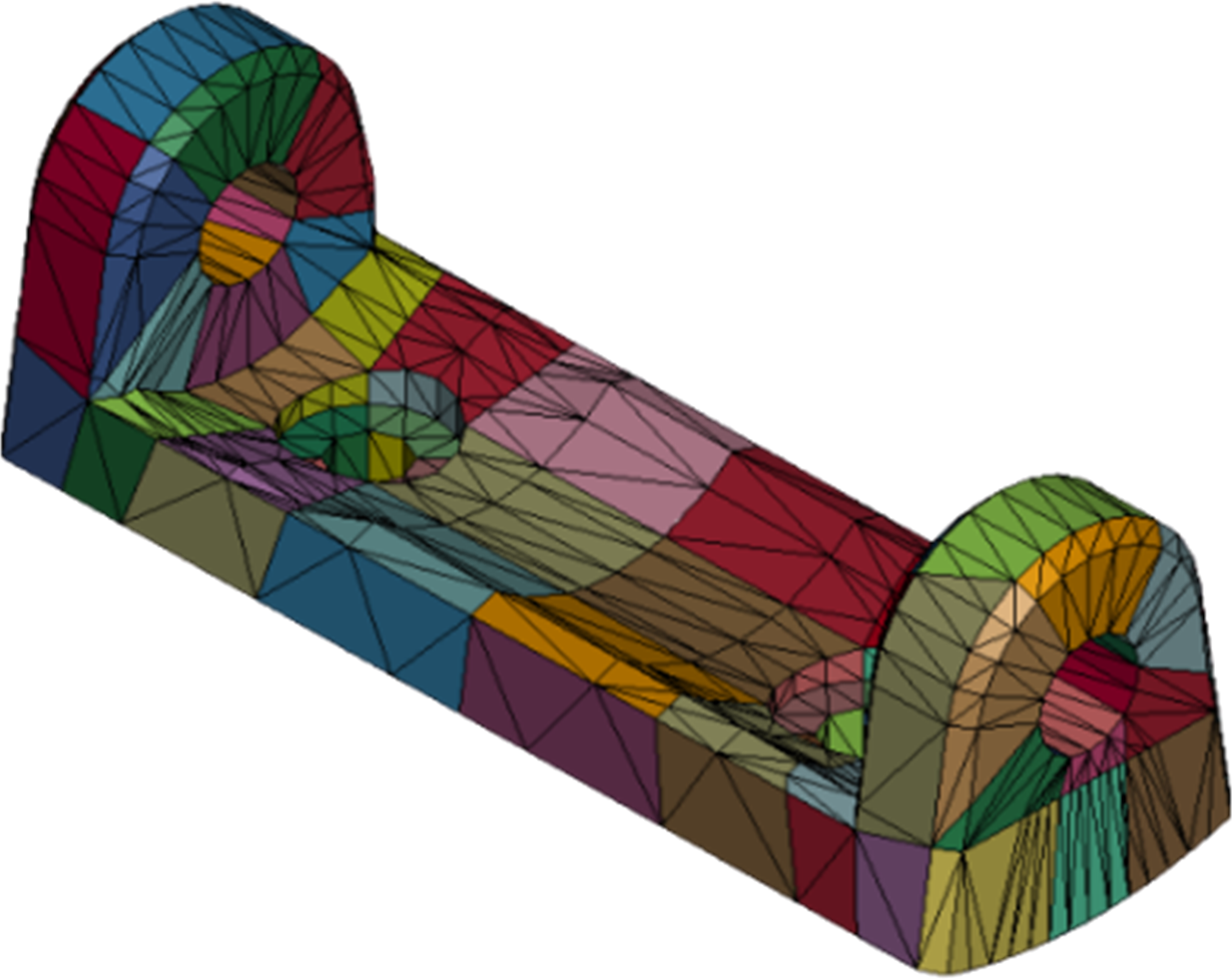}
  & \includegraphics[height=0.28\linewidth]{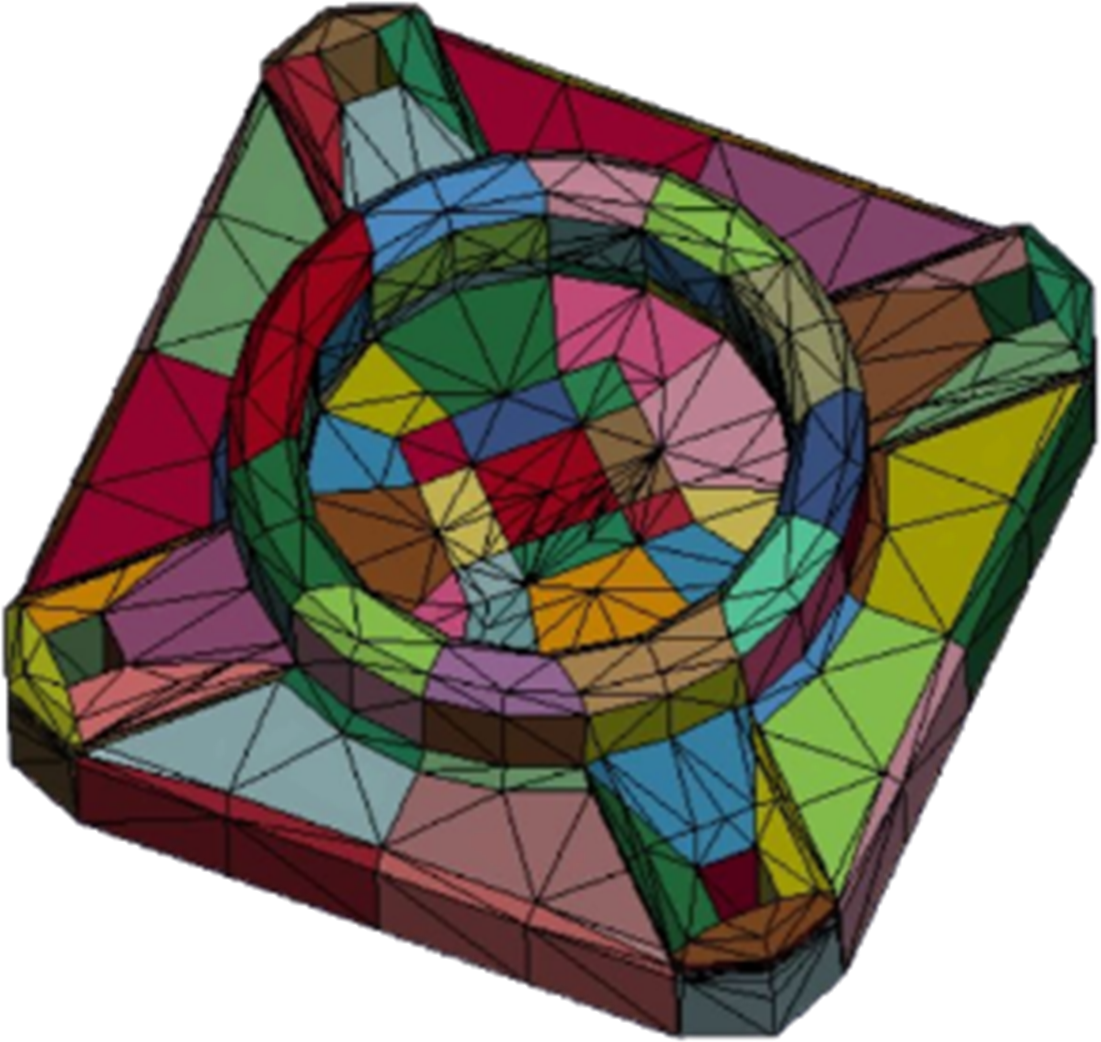}
    & \includegraphics[height=0.26\linewidth]{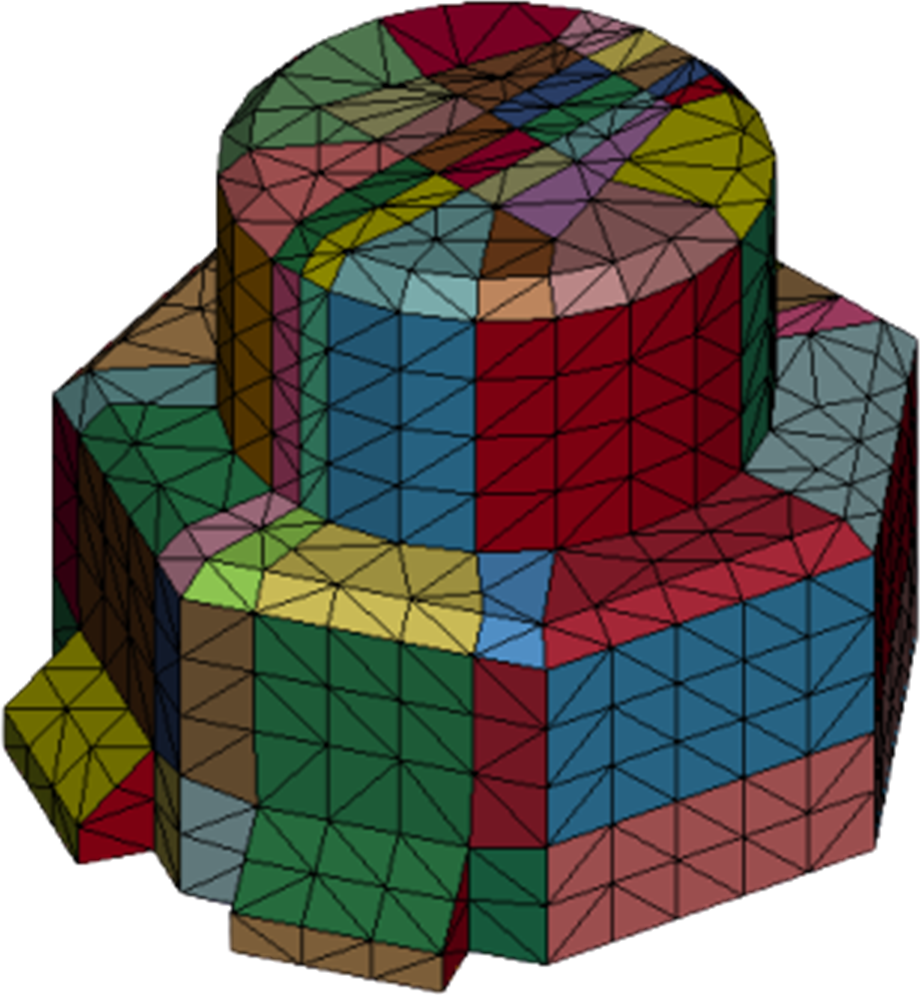}\\
        & (a) &\\
\includegraphics[height=0.28\linewidth]{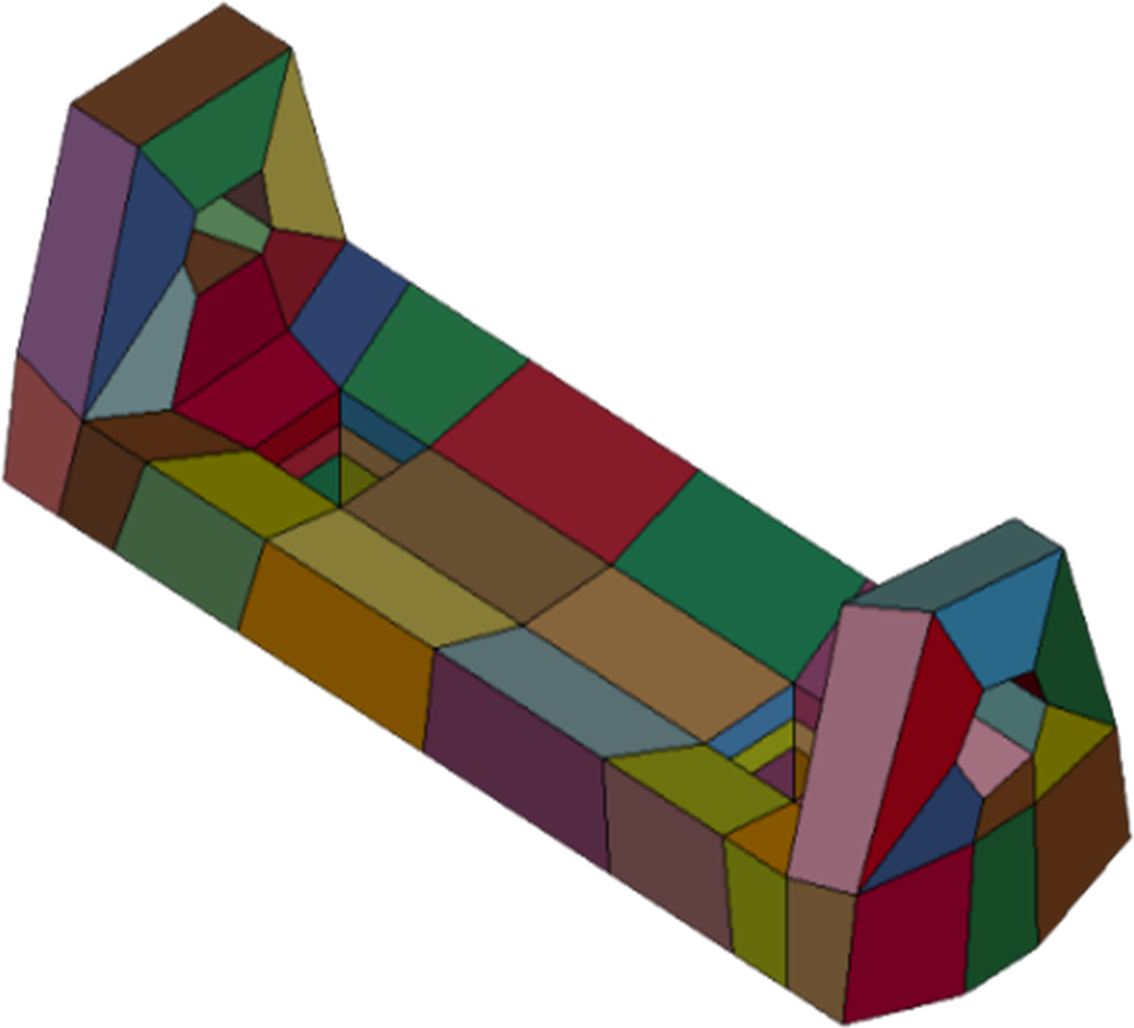}
  & \includegraphics[height=0.28\linewidth]{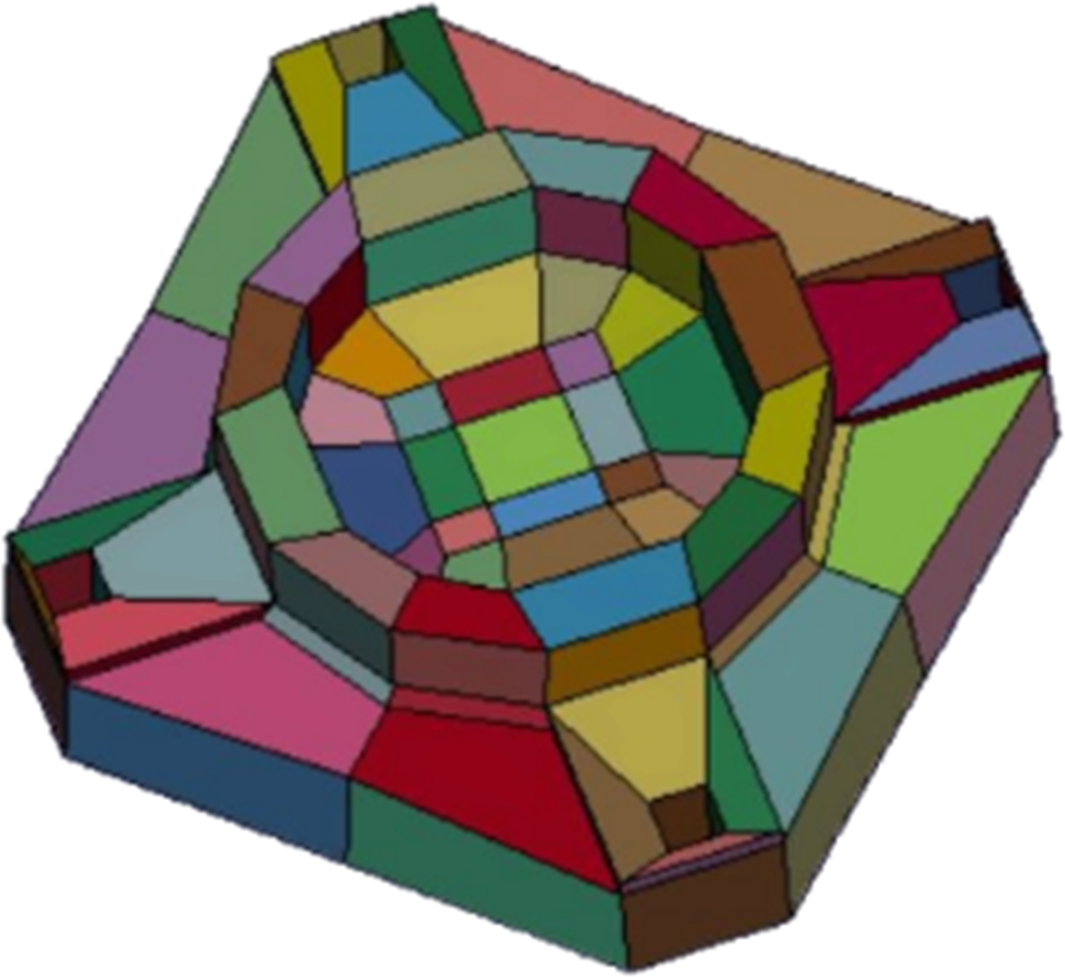}
    & \includegraphics[height=0.26\linewidth]{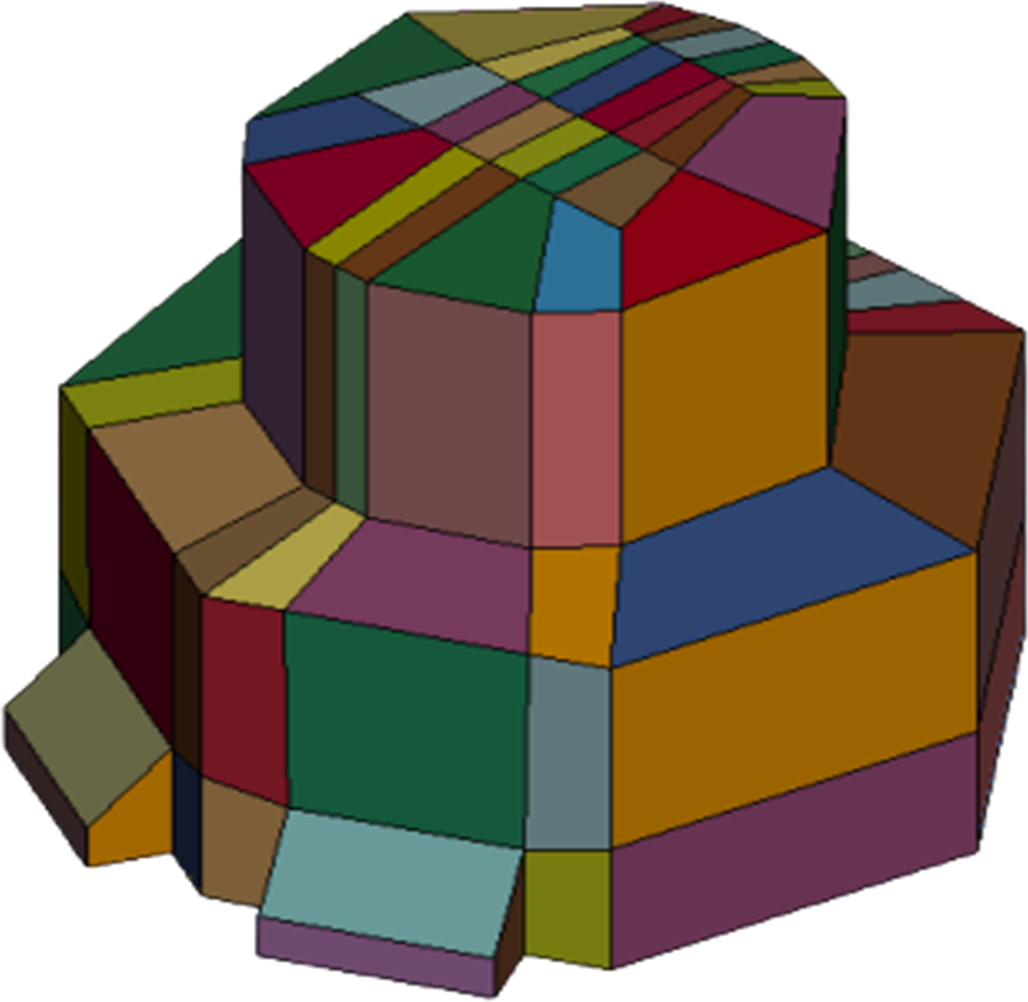}\\
        & (b) &\\
\includegraphics[height=0.28\linewidth]{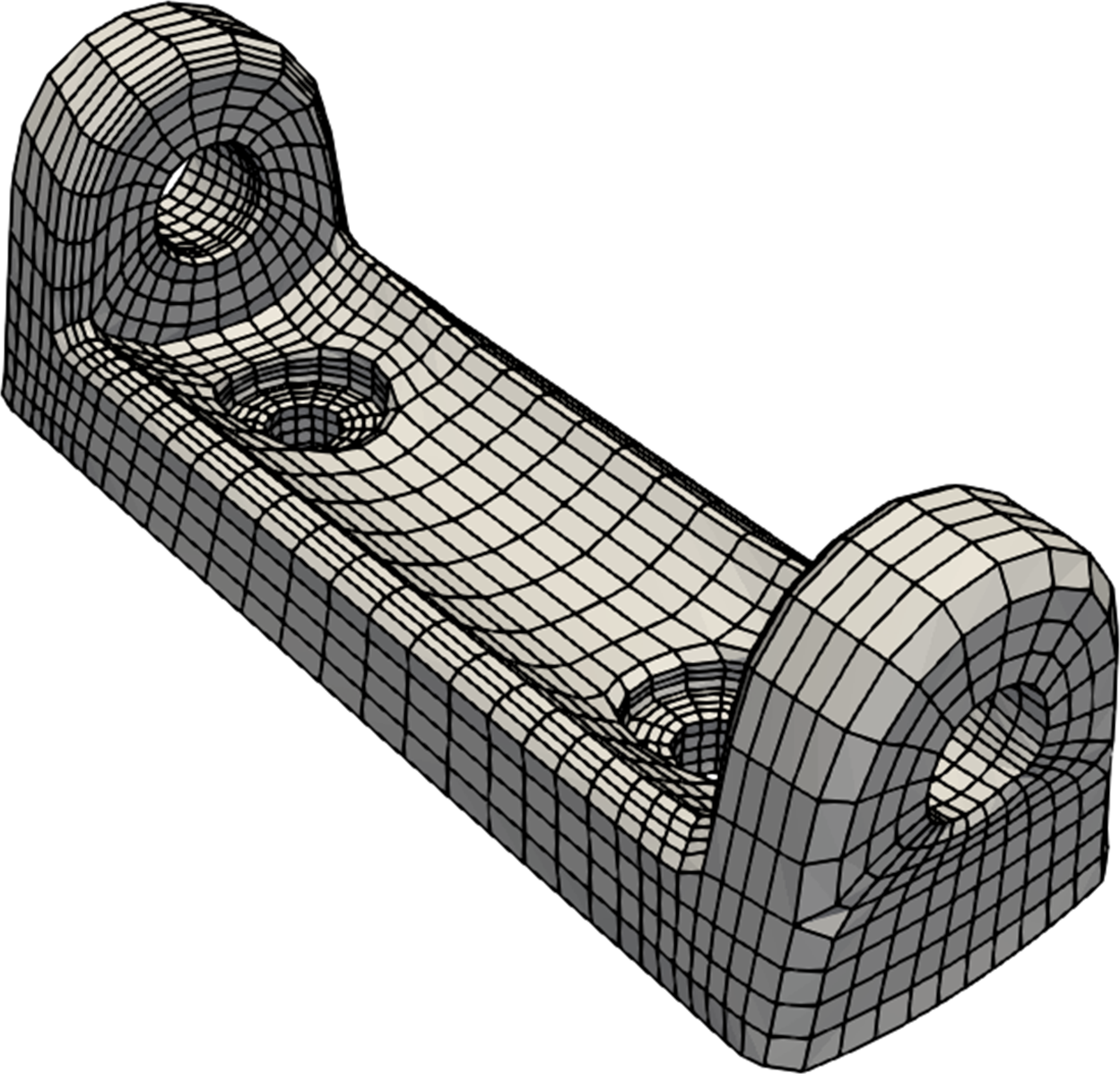}
  & \includegraphics[height=0.28\linewidth]{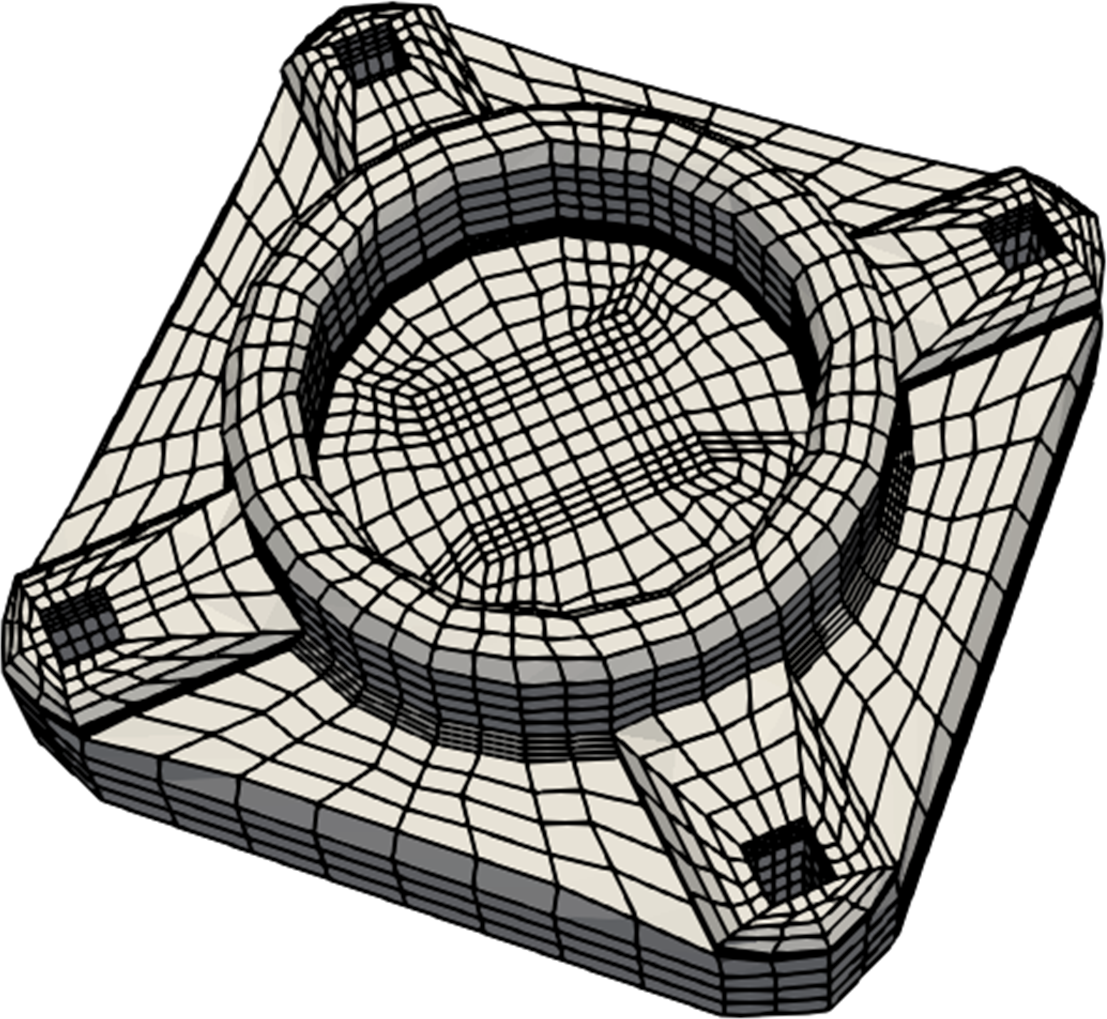}
    & \includegraphics[height=0.28\linewidth]{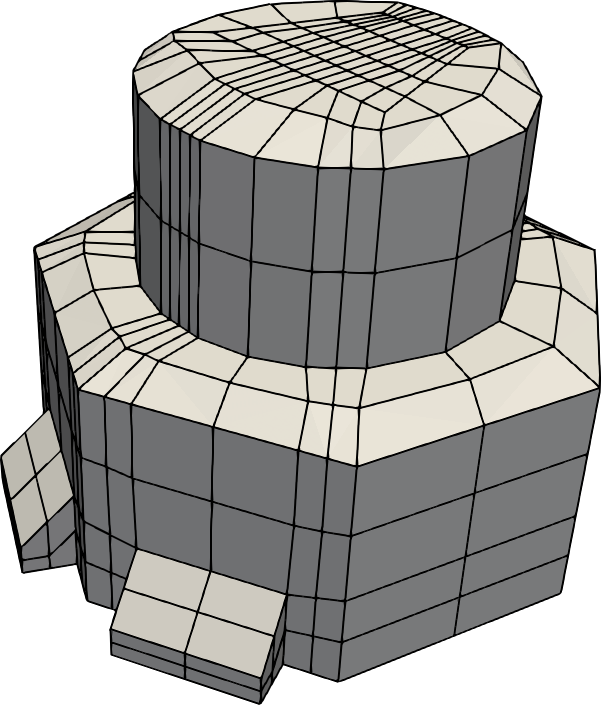}\\
        & (c) &\\
\includegraphics[height=0.28\linewidth]{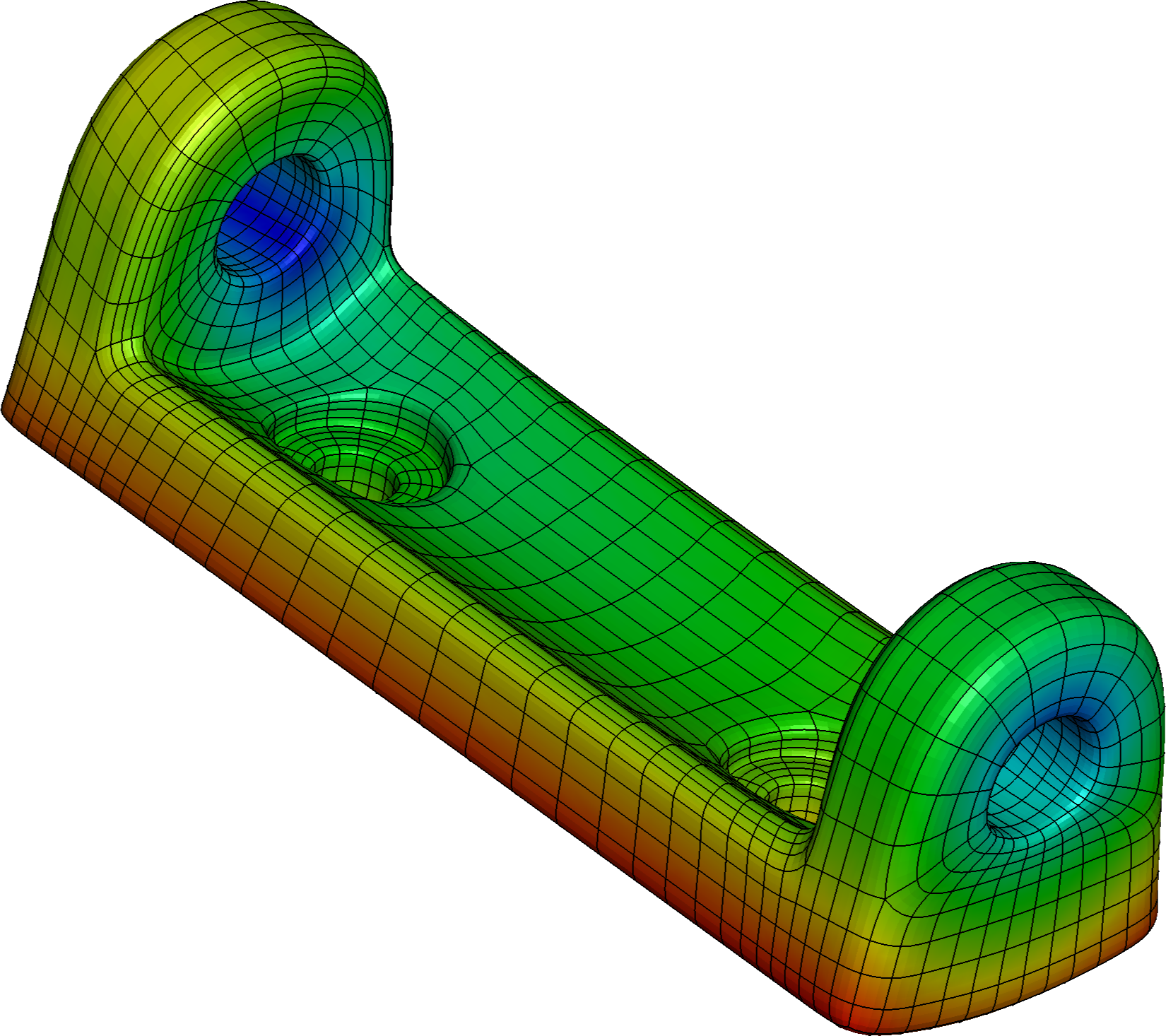}
  & \includegraphics[height=0.28\linewidth]{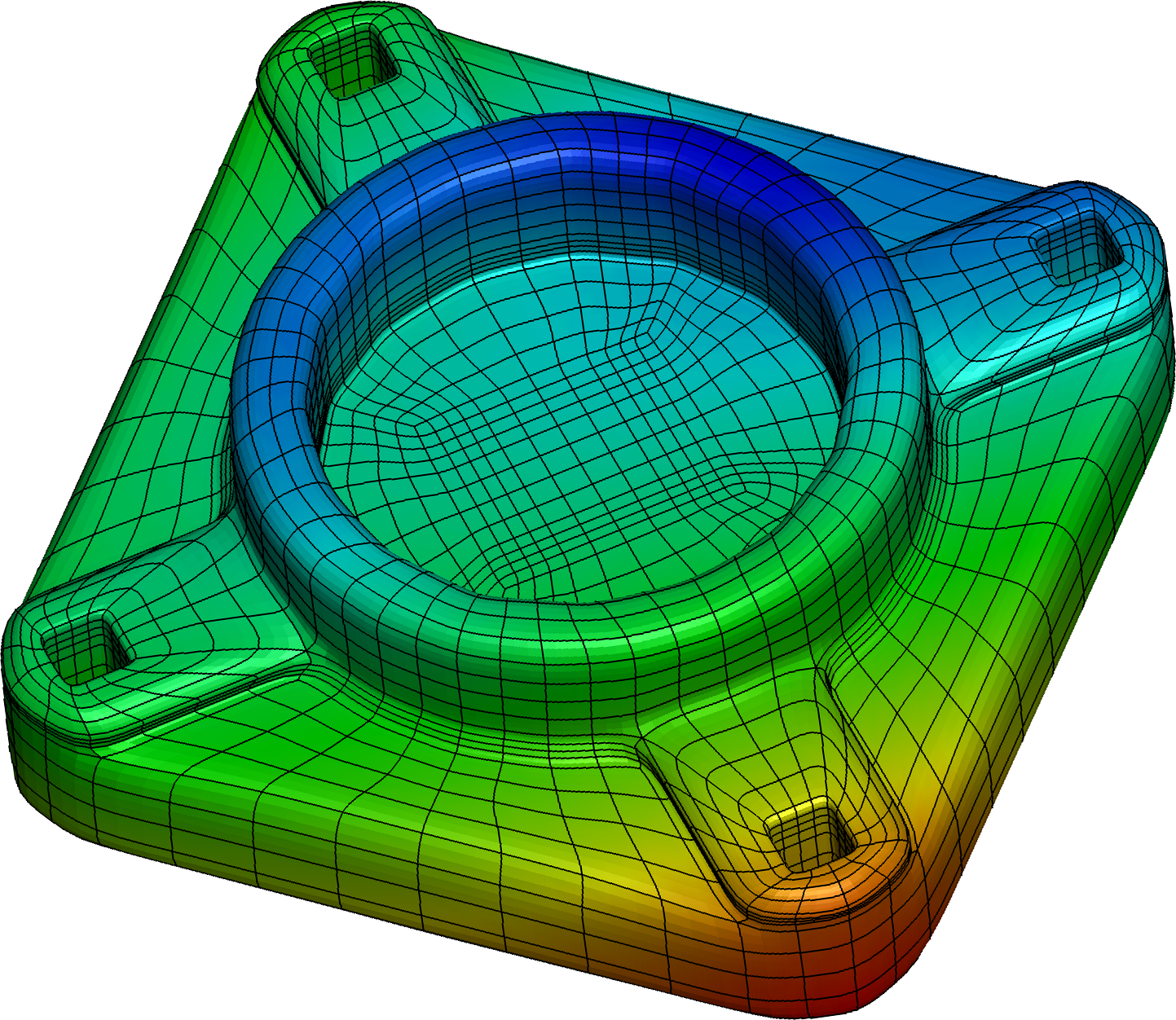}
    & \includegraphics[height=0.28\linewidth]{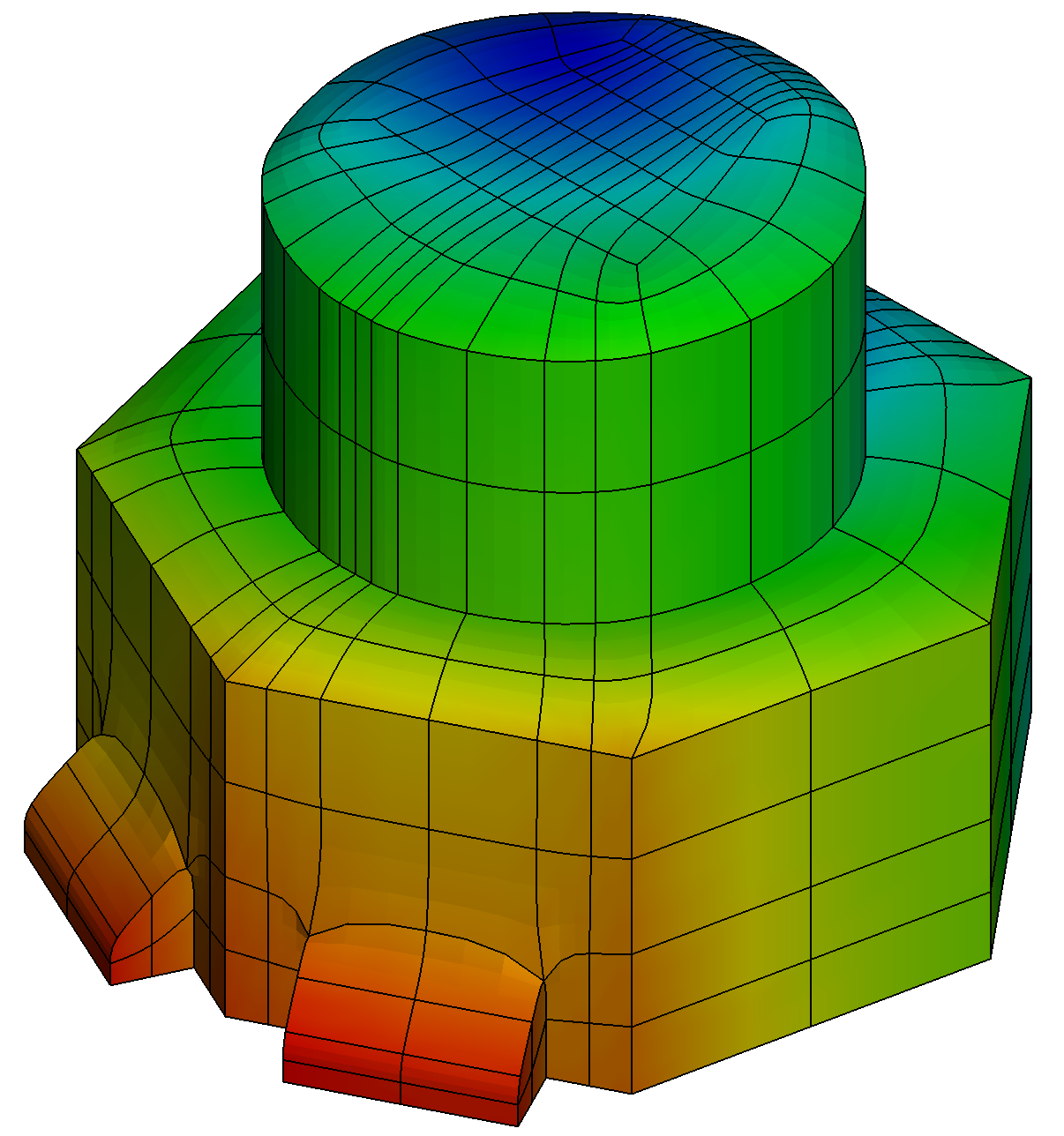}\\
    & (d) &\\
\end{tabular}
\caption{ Results of two types of mount and hepta models. (a) Surface
  triangle meshes and segmentation results; (b) Polycube structures;
  (c) All-hex control meshes; (d) Volumetric splines with IGA results of eigenvalue analysis in LS-DYNA. }
    \label{fig:model1}
\end{figure}

\begin{figure}[htp]

\begin{tabular}{lcr}
  \includegraphics[height=0.37\linewidth]{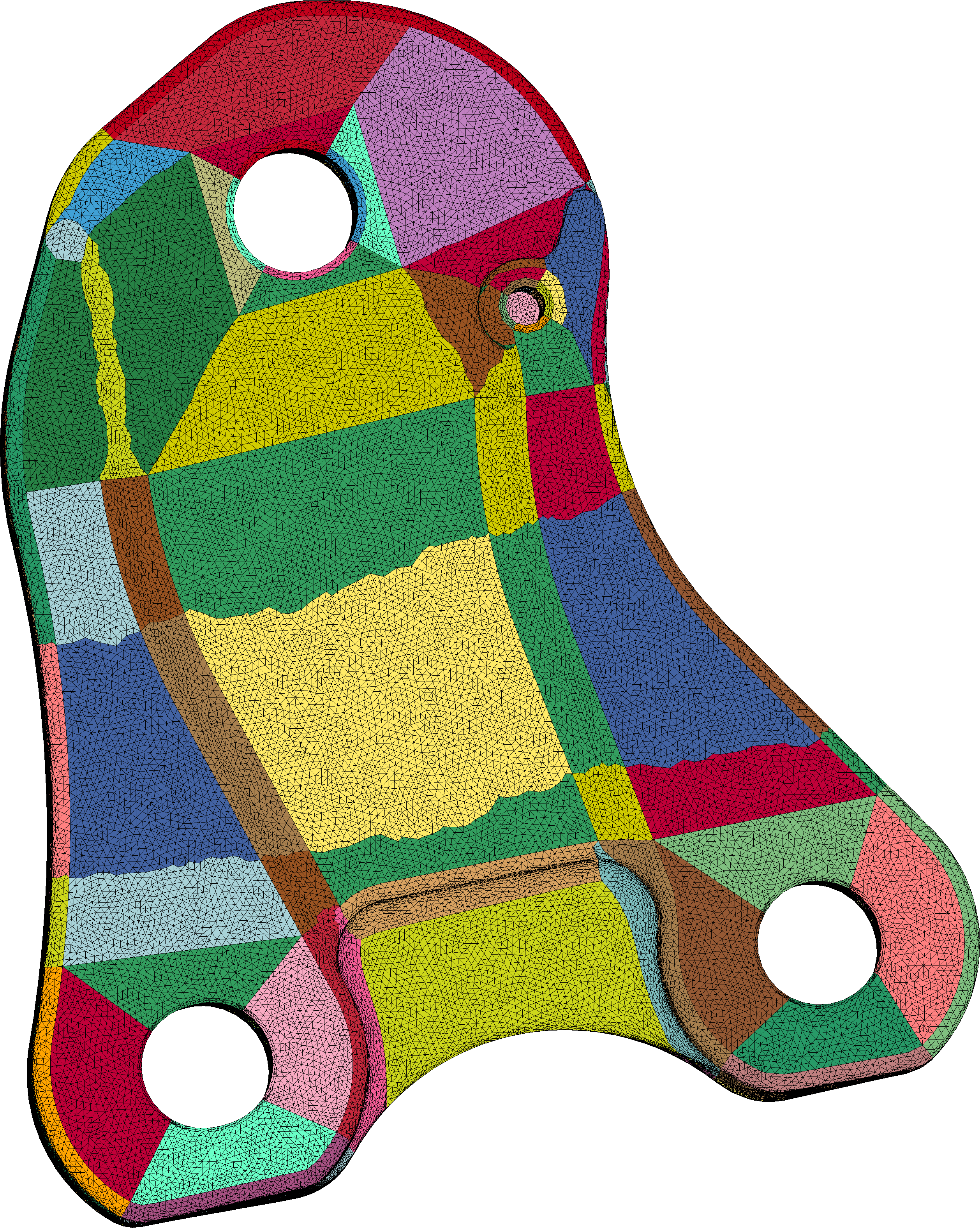} \hspace{0.04\linewidth}
  & \includegraphics[height=0.37\linewidth]{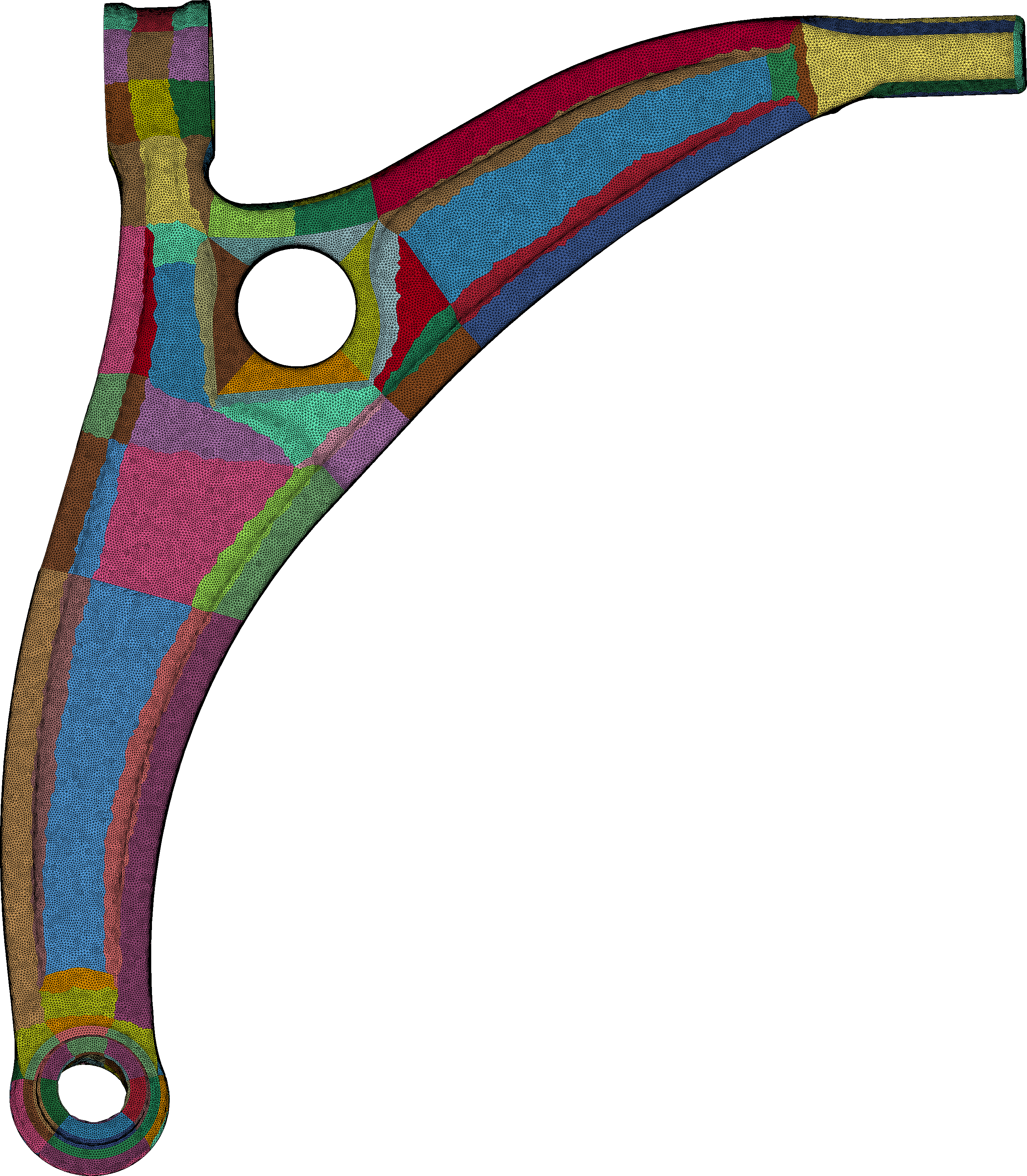} \hspace{0.1\linewidth}
    & \includegraphics[height=0.37\linewidth]{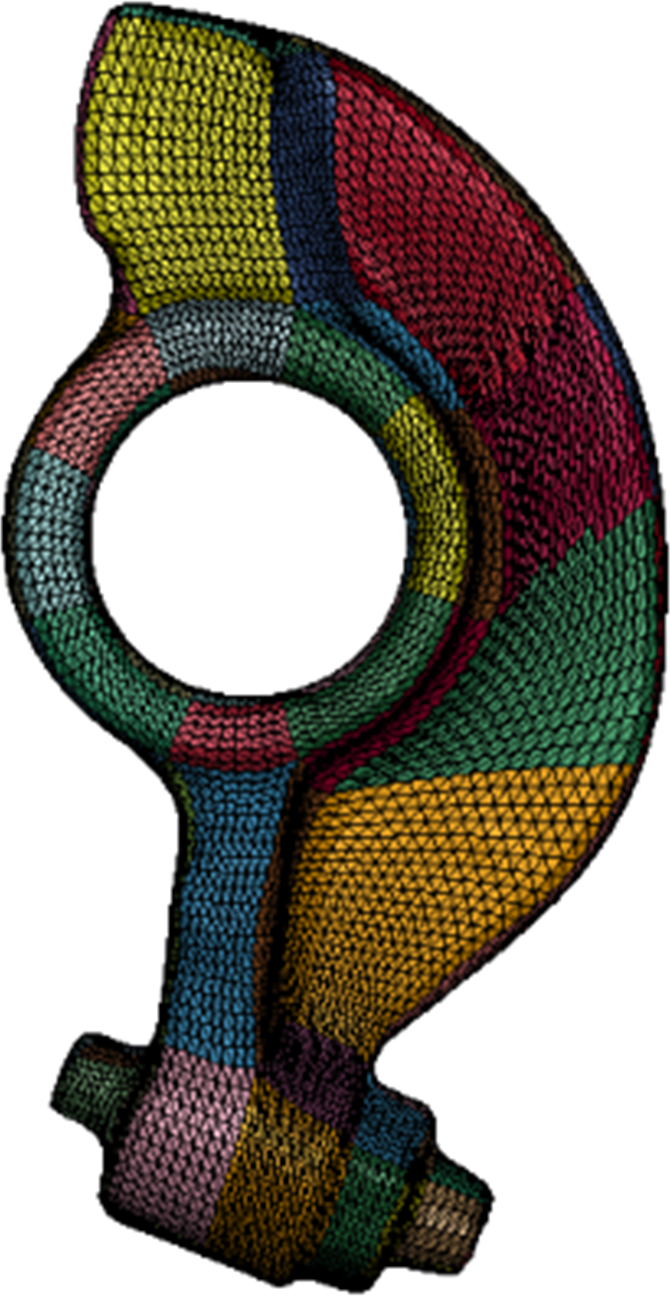}\\
    & (a) &\\
  \includegraphics[height=0.37\linewidth]{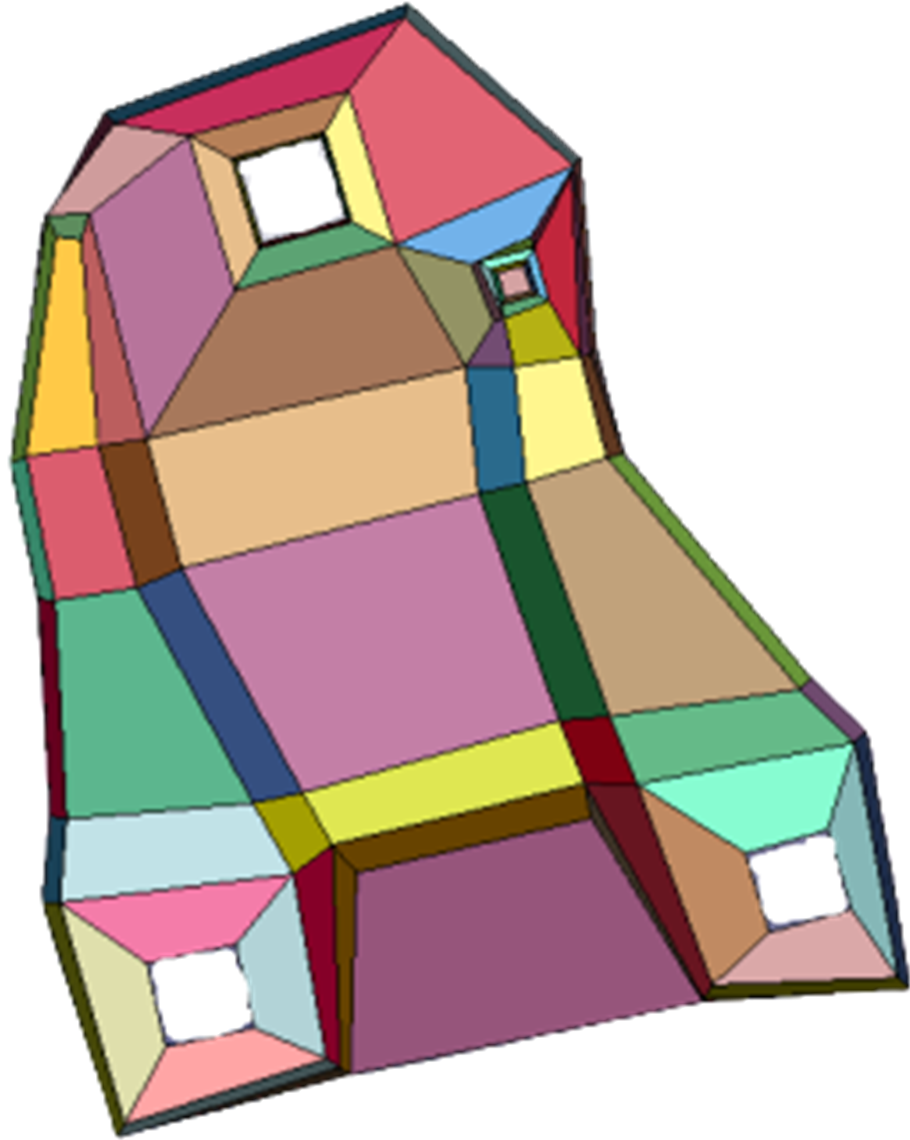}
  & \includegraphics[height=0.37\linewidth]{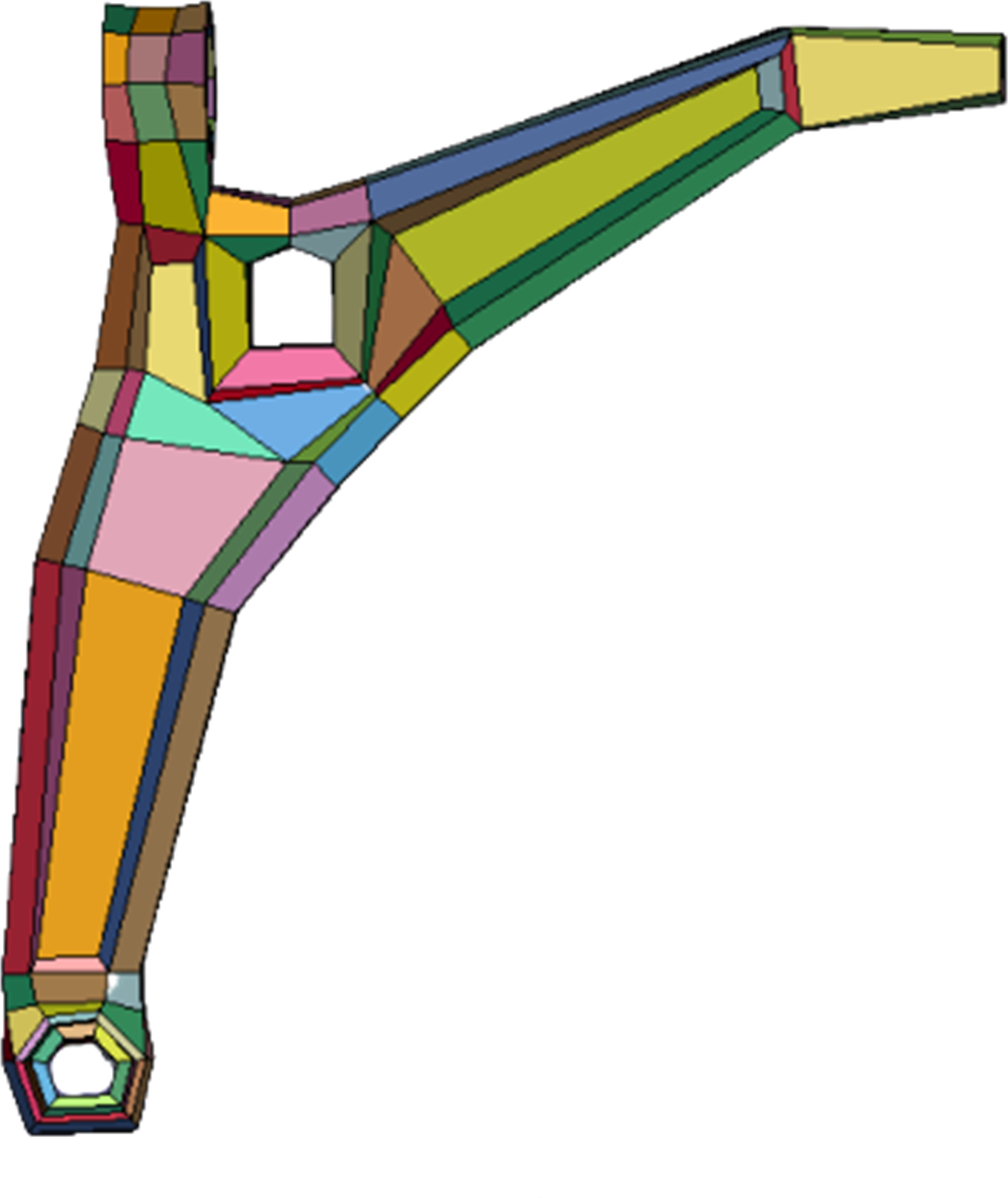}
    & \includegraphics[height=0.37\linewidth]{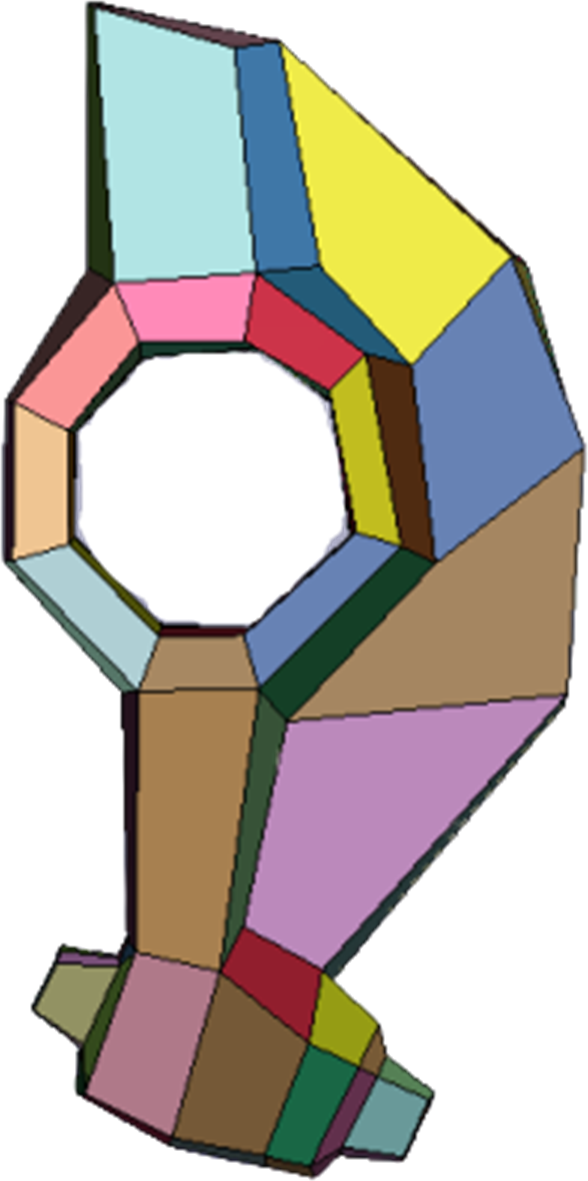} \\
    & (b) &\\
  \includegraphics[height=0.37\linewidth]{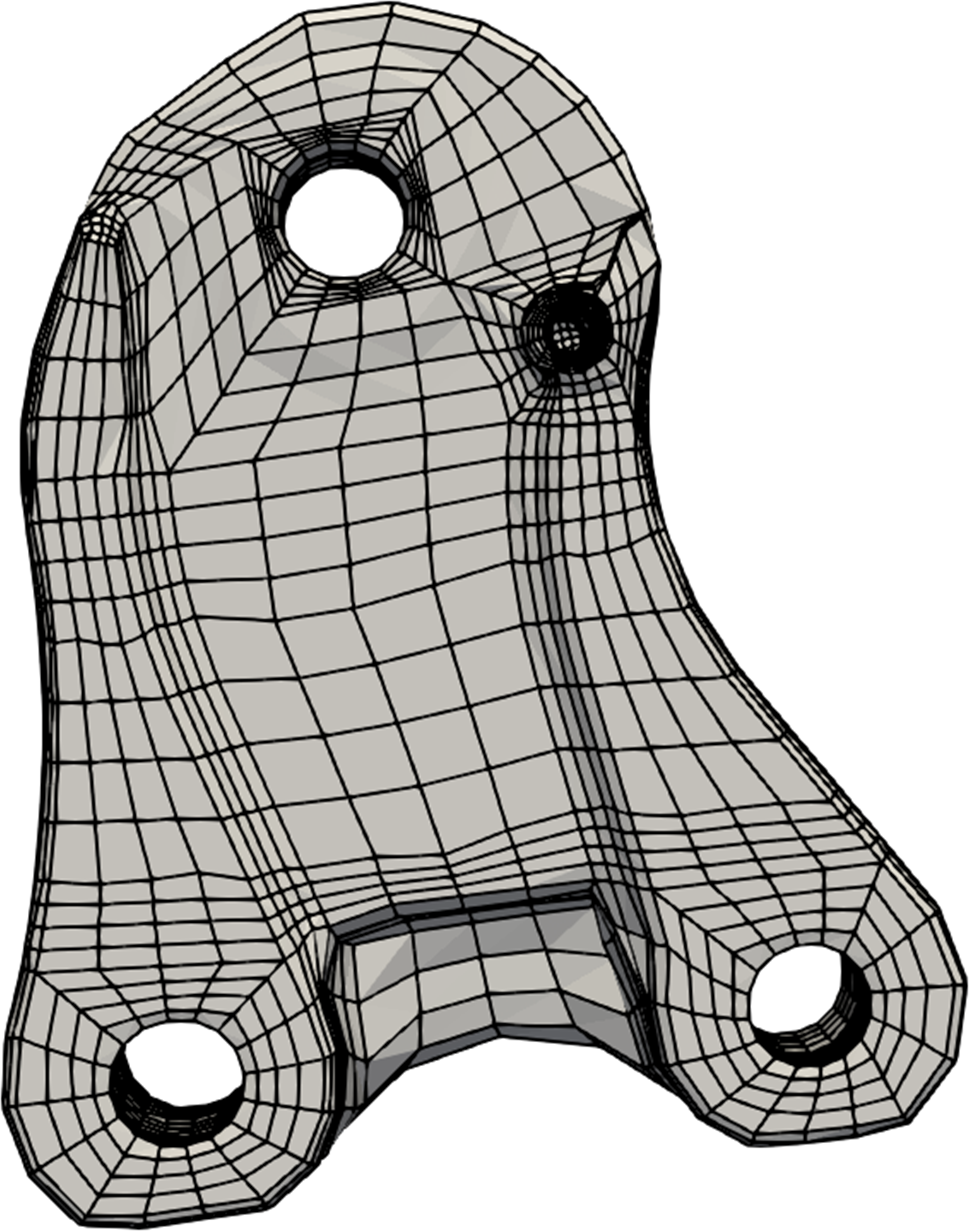}
  & \includegraphics[height=0.37\linewidth]{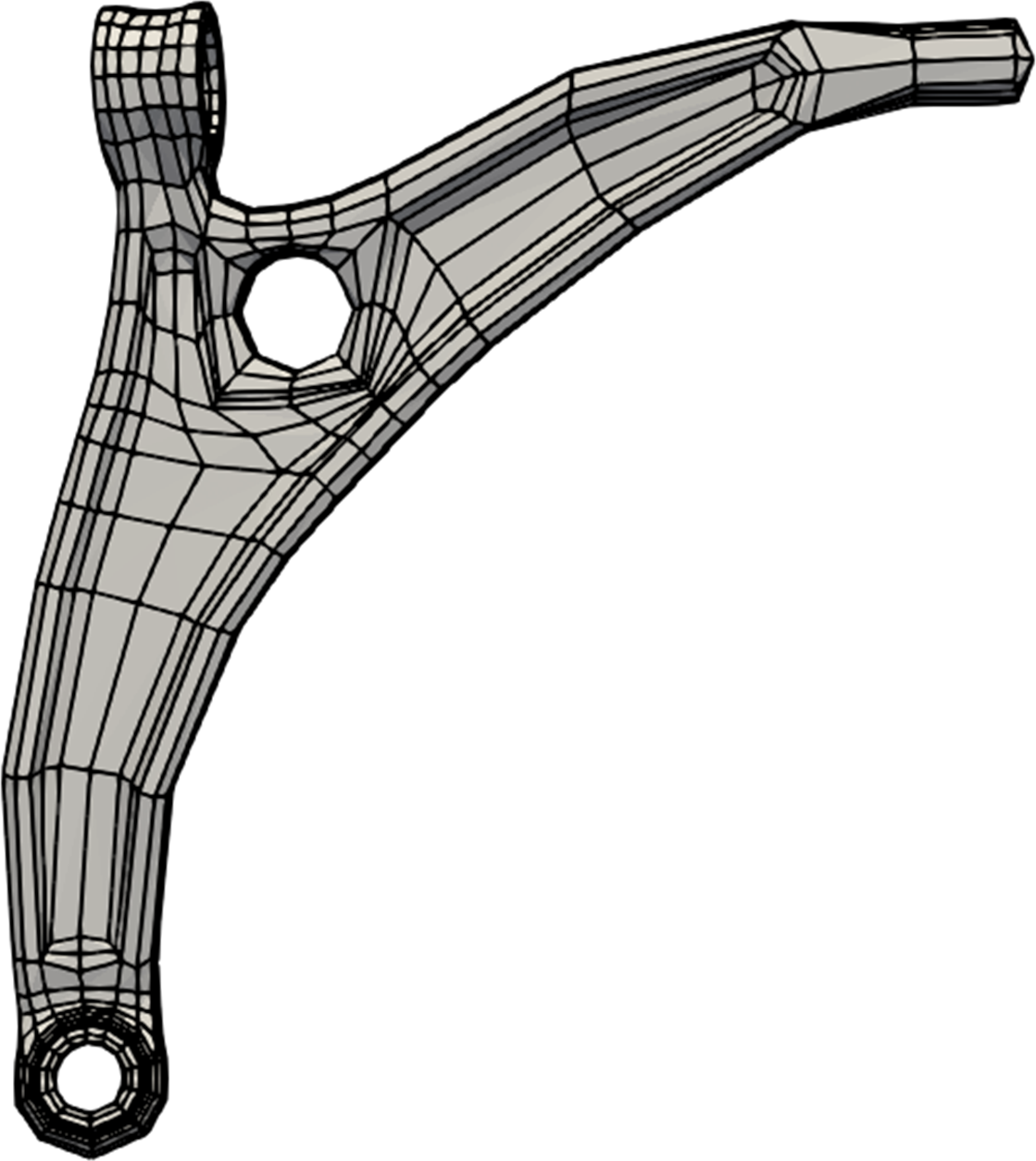}
    & \includegraphics[height=0.37\linewidth]{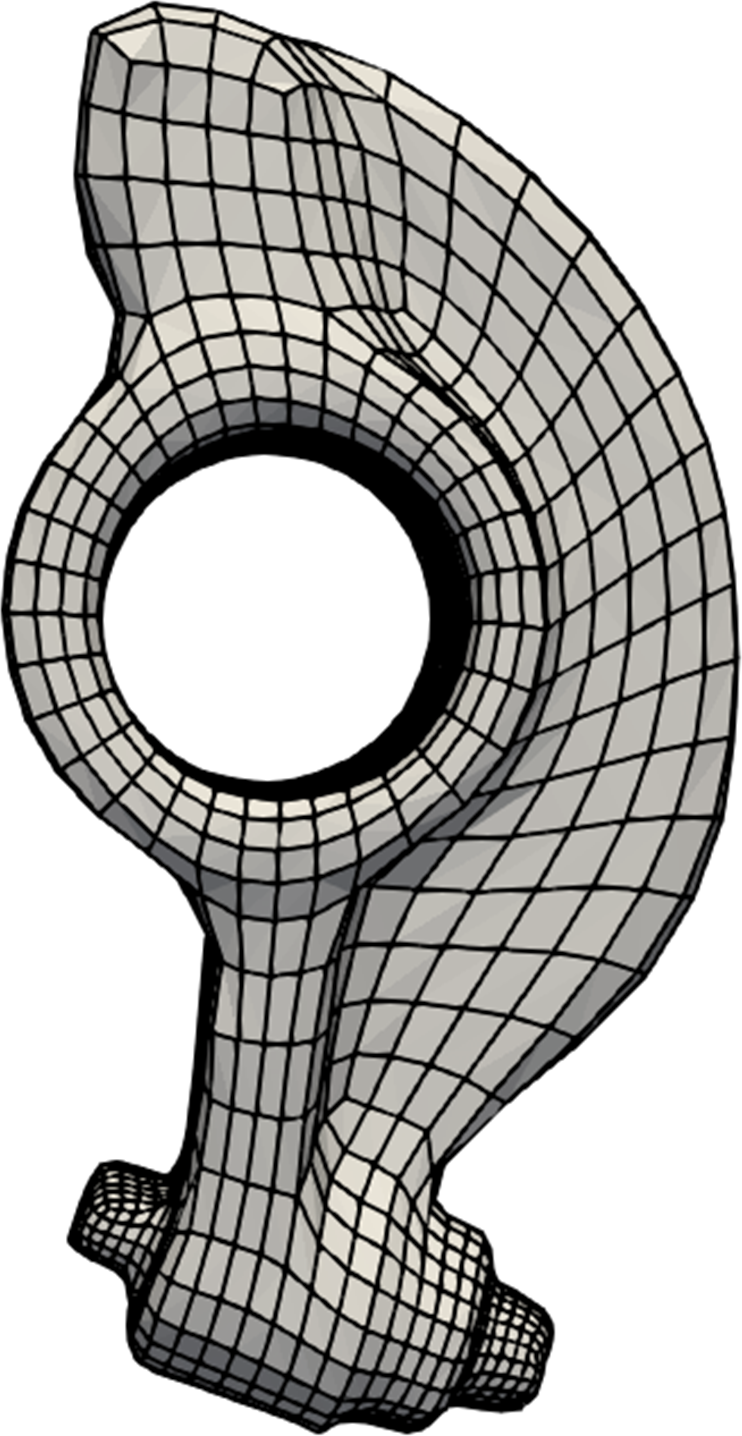}\\
    & (c) &\\
  \includegraphics[height=0.37\linewidth]{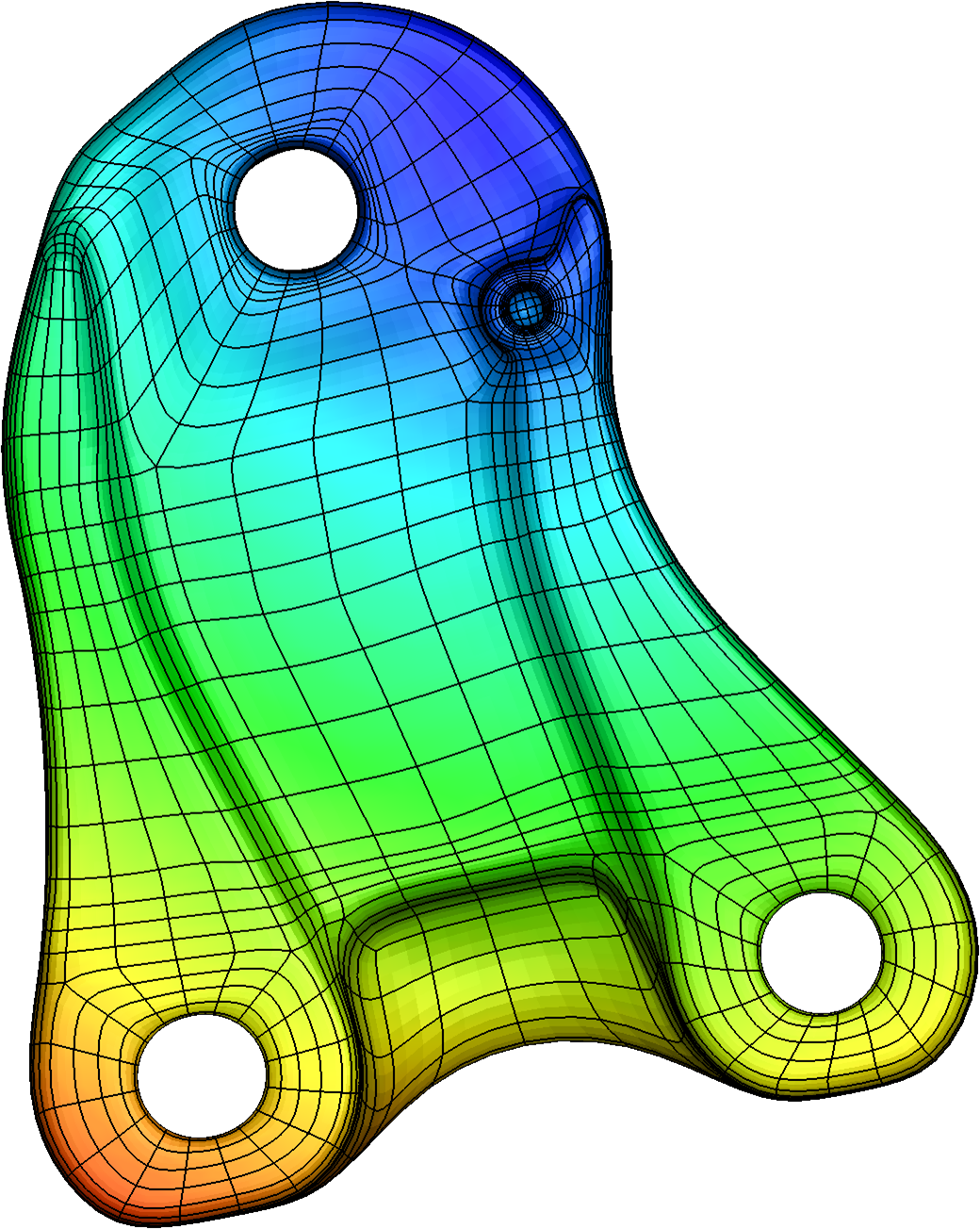}
  & \includegraphics[height=0.37\linewidth]{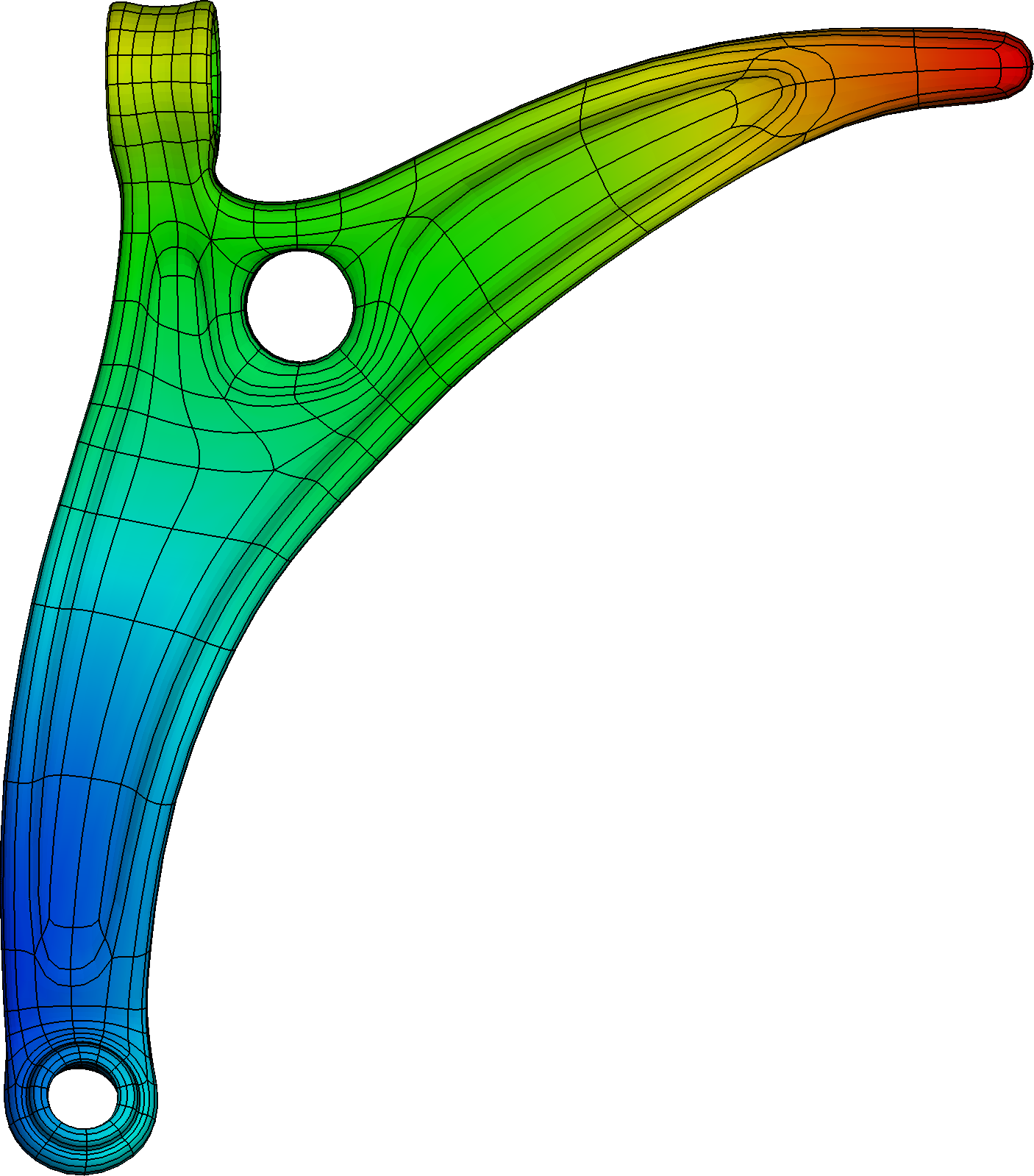}
    & \includegraphics[height=0.37\linewidth]{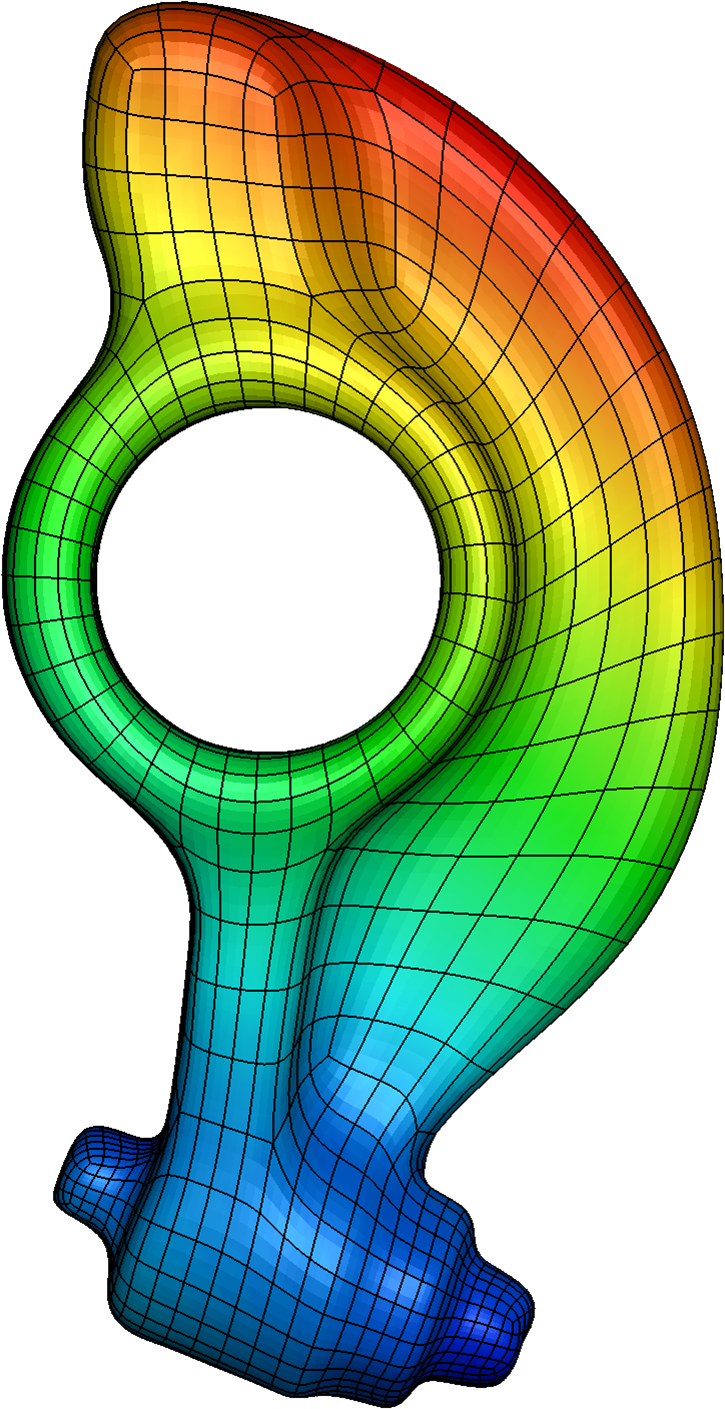}\\
    & (d) &
\end{tabular}
\caption{ Results of engine mount, lower arm, and rockerarm model. (a) Surface
  triangle meshes and segmentation results; (b) Polycube structures;
  (c) All-hex control meshes; (d) Volumetric splines with IGA results of eigenvalue analysis in LS-DYNA. }
    \label{fig:model2}
\end{figure}

\begin{figure}[htp]
\centering
\begin{tabular}{ccc}
  \includegraphics[height=0.3\linewidth]{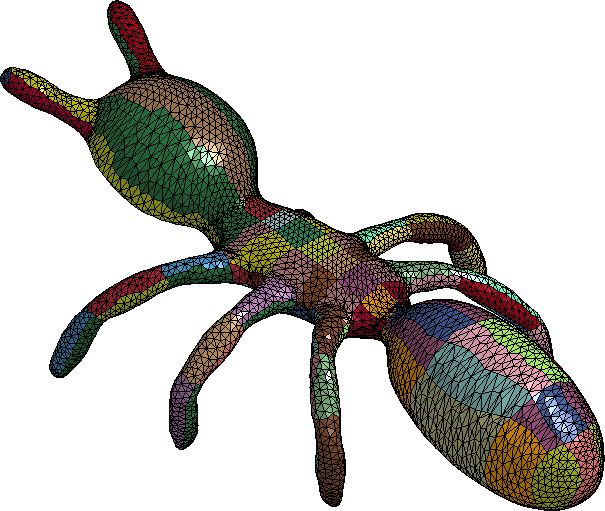}
  & \includegraphics[height=0.3\linewidth]{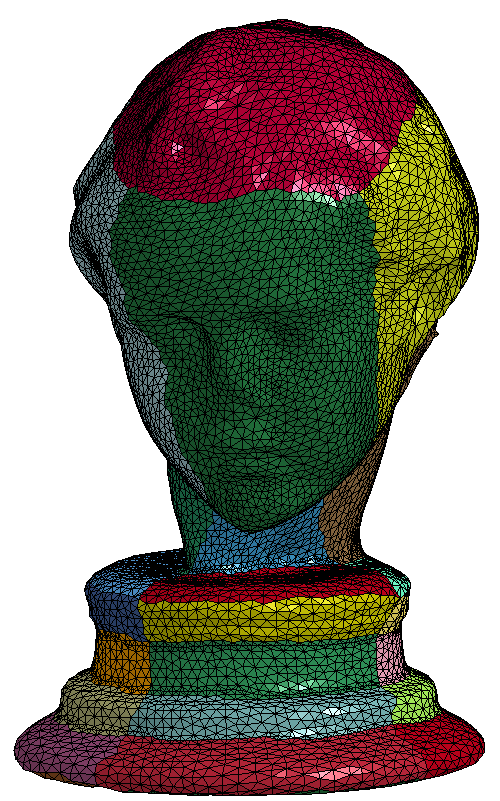}
    & \includegraphics[height=0.3\linewidth]{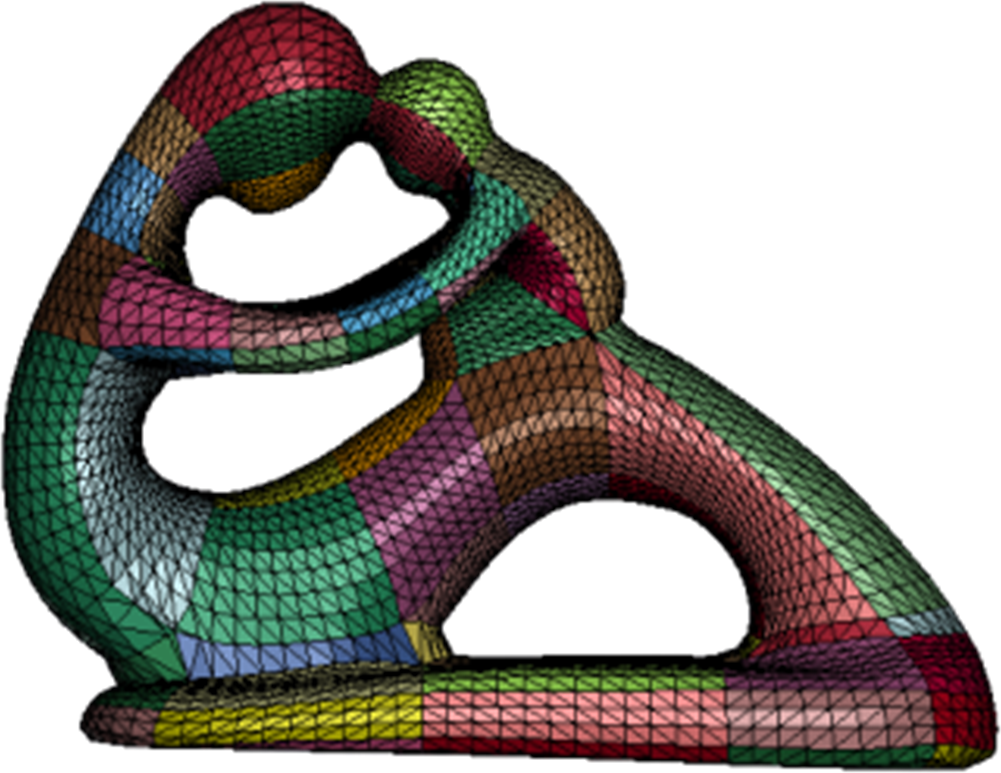}\\
    & (a) &\\
  \includegraphics[height=0.3\linewidth]{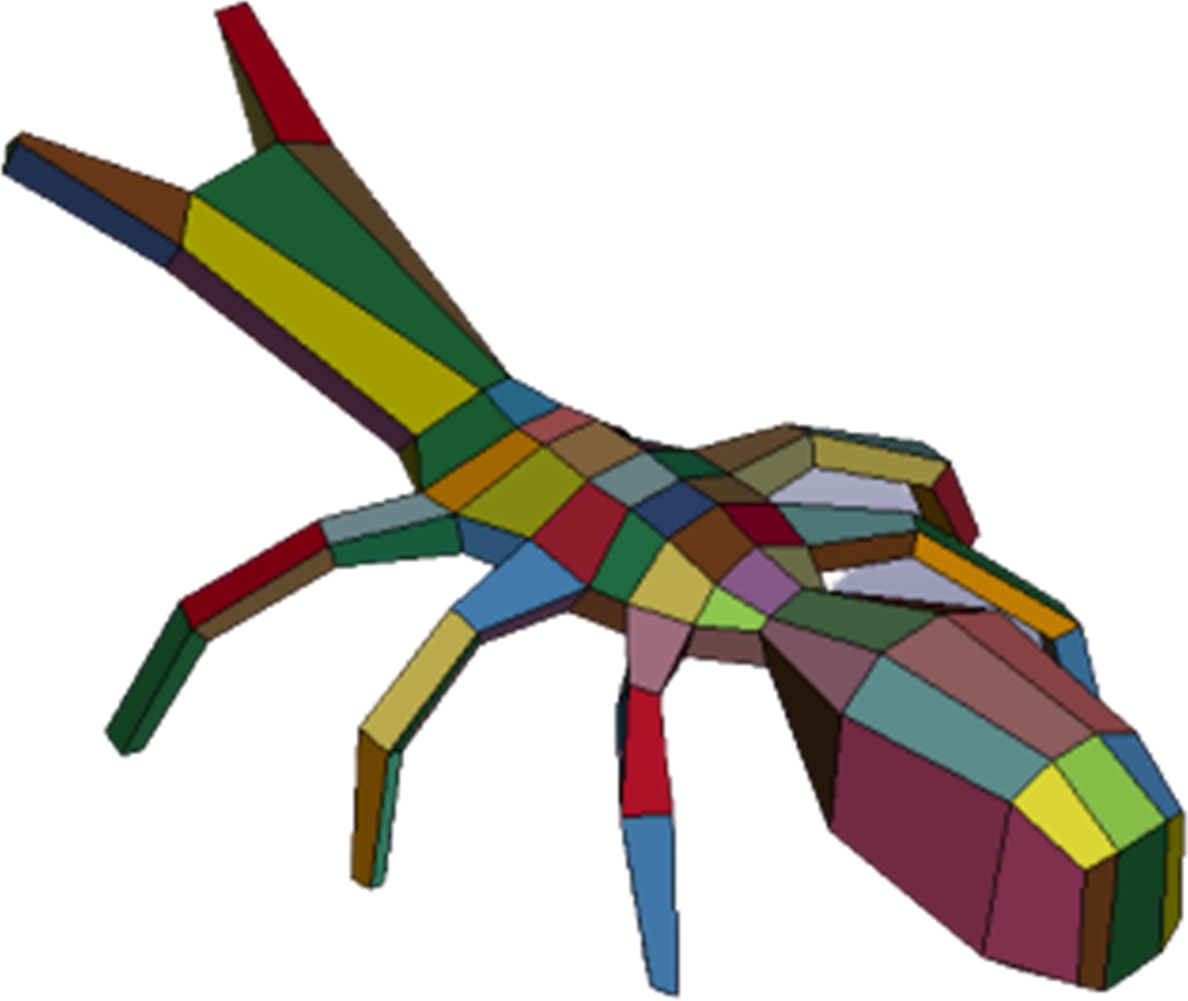}
  & \includegraphics[height=0.3\linewidth]{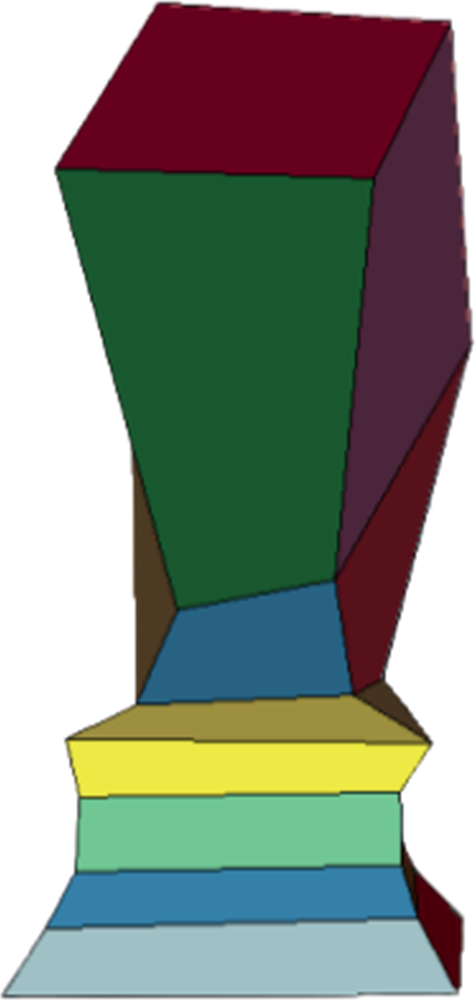}
    & \includegraphics[height=0.3\linewidth]{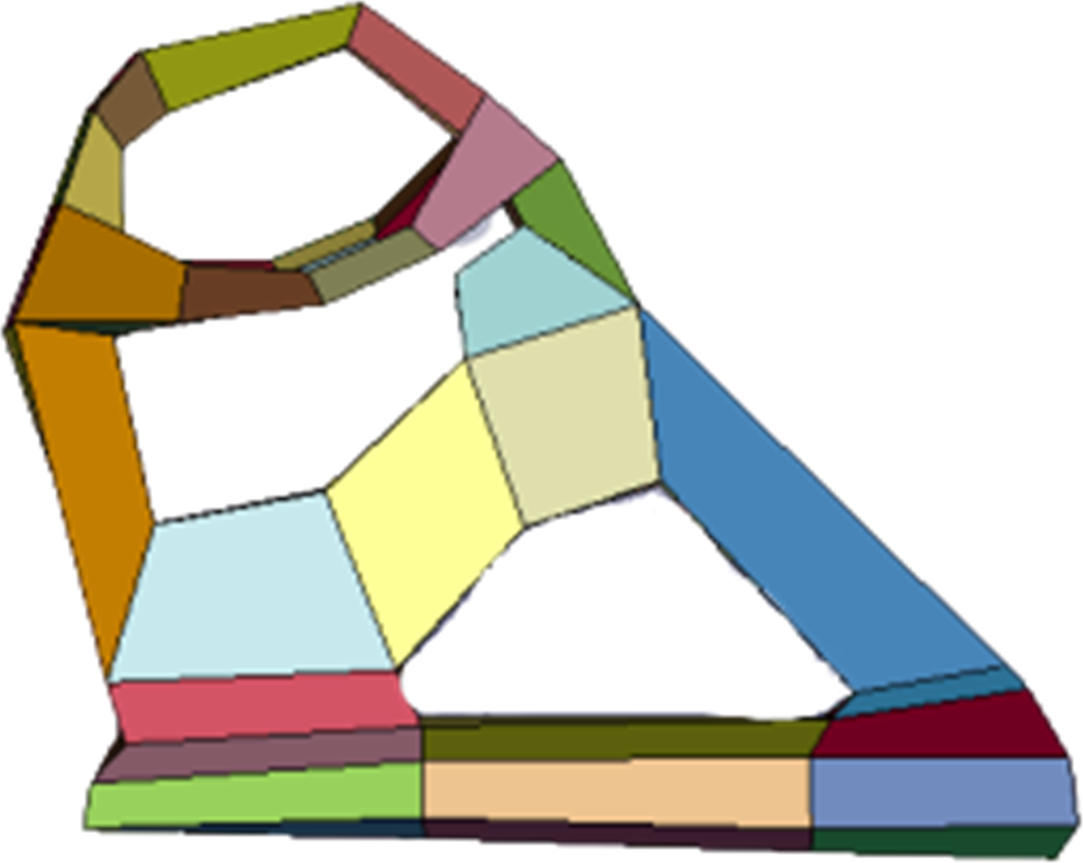}\\
    & (b) &\\
  \includegraphics[height=0.3\linewidth]{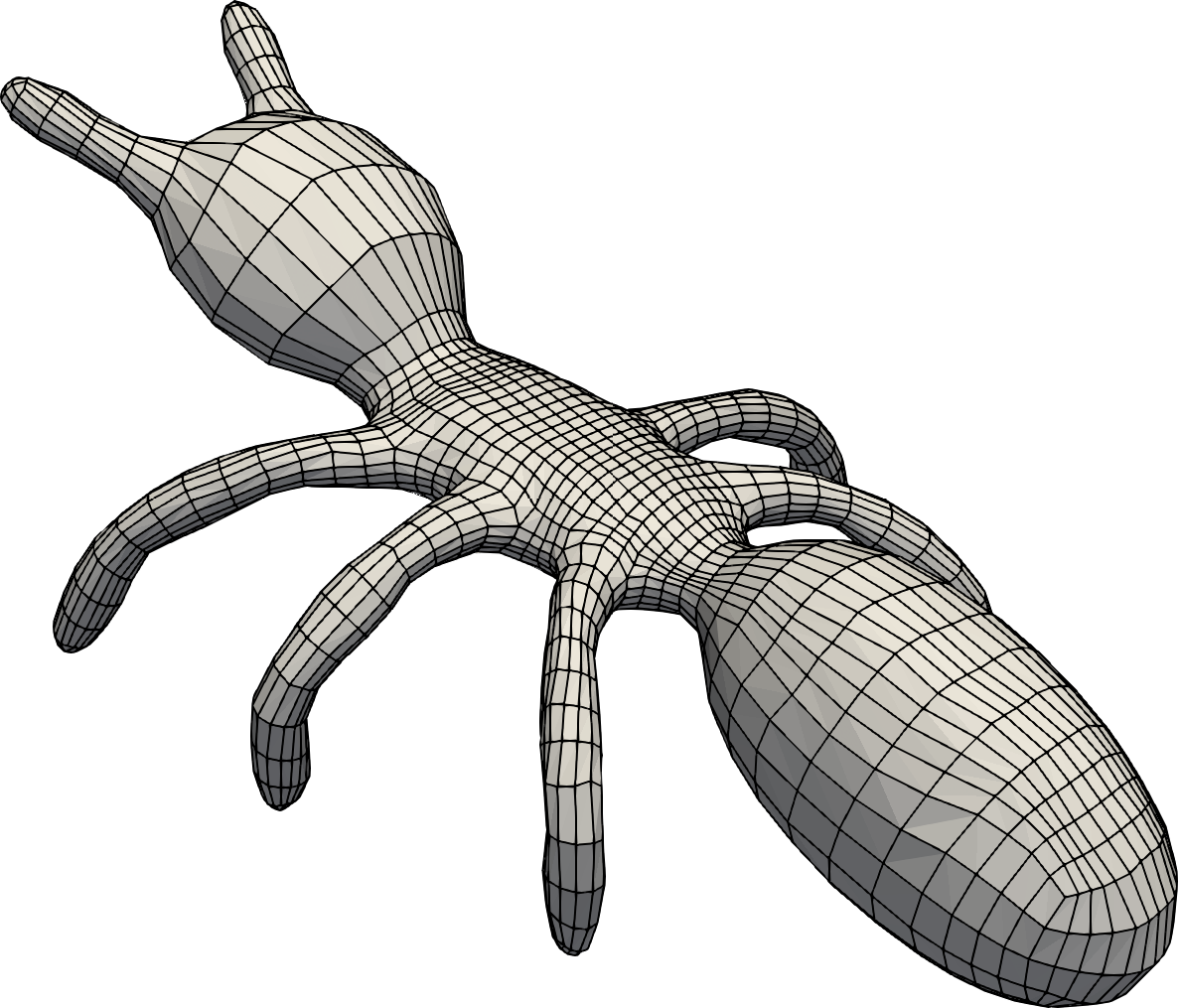}
  & \includegraphics[height=0.3\linewidth]{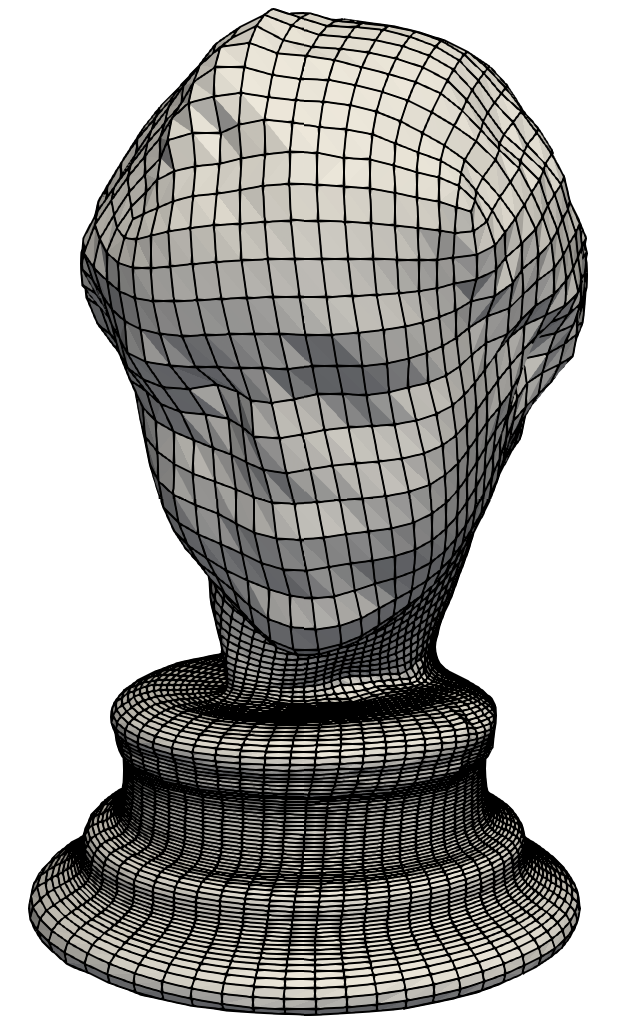}
    & \includegraphics[height=0.3\linewidth]{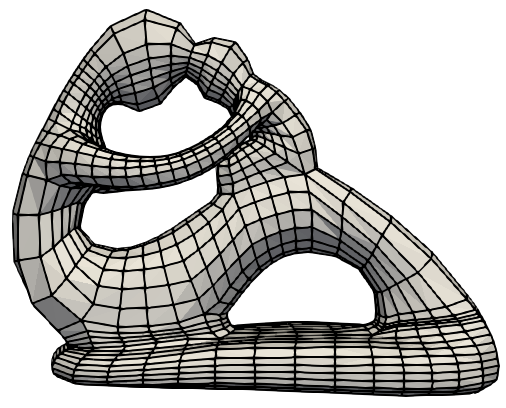}\\
    & (c) &\\
  \includegraphics[height=0.3\linewidth]{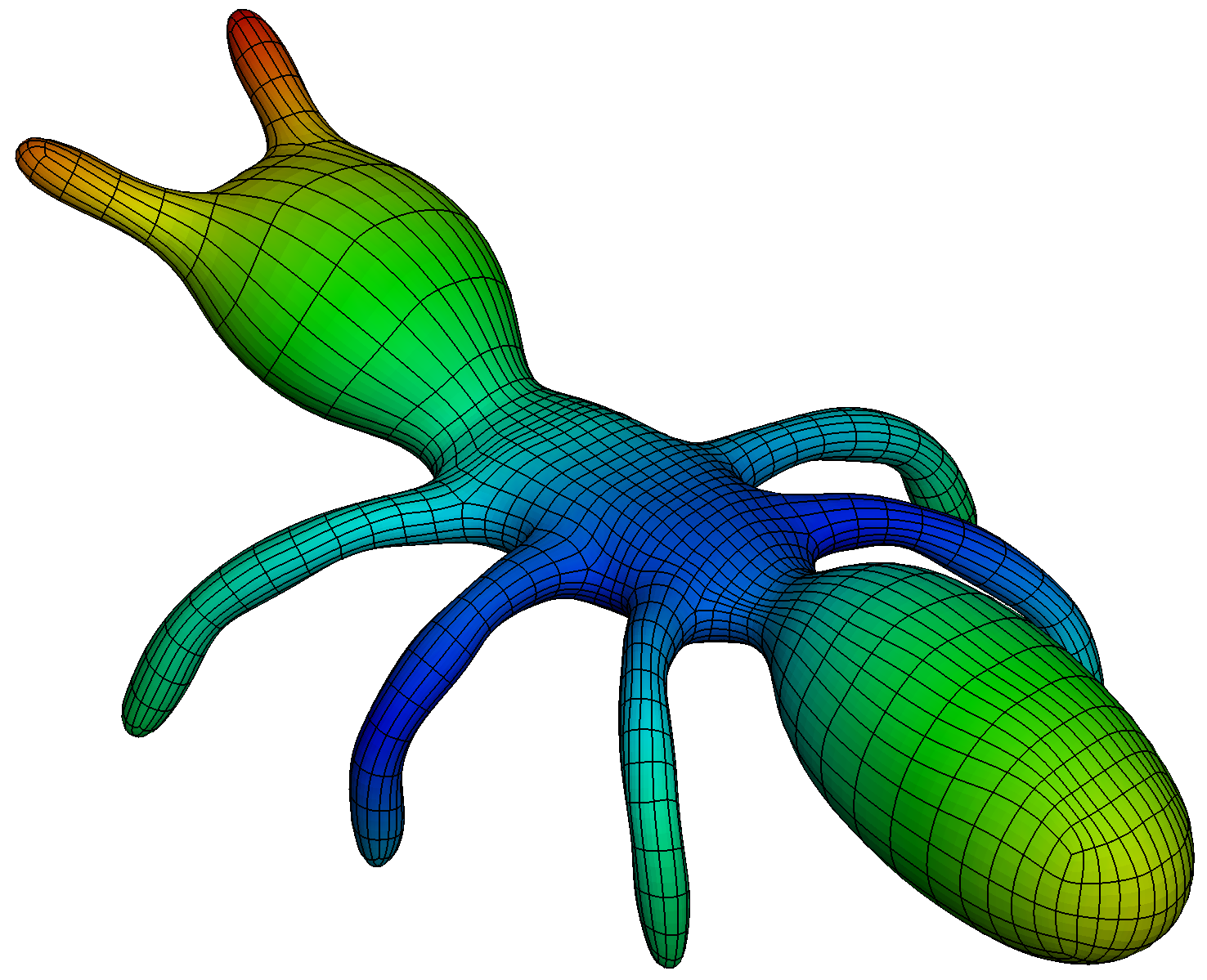}
  & \includegraphics[height=0.3\linewidth]{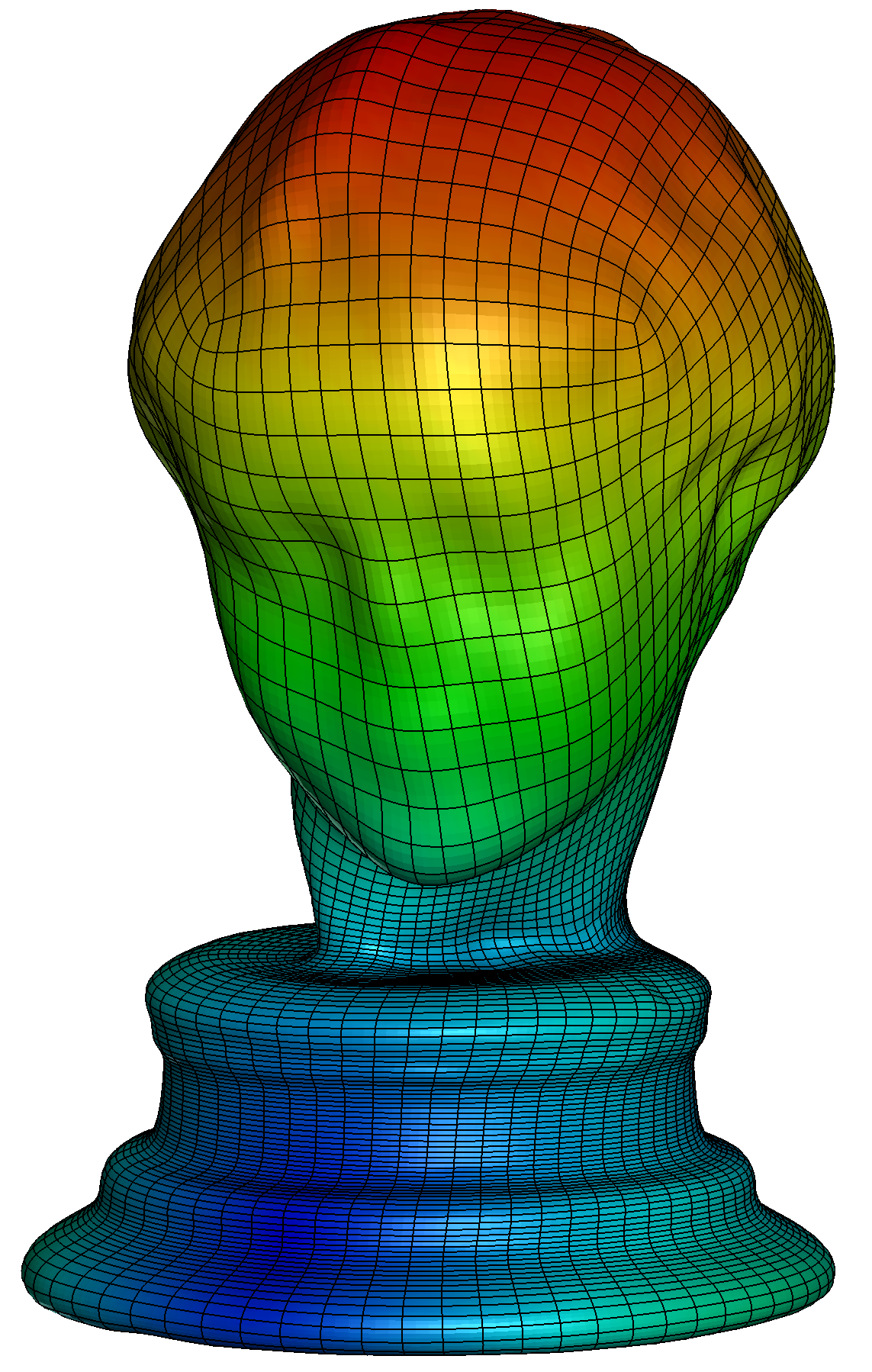}
    & \includegraphics[height=0.3\linewidth]{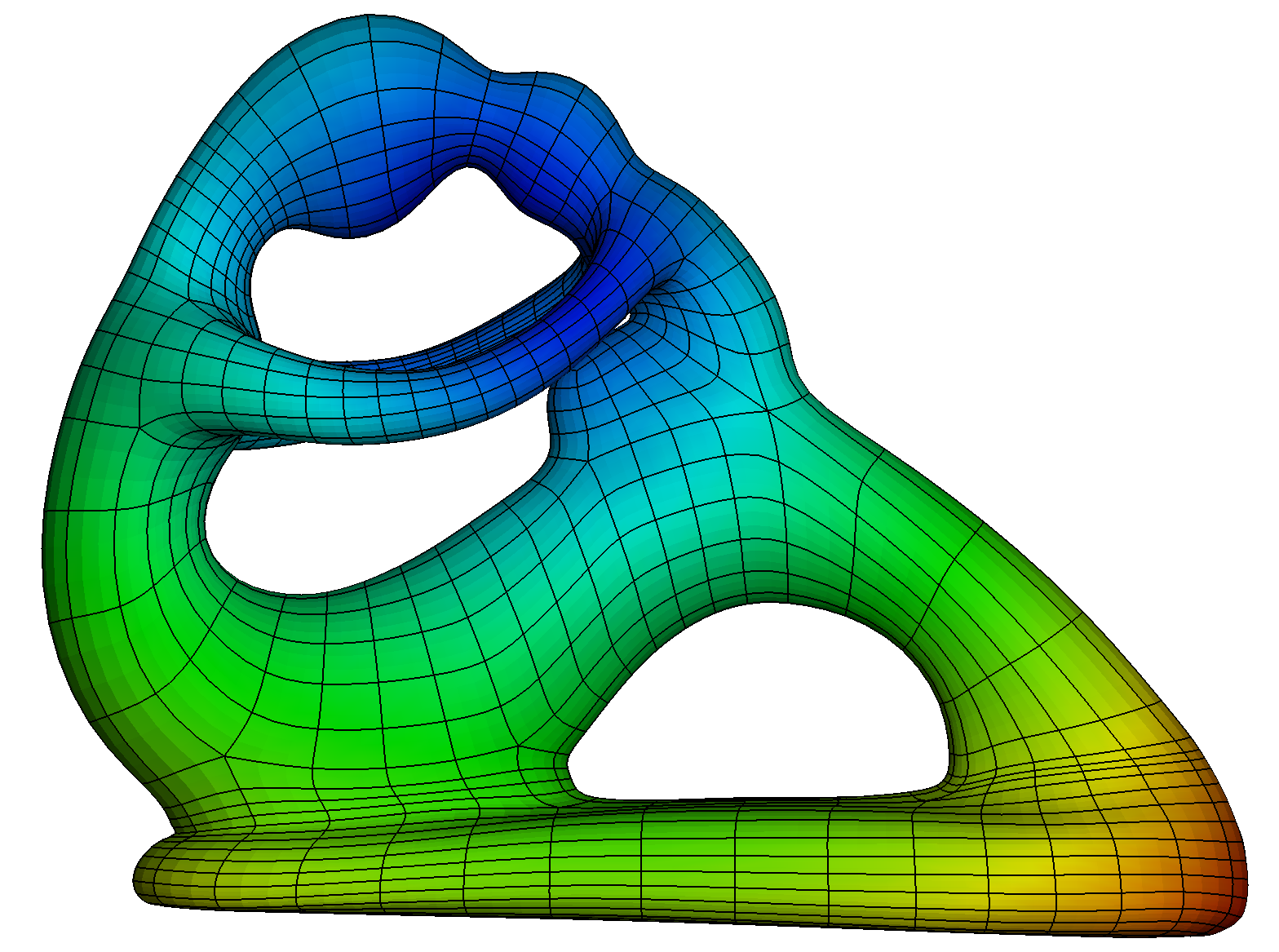}\\
    & (d) &\\
\end{tabular}
\caption{ Results of ant, bust, and fertility model. (a) Surface
  triangle meshes and segmentation results; (b) Polycube structures;
  (c) All-hex control meshes; (d) Volumetric splines with IGA results of eigenvalue analysis in LS-DYNA. }
    \label{fig:model3}
  \end{figure}

\begin{figure}[htp]
\centering
\begin{tabular}{cccc}
  \includegraphics[width=0.24\linewidth]{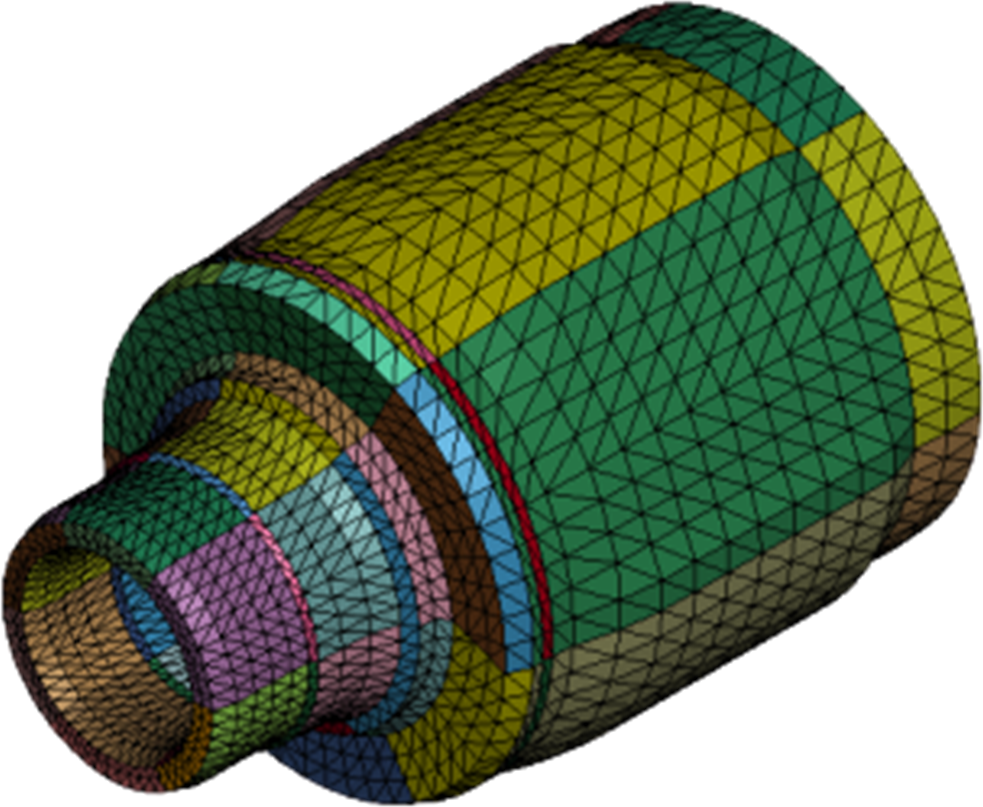}
  & \includegraphics[width=0.24\linewidth]{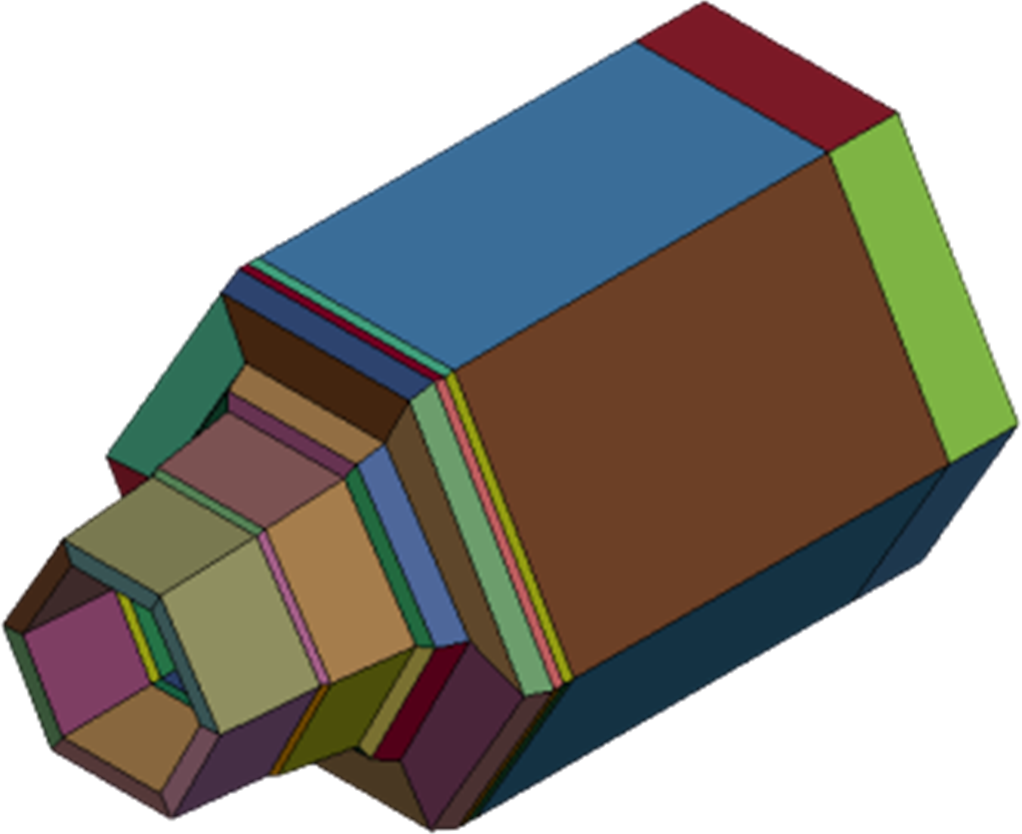}
    &    \includegraphics[width=0.24\linewidth]{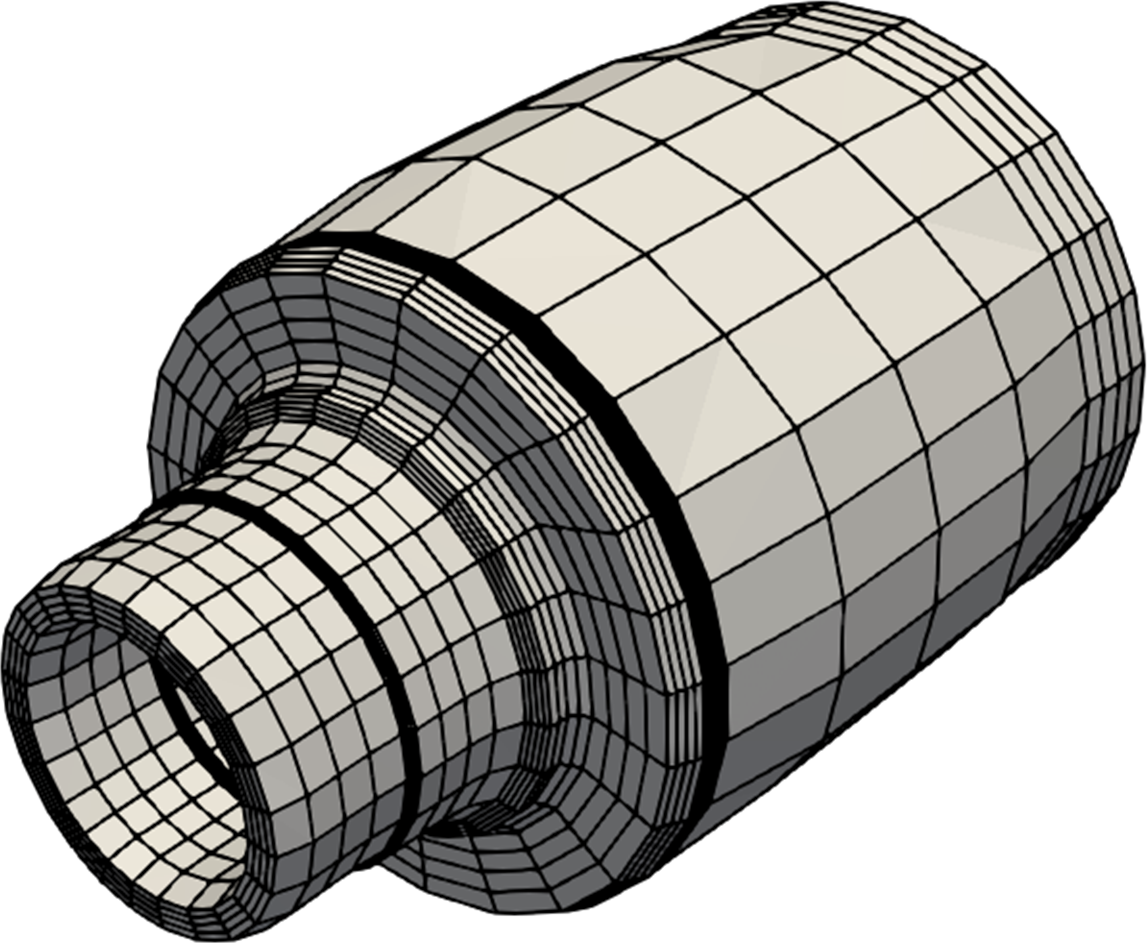}
  & \includegraphics[width=0.24\linewidth]{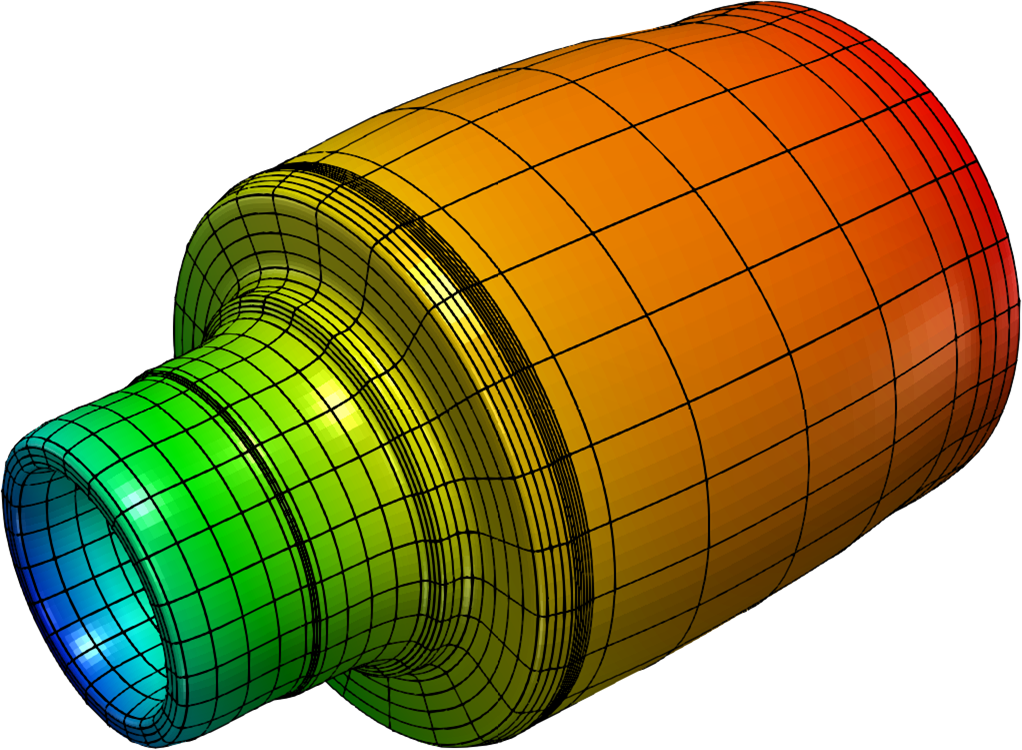}\\
  (a)&(b)&  (c)&(d)\\
\end{tabular}
\caption{ Results of joint model. (a) Surface
  triangle meshes and segmentation results; (b) Polycube structures;
  (c) All-hex control meshes; (d) Volumetric splines with IGA results of solving Poisson equation in LS-DYNA. }
    \label{fig:model4}
\end{figure}

We have applied the software packages to several models and generated
all-hex meshes with good quality. For each model, we show the HBECVT
based segmentation result, further segmentation result, corresponding
polycube structure, and the all-hex mesh. These models include: two
types of mount and hepta models (Fig.~\ref{fig:model1}); engine
mount and lower arm from Honda
Co. along with rockerarm (Fig.~\ref{fig:model2}); ant, bust, and fertility models
(Fig.~\ref{fig:model3}); and the joint model from Honda
Co. (Fig.~\ref{fig:model4}).
Table~\ref{Polycube_Table_1} shows the statistics of all testing
models.  We use the scaled Jacobian to evaluate the quality of all-hex
meshes. From Table~\ref{Polycube_Table_1}, we can observe that the
obtained all-hex meshes have good quality (minimal Jacobian $>$
0.1). Figs.~\ref{fig:model1}-\ref{fig:model4}(a) show HBECVT segmentation
results of testing models, we can observe that the initial
segmentation results generated by the HBECVT do not satisfy the
topological constraints for polycube construction. We need to further
segment each Voronoi region into several patches. The generated polycubes
(Figs.~\ref{fig:model1}-\ref{fig:model4}(b)) align with the given
geometry better, which in turn induces less mesh distortion and yields
a mesh of better quality.

After generating all-hex meshes
(Figs.~\ref{fig:model1}-\ref{fig:model4}(c)), we tested all the models
for IGA by using TH-spline3D. B\'ezier elements are
extracted for the IGA analysis
(Figs.~\ref{fig:model1}-\ref{fig:model4}(d)). For each testing model in Figs.~\ref{fig:model1}-\ref{fig:model3} , we use LS-DYNA to perform eigenvalue analysis and show the first mode result. For the testing model in Fig.~\ref{fig:model4}, we show the result of solving a Poisson problem in LS-DYNA. From the results we can observe that our algorithm
yields valid TH-spline3D for IGA applications in LS-DYNA.

\begin{table}[htp]
\caption{Statistics of all the tested models.}
\label{Polycube_Table_1}
\centering
\setlength{\tabcolsep}{3.7pt}
\scriptsize
\begin{tabular}{|c|c|ccc|}
\hline
 Model &Input triangle mesh &Octree &Output hex mesh&Jacobian\\
       &(vertices, elements)&levels&(vertices, elements)&worst\\
  \hline
   rod (Fig.~\ref{fig:Pipeline_Structure})&(2,238, 4,480)&2&(1,815, 1,280)&0.32\\
  mount1 (Fig.~\ref{fig:model1})&(886, 1,782)&2&(5,849, 4,480)&0.16\\
   mount2 (Fig.~\ref{fig:model1})&(895, 1,802)&2&(7,945, 6,208)&0.14\\
   hepta (Fig.~\ref{fig:model1})&(676, 1,348)&1&(1,259, 944)&0.24\\
  engine mount (Fig.~\ref{fig:model2})&(57,487, 114,982)&2&(7,338, 5,888)&0.11\\
  lower arm (Fig.~\ref{fig:model2})&(165,201, 330,410)&1&(1,996, 1,328)&0.10\\
    rockerarm (Fig.~\ref{fig:model2})&(11,655, 23,310&2&(6,880, 5,696)&0.10\\
  ant (Fig.~\ref{fig:model3})&(7,216, 14,428)&2&(7,711, 6,176)&0.17\\
  bust (Fig.~\ref{fig:model3})&(12,596, 25,188)&4&(103,299, 98,816)&0.11\\
  fertility (Fig.~\ref{fig:model3})&(6,622, 13,256)&2&(4,983, 4,000)&0.20\\
  joint (Fig.~\ref{fig:model4})&(3,806, 7,612)&2&(7,512, 6,144)&0.10\\
  \hline
\end{tabular}
\end{table}


\section{Conclusion and future work}

In this paper, we present two software packages (HexGen and
Hex2Spline) for IGA applications in LS-DYNA. The main goal of HexGen
and Hex2Spline is to make our pipeline accessible to industrial and
academic communities who are interested in real-world engineering
applications. The all-hex mesh generation program (HexGen) can
generate all-hex meshes. It consists of four executable files, namely
segmentation module (Segmentation.exe), polycube construction module
(Polycube.exe), all-hex mesh generation module (ParametricMapping.exe)
and quality improvement module (Quality.exe). The volumetric spline
construction program (Hex2Spline.exe) is developed based on the spline
construction method in~\cite{wei17a}. Users can generate a volumetric
spline model given any unstructured hex mesh and output a BEXT file to
perform IGA in LS-DYNA. Both programs are compiled in executable files
and can be easily run in the Command Prompt (cmd) in platform. The rod
model is used to explain how to use these two programs in detail. We
also tested our software package using several other models.

In conclusion, we integrate our hex mesh generation and volumetric spline construction techniques and develop a software platform to create IGA models for LS-DYNA. Our software also has limitations that we will address in our future work. First, the hex mesh generation module is semi-automatic and needs user intervention to create polycube structure. We will improve the underneath algorithm and make polycube construction more automatic. In addition, our software cannot guarantee to generate good quality hex mesh for complex geometry. Therefore, in the future we will expand our software package to use hex-dominant meshing methods to create hybrid meshes for IGA applications.

\section*{Acknowledgment}
Y. Yu, A. Li, J. Liu and Y. Zhang were supported in part by Honda
funds. X. Wei is partially supported by the ERC AdG project CHANGE
n. 694515, as well as the Swiss National Science Foundation project
HOGAEMS n.200021\_188589. We also acknowledge the open source
scientific library Eigen and its developers. The authors would like to
thank Kenji Takada for providing the CAD geometries of engine mount,
lower arm and joint. The authors would also like to thank
Attila P. Nagy and David J. Benson for various fruitful discussions
about the commercial software LS-DYNA.

\renewcommand{\lstlistingname}{List}
\setcounter{section}{0}
\renewcommand{\thesection}{A\arabic{section}}
\setcounter{table}{0}
\renewcommand{\thetable}{A\arabic{table}}
\setcounter{figure}{0}
\renewcommand{\thefigure}{A\arabic{figure}}
\section*{\Large Appendix}
\section{Input text file to correct segmentation result from Segmentation.exe}
In this section, we describe the data format of the input text file used in Segmentation.exe to correct the segmentation result. One can prepare this file to move elements on the wrong patch to the desired patch. In this text file, each row has two values to define this modification of one element (see List \ref{lst:1}). The first value indicates the element index in the triangular mesh and the second value is the desired patch index. Segmentation.exe can read this file through option \textbf{-m} to improve the segmentation
result.
\begin{lstlisting}[language=Python, caption=Snippets of the input text
file,label={lst:1}]
     355       1
     356       1
     361       1
     362       1
     365       1
     366       1
     369       1
     370       1
     495       6
     496       6
     499       6
     500       6
\end{lstlisting}
\section{Further segmentation in LS-PrePost}
\begin{figure}[!htb]
    \centering
    \includegraphics[width = 0.8\linewidth]{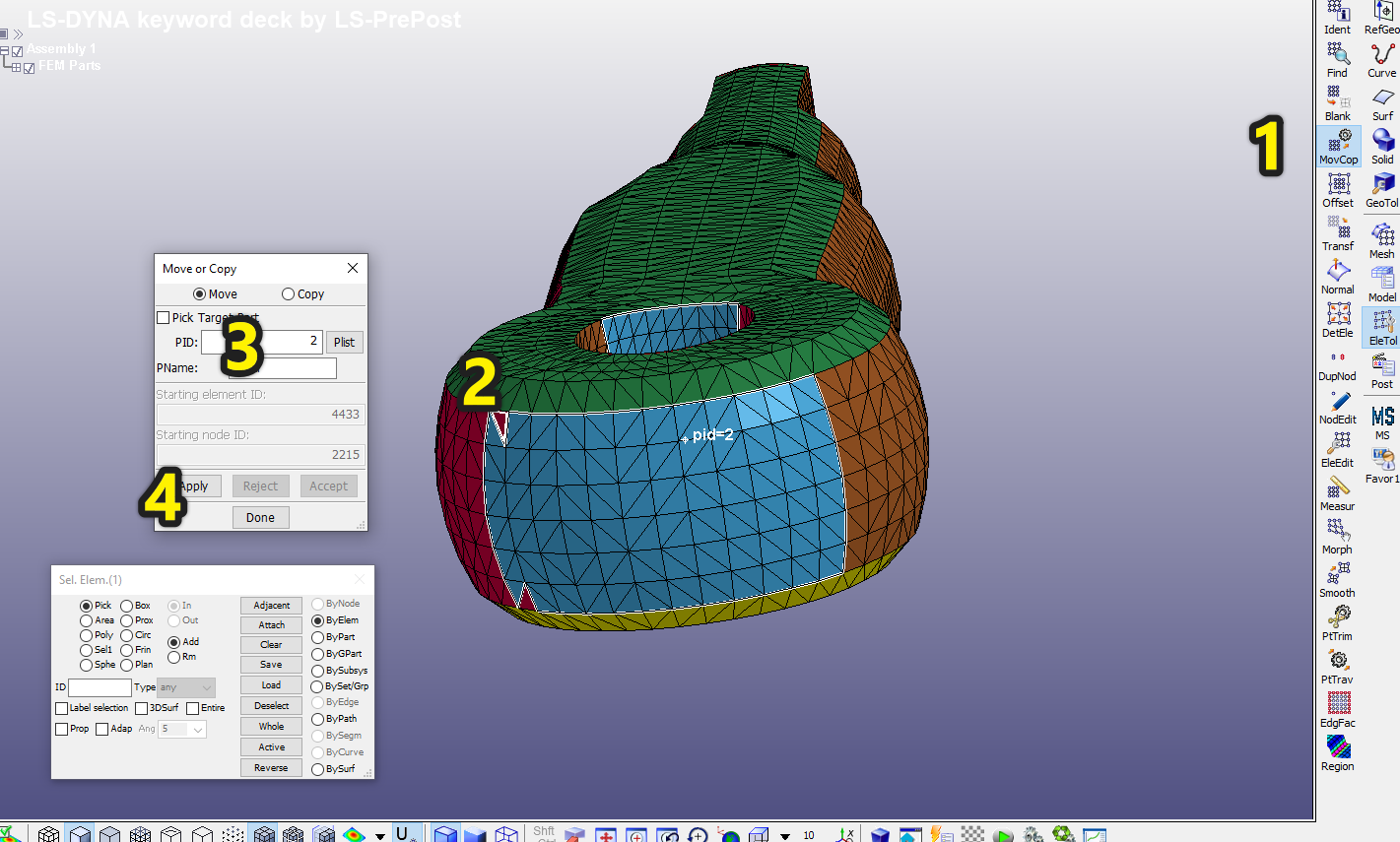}
    \caption{The detailed operations for further segmentation.}
    \label{fig:further_segmentation}
\end{figure}

In this section, we introduce how to perform further segmentation in LS-PrePost and obtain an admissible segmentation for polycube construction. It mainly involves reassigning elements to different patches and there are four steps to achieve this (see Fig.~\ref{fig:further_segmentation}): 1) click \textbf{move/copy} tab; 2) select elements; 3) reassign the patch ID; and 4) click the \textbf{Apply} button to finish.

\section{Building the interior connectivity of polycube structure in LS-PrePost}
\begin{figure}[!htb]
    \centering
    \includegraphics[width = 0.8\linewidth]{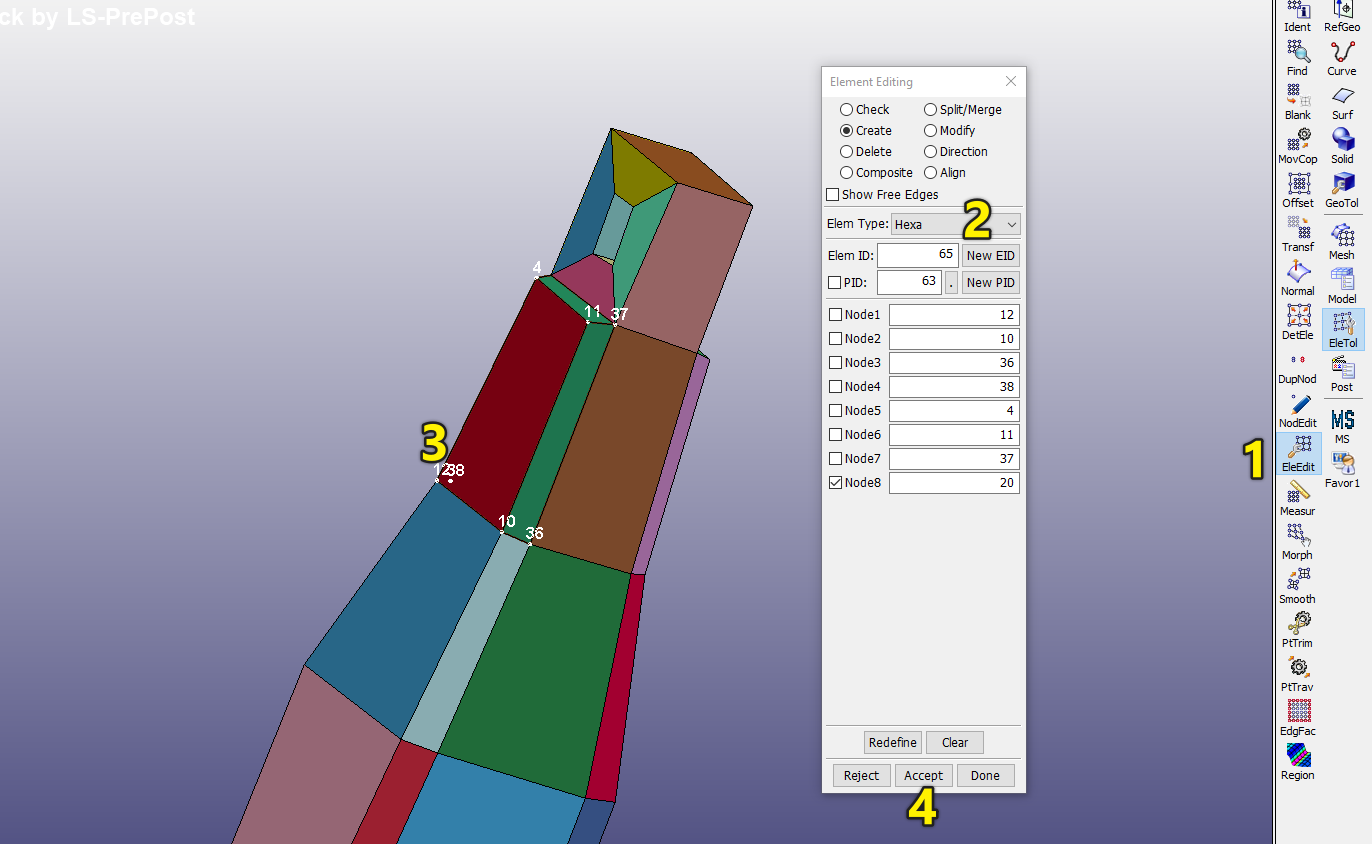}
    \caption{The detailed operations to build the interior connectivity of polycube structure.}
    \label{fig:building_interior}
\end{figure}

In this section, we use the rod model to introduce how to build the interior connectivity of polycube structure in LS-PrePost. There are four steps to create one cubic region ( see Fig.~\ref{fig:building_interior}): 1) click \textbf{EleEdt} tab; 2) select \textbf{Elem Type} as Hexa; 3) select eight nodes to define cubic regions, you can also check the selection on the float box; and 4) click the \textbf{Accept} button to finish.
By repeating the same operation, we generate the polycube structure with multiple cubic regions to split the volumetric domain of the geometry.

\section{Output text file from Polycube.exe}
In CLI program (PolyCube.exe), we output a .k file which contains the corners and their connectivity for the
boundary surface of the polycube. It can be directly opened by LS-PrePost. If one intends to use other software to build a polycube structure, we also provide option to output the corners, edges, and faces of the polycube in three separate text files (see Lists \ref{lst:2}-\ref{lst:4}).
In the corner file, each row depicts the associated vertex
($v_i$) information. The first value indicates the index of the vertex in the triangles mesh, the last three values are its x, y, z coordinates
($x_i, y_i, z_i$).
In the edge file, each row uses the indices of two corners to define the edge between them. These indices should agree with the corner file. The face file stores the information of boundary faces on polycube structure. Each row contains four vertex indices in counter-clockwise order to define the connectivity of one face.

\begin{lstlisting}[language=Python, caption=Snippets of the corner
file,label={lst:2}]
21 4.06622 0.0052 2.31336
143 16.4588 0.7317 2.29912
153 4.0543 3.3803 2.31406
371 16.4604 2.67402 4.12666
391 13.5206 0.65268 2.07416
\end{lstlisting}

\begin{lstlisting}[language=Python, caption=Snippets of the edge
file,label={lst:3}]
71,446
436,446
436,442
371,442
1464,1639
1601,1639
\end{lstlisting}

\begin{lstlisting}[language=Python, caption=Snippets of the face
file,label={lst:4}]
71,446,436,442
1464,1639,1601,1503
1439,1664,1639,1464
1744,1917,1877,1784
1246,1784,1877,153
\end{lstlisting}

\section{Sharp feature file for Quality.exe and Hex2Spline.exe}
\begin{figure}[!htb]
    \centering
    \includegraphics[width = \linewidth]{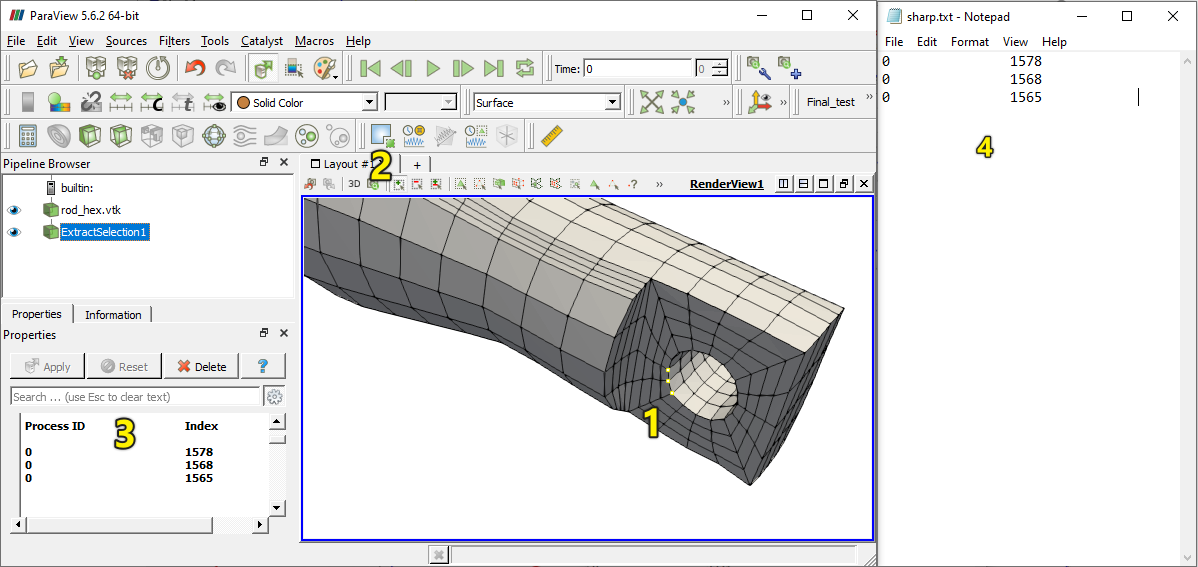}
    \caption{The detailed operations to create user-defined sharp features.}
    \label{fig:sharp_feature_file}
\end{figure}

This section describes how to manually define sharp features for Quality.exe and Hex2Spline.exe. One can prepare an input file including the sharp feature information with the help of Paraview. There are four steps (see Fig.~\ref{fig:sharp_feature_file}): 1) select the points along the sharp feature; 2) use \textbf{Extract selection} function to extract the indices of these nodes; 3) check the selected points information under "Properties"; and 4) copy the index to "sharp.txt" file and use it as the input sharp feature file for Quality.exe and Hex2Spline.exe.

\section{Input BEXT file for LS-DYNA}
\begin{figure}[!htb]
    \centering
    \includegraphics[width = \linewidth]{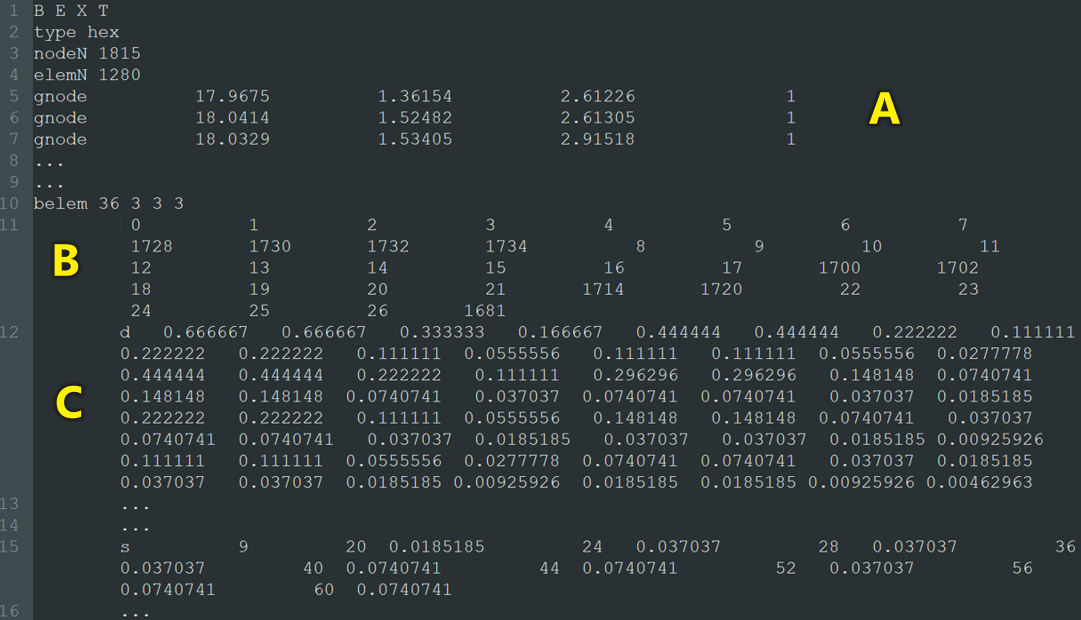}
    \caption{Snippets of the BEXT file.}
    \label{fig:bext_file}
\end{figure}

In this section, we describe the data format of BEXT file for IGA in LS-DYNA. The BEXT file consists of two parts to store the spline information: 1) control point (Fig.~\ref{fig:bext_file}A); and 2) B\'{e}zier element including the indices of control points supported by this B\'{e}zier element (Fig.~\ref{fig:bext_file}B) and the B\'{e}zier extraction matrix (Fig.~\ref{fig:bext_file}C). The B\'{e}zier extraction matrix is output row by row and the format of each row depends on the number of non-zero values in this row. If this row has less than 20 non-zeros, a sparse format is used to store their column indices and values; otherwise, a dense format is used to store all values in this row. To distinguish between two formats, a sparse row begins with \textbf{s} while a dense row begins with \textbf{d}.

\clearpage

\bibliography{INdAM_Software_Hex} \bibliographystyle{spmpsci}

\end{document}